\documentclass[useAMS,usenatbib]{mn2e}
\usepackage{myaasmacros,graphicx}

\def\ltsima{$\; \buildrel < \over \sim \;$}
\def\simlt{\lower.5ex\hbox{\ltsima}}   
\def\gtsima{$\; \buildrel > \over \sim \;$}
\def\simgt{\lower.5ex\hbox{\gtsima}}
\newcommand\bcite[1]{\citeauthor{#1} \citeyear{#1}}
\def\meanFeH{\left<[\mathrm{Fe}/\mathrm{H}]\right>}

\def\rhothick{\rho_\mathrm{THICK}}
\def\rhohalo{\rho_\mathrm{HALO}}
\def\rhothin{\rho_\mathrm{THIN}}
\def\rhodd{\rho_\mathrm{DDISC}}
\def\rhoh{\rho_\mathrm{HALO}}

\def\LMCten{LMC-10$^\mathrm{o}$}
\def\LMCtwe{LMC-20$^\mathrm{o}$}
\def\LMCfor{LMC-40$^\mathrm{o}$}
\def\LMCsix{LMC-60$^\mathrm{o}$}
\def\LLMC{LLMC-10$^\mathrm{o}$}
\def\LLMCMWB{LLMC-10$^\mathrm{o}$-MWB}
\def\LLMCBMWB{LLMCB-10$^\mathrm{o}$-MWB}
\def\LMCRet{LMC-Ret-10$^\mathrm{o}$}
\def\Fornax{Fornax-10$^\mathrm{o}$}
\def\LMCabig{LMC-10$^\mathrm{o}$-abig}
\def\LMCeless{LMC-10$^\mathrm{o}$-eless}

\def\vmaxd{v_\mathrm{max}}
\def\vmaxt{v_\mathrm{max}^t}
\def\vmaxdhost{v_\mathrm{max}^\mathrm{host}}

\title[Thin, thick and dark discs]{Thin, thick and dark discs in $\Lambda$CDM}

\author[Read et al.]{J. I. Read\thanks{Email: justin@physik.unizh.ch}$^1$, G. Lake$^1$, O. Agertz$^1$  \& Victor P. Debattista$^{2}$\\
$^1$Institute of Theoretical Physics, University of Z\"urich,  
Winterthurerstrasse 190, 8057 Z\"urich, Switzerland. \\
$^2$RCUK Fellow; Centre For Astrophysics, University of Central 
Lancashire, Preston, PR1 2HE, UK.
}

\begin{document}

\maketitle

\begin{abstract}
In a $\Lambda$CDM cosmology, the Milky Way accretes satellites into the stellar disc. We use cosmological simulations to assess the frequency of near disc plane and higher inclination accretion events, and collisionless simulations of satellite mergers to quantify the final state of the accreted material and the effect on the thin disc. 

On average, a Milky Way-sized galaxy has 3 subhalos with $\vmaxd>80$\,km/s; 7 with $\vmaxd>60$\,km/s; and 15 with $\vmaxd>40$\,km/s merge at redshift $z \simgt1$. Assuming isotropic accretion, a third of these merge at an impact angle $\theta <20^\mathrm{o}$ and are dragged into the disc plane by dynamical friction. Their accreted stars {\it and dark matter} settle into a thick disc. The stellar thick disc qualitatively reproduces the observed thick disc at the solar neighbourhood, but is less massive by a factor $\sim 2-10$. The dark matter disc contributes $\rhodd = 0.25 - 1$\,$\rhoh$ at the solar position. Although not likely to be dynamically interesting, the dark disc has important implications for the direct detection of dark matter because of its low velocity with respect to the Earth. 

Higher inclination encounters $\theta > 20^\mathrm{o}$ are twice as likely as low inclination ones. These lead to structures that closely resemble the inner/outer stellar halos recently discovered by \citet{2007arXiv0706.3005C}. They also do more damage to the Milky Way stellar disc creating a more pronounced flare, and warp; both long-lived and consistent with current observations. The most massive mergers ($\vmaxd\simgt 80$\,km/s) heat the thin disc enough to produce a thick disc. These heated thin disc stars are essential for obtaining a thick disc as massive as that seen in the Milky Way; they likely comprise some $\sim 50-90$\% of the thick disc stars. The Milky Way thin disc must reform from fresh gas after $z=1$.

Only one in four of our sample Milky Way halos experiences mergers massive and late enough to fully destroy the thin disc. We conclude that thick, thin and dark discs occur naturally within a $\Lambda$CDM cosmology. 

\end{abstract}


\section{Introduction}\label{sec:introduction}

The energy density of the Universe is primarily composed of a cosmological constant $\Lambda$ (72\%) and cold dark matter (22\%) ($\Lambda$CDM; \bcite{2006astro.ph..3449S}; \bcite{2006astro.ph..4335S}). Cosmological N-body simulations make accurate predictions for the evolution of the dark matter component, finding self similar dark matter halos that merge hierarchically, resulting in a complex phase space distribution criss-crossed by streams (\bcite{1999ApJ...524L..19M}; \bcite{2008MNRAS.tmp..143V}). 

The Milky Way provides a natural laboratory for testing such predictions. The Milky Way is composed of a stellar disc, an old stellar halo, a central bar/bulge of stars, and a dark matter halo \citep{2002ARA&A..40..487F}, as summarised in Table \ref{tab:milkyway}. Many of these components separate into finer substructures that are likely evidence of past accretions. The most visible of these are stellar streams: the Sagittarius stream that completes nearly two wraps in the plane perpendicular to the Galactic disc (\bcite{1994Natur.370..194I}; \bcite{2003ApJ...599.1082M}); the `Monoceros' ring that surrounds the Galactic disc (\bcite{2003MNRAS.340L..21I}; \bcite{2007MNRAS.376..939C}); and many smaller streams that have recently been found in the Galactic halo \citep{2006ApJ...642L.137B}. 

More subtle accretion signatures may also be found in the Milky Way disc and stellar halo. \citet{2007arXiv0706.3005C} find an inner and outer stellar halo that are kinematically distinct, suggestive of separate accretion events, while \citet{1983MNRAS.202.1025G} were the first to find that the Milky Way disc separates into thin and thick components. The thick disc is older ($>8$\,Gyrs), hotter ($\sigma_z \sim 42$\,km/s), more metal poor ($\meanFeH \sim -0.6$), and more slowly rotating (by $\sim 40$\,km/s) than the thin disc (see Table \ref{tab:milkyway}). Its age separation from the thin disc corresponds to redshift $z=1$ in $\Lambda$CDM, which is the epoch when the merger rate is calculated and observed to drop rapidly as the Universe changes from matter to dark energy domination (\bcite{2002AJ....124.1328D}; \bcite{2007ApJ...663L..13B}; \bcite{2001ApJ...546..223G}; \bcite{2006astro.ph.11187K}). This is certainly indicative of a merger origin for the thick disc, though there may not be any single well-defined formation epoch \citep{2004A&A...418..989N}\footnote{A connection with mergers for the formation of the thick disc is compelling, but not certain. \citet{2002MNRAS.330..707K}, for example, suggest an alternative mechanism where the thick disc forms from massive star clusters produced in a phase of very high star formation rate.}. Finally, the Milky Way thin disc presents evidence for past accretions in its spiral arms, warp, flare and bar \citep{2002SSRv..100..129G} -- all of which may be induced by satellite-disc interactions (see e.g. \bcite{2008arXiv0802.3997D}; \bcite{2007arXiv0708.1949K}). 

The Milky Way is not alone in these properties. Streams are now being found in extra-galactic systems -- most notably our nearest neighbour Andromeda (\bcite{2007arXiv0704.1318I}; \bcite{2008arXiv0801.4657M}), while thick discs appear to be ubiquitous (\bcite{1977PhDT.........4T}; \bcite{1979ApJ...234..829B}; \bcite{1994A&A...286..415V}; \bcite{2000AJ....120.1764M}; \bcite{2002AJ....124.1328D}; \bcite{2004A&A...422..465P}; \citet{2005ApJ...634..287I}; \bcite{2005ApJ...624..701Y}; \bcite{2006AJ....131..226Y}; \bcite{2006ApJ...650..644E}). 

In this paper, we study how the Milky Way disc affects the accretion of satellite galaxies in a $\Lambda$CDM cosmology, and how these satellites in turn affect the Milky Way disc. The Milky Way disc is the dominant mass component of the Milky Way interior to the solar circle. It is important because dynamical friction against the disc causes satellites to be preferentially dragged into the disc plane (\bcite{1986ApJ...309..472Q}; \bcite{1993ApJ...403...74Q}). As satellites are torn apart by tidal forces, they deposit both their stars {\it and their dark matter} into a thick disc \citep{1989AJ.....98.1554L}. The latter point is the key new idea presented in this work: a dark matter disc must form in a $\Lambda$CDM cosmology and we set out to quantify its mass and kinematic properties. At the same time, satellite accretions leave interesting morphological signatures in the stars and we quantify these too. Low inclination events give an accreted thick disc (\bcite{1996ApJ...460..121W}; \bcite{2006ApJ...650L..33P}); high inclination ones contribute to the Milky Way stellar halo (\bcite{2005ApJ...635..931B}; \bcite{2006ApJ...653.1180G}); and both heat and distort the Milky Way thin disc (\bcite{1986ApJ...309..472Q}; \bcite{1996ApJ...460..121W}; \bcite{1997ApJ...480..503H}; \bcite{1998ApJ...506..590S}; \bcite{1999MNRAS.304..254V}; \bcite{2001ApJ...563L...1F}; \bcite{2003ApJ...597...21A}; \bcite{2003ApJ...596..204A}; \bcite{2004MNRAS.351.1215B}; \bcite{2006ApJ...653.1180G}; \bcite{2006PASJ...58..835H}; \bcite{2007arXiv0708.1949K}; \bcite{Villalobos:2008rw}). 

Satellite heating of the thin disc has been a subject of particular interest in many previous papers. \citet{2007arXiv0708.1949K} were the first to be able to heat the thin disc enough to form a thick disc\footnotemark. This is because they measured and used the high-redshift satellite distribution that is both more massive and more destructive (because the orbits are more radial) than the surviving distribution (see also \bcite{2004MNRAS.352..535D}). Using the redshift zero satellite distribution will not give enough heating to form a thick disc (\bcite{2001ApJ...563L...1F}; \bcite{2008arXiv0802.3997D}). Here we also calculate and use the high-redshift satellite distribution. We find similarly to \citet{2007arXiv0708.1949K} that only the most massive of these satellites can heat the thin disc enough to form a thick disc. In addition, we consider what happens to the material accreted in such encounters -- both the stars and the dark matter,  and we quantify the fraction of the thick disc that can be accreted. This is particularly relevant for extra-galactic systems, where one counter-rotating thick disc has been found that cannot have formed from thin disc heating alone \citep{2005ApJ...624..701Y}.

\footnotetext{In a study concurrent with our own, \citet{Villalobos:2008rw} also consider the formation of thick discs from satellite heating and manage to successfully form heated thick discs. Their focus is on finding unique morphological and kinematic signatures of such accretions; ours is on calculating the expected number and mass of mergers in $\Lambda$CDM and the effect these have on an average Milky Way galaxy. In this sense our studies are complementary.} 

Our strategy is to use a cosmological structure formation simulation to determine the frequency of satellite-disc encounters, and collisionless simulations of satellite mergers to quantify the final state of both the accreted material and the stellar disc. We do not model the gaseous component of the disc, since we are primarily interested in collisionless mergers. Models for the Milky Way suggest that stars have continued to form at a constant rate since $z=1$, yet the surface density of gas likely remained constant with time due to fresh gas inflow (\bcite{2000A&A...358..869R}; \bcite{2006MNRAS.366..899N}; \bcite{2007arXiv0706.3850J}). The gas is important because it can cool and re-form a thin disc in the 8\,Gyrs between $z=1$ and the present. But it is also important because, unlike the stellar disc that is continually heated by mergers, the gas disc cools and maintains a low vertical dispersion. This keeps the density in the disc plane high, producing more significant dynamical friction. As a result, our gas free simulations likely under-estimate the effect of disc plane dragging. To compensate for this, we consider a range of disc masses from half of the present Milky Way disc mass up to its present mass. We will study the effect of gaseous Milky Way discs in future work. 

A complementary approach is to run full gas-hydrodynamical simulations of the formation of the Milky Way that include cooling, star formation and feedback physics. Such simulations remain a significant challenge both because of the computational cost of including gas physics, but also owing to uncertainties as to which physics is most important (see e.g. \bcite{2000ApJ...545..728T}; \bcite{2002ApJ...575...33R}; \bcite{2003ApJ...597...21A}; \bcite{2004ApJ...606...32R}; \bcite{2007arXiv0712.3285C}; \bcite{2007MNRAS.374.1479G}). Despite these difficulties, \citet{2003ApJ...597...21A} found a thick disc in their simulated Milky Way that formed from a mix of accreted and heated stars, while in a series of papers, \citet{2004ApJ...612..894B}, \citet{2005ApJ...630..298B} and \citet{2007ApJ...658...60B} find thick discs form directly from heated gas in their simulations. Our approach is complementary to these studies. Predictions for the evolution of the dark component rely only on getting the gravitational clustering right and are now robust \citep{2007arXiv0706.1270H}. We combine these with our knowledge of what galaxies look like in the Universe today to largely side-step the unknown physics of galaxy formation.  

This paper is organised as follows. In \S\ref{sec:cosmology} we use a concordance cosmology simulation to assess the frequency of high and low inclination satellite accretions for a typical Milky Way galaxy. In \S\ref{sec:simulations}, we present a suite of collisionless simulations of disc-satellite mergers, where we vary the satellite impact angle, orbit and mass. Finally, in \S \ref{sec:conclusions} we present our conclusions. 

\newlength{\myspace}
\setlength{\myspace}{0.013\textwidth}
\begin{table*}
\begin{center}
\setlength{\arrayrulewidth}{0.3mm}
\begin{tabular*}{\textwidth}{@{\hspace{\myspace}}c@{\hspace{\myspace}}c@{\hspace{\myspace}}c@{\hspace{\myspace}}c@{\hspace{\myspace}}c@{\hspace{\myspace}}c@{\hspace{\myspace}}c@{\hspace{\myspace}}c@{\hspace{\myspace}}c@{\hspace{\myspace}}c@{\hspace{\myspace}}c@{\hspace{\myspace}}}
& M & $\rho(R_\odot)$ &  $\Sigma(R_\odot)$ & $R_{1/2}$ & $z_{1/2}$ & $(\sigma_R$,$\sigma_\phi$,$\sigma_z$) & $v_c$ & $\meanFeH$ & $T$ & Ref \\
 & $(10^{10}M_\odot)$ & (M$_\odot$pc$^{-3}$) & (M$_\odot$pc$^{-2}$) & (kpc) & (kpc) & (km/s) & (km/s) & dex & (Gyrs) & \\
\hline
{\it Thin disc} & $4.8 \pm 0.4$ & 0.09 & 50 & $4.7\pm 0.5$ & $0.2; 0.25$ & $39$,$20$,$20 \pm 4$ & $220 \pm 3$ & $\sim 0$ & $\simlt 8$ &1,3,5,7--9,13\\
{\it Thick disc}	&$\sim 1$ & 0.011 & 5 -- 16 & $5.9 \pm 0.8$ & $0.63\pm 0.18$ & $63$,$39$,$39\pm 4$ & $180\pm 10$ & $\sim -0.6$ & $\simgt 8$ & 1,3,5--9,11\\ 
{\it Bulge}		&$0.8 \pm 0.4$ & -- & -- & $0.8$ & $0.25$ & $-$,$-$,$117\pm 5$ & $-$ & $-0.1 \pm 0.04 $ & $-$ & 3,10\\

{\it Inner halo}	&\raisebox{-1.5ex}{{$\left. \right\}$} $\sim 0.1$} & -- & -- & $10-15^*$ & $6-9$ & $-$,$-$,$\sim 100$ & $\sim 50$ & $< -1.6$ & $> 10$ & 2,12\\ [-1.5ex]
{\it Outer halo}	&  & -- & -- & $15-20^*$ & $-$ & $-$,$-$,$\sim 100$ & $\sim -40$;$-70$ & $< -2.2$ & $> 10$ & 2,12\\

{\it Dark halo}	& $\sim 100$ & $\sim0.01$ & $\sim15$ & $50-70^+$ & $-$ & $-$,$-$,$-$ & $-$ & $-$ & $-$ & 3,4\\
\end{tabular*}
\end{center}
\begin{flushleft}{\tiny $*$ These numbers are not half mass scale lengths, but rather radii at which the inner and outer halo dominate. \\\vspace{-1mm}
$+$ This is the half mass radius in spherical polar coordinates, $r_{1/2}$.}
\end{flushleft}
\caption[]{The distinct components of the Milky Way. From left to right the columns give: mass; local volume density; local surface density for $|z|<1.1$\,kpc; half mass scale length and height (see \S\ref{sec:defines}); $R,\phi,z$ velocity dispersion; rotation velocity; mean metallicity; mean age; and references. Where two values are given this reflects the fact that parameters often depend on the magnitude cut, or in which direction one looks (see e.g. \bcite{2004MNRAS.355..307K}; \bcite{2008ApJ...673..864J}; \bcite{2007A&A...464..565C}). Similar results have been observed for nearby extra-galactic systems (see text). Data are taken from: 1:\bcite{1989MNRAS.239..651K}; 2:\bcite{1993AJ....106..578M}; 3:\bcite{1998MNRAS.294..429D}; 4:\bcite{1999MNRAS.310..645W}; 5:\bcite{2001MNRAS.322..426O}; 6:\bcite{2003A&A...398..141S}; 7:\bcite{2004MNRAS.355..307K}; 8:\bcite{2004MNRAS.352..440H}; 9:\bcite{2008ApJ...673..864J}; 10:\bcite{2006ApJ...636..821F}; 11:\bcite{2007ApJ...663L..13B}; 12:\bcite{2007arXiv0706.3005C}; 13:\bcite{2007arXiv0707.1027S}.}
\label{tab:milkyway}
\end{table*}

\section{Cosmological simulations: the likelihood of disc encounters}\label{sec:cosmology} 

In this section we use a cosmological simulation to assess the likelihood of near-disc plane mergers in the Milky Way. Some movies of the simulation and the subhalo orbits can be seen at: {\tt http://justinread.net}. 

\begin{figure*}
\begin{center}
\includegraphics[height=0.32\textwidth]{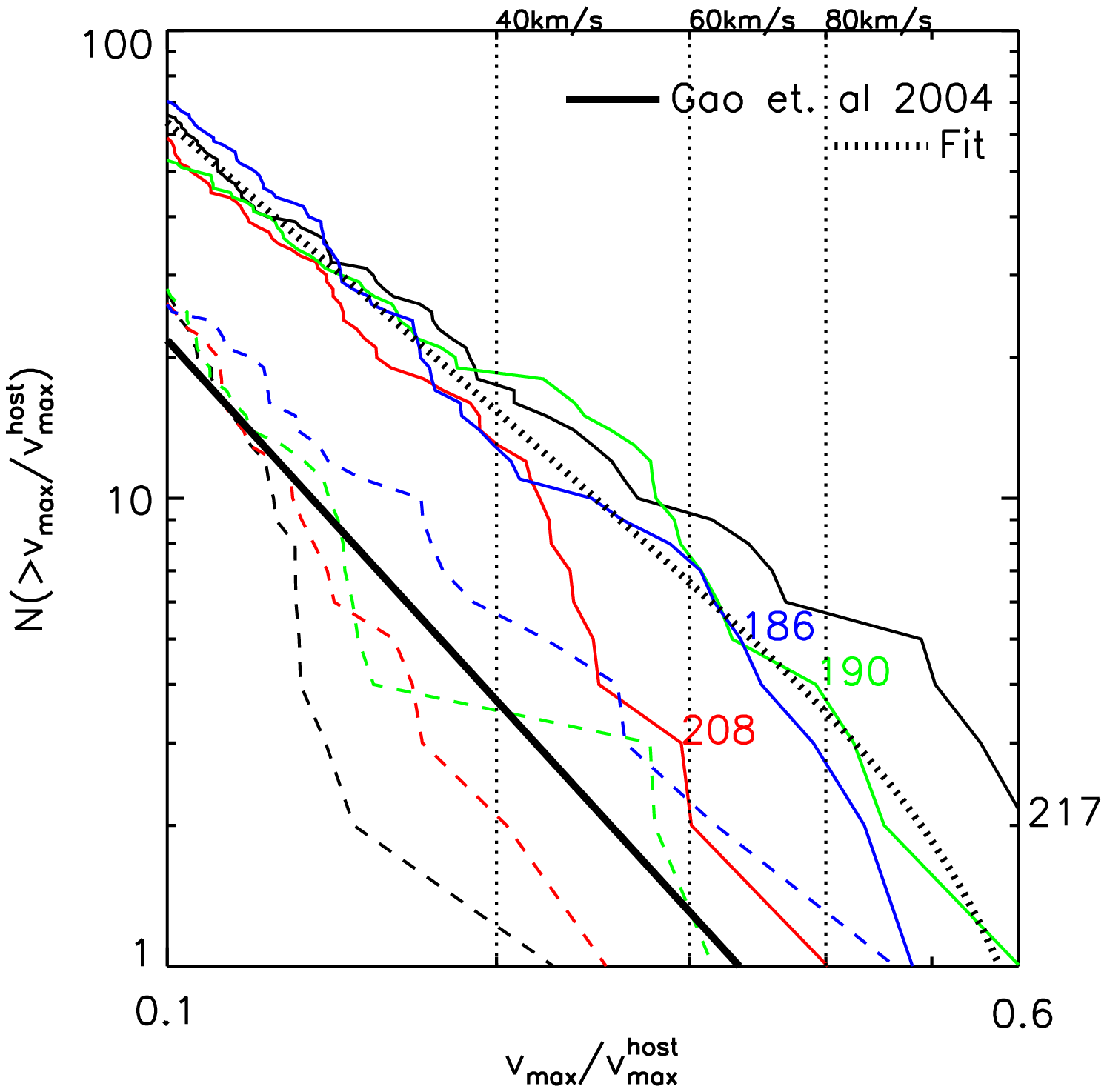}
\includegraphics[height=0.32\textwidth]{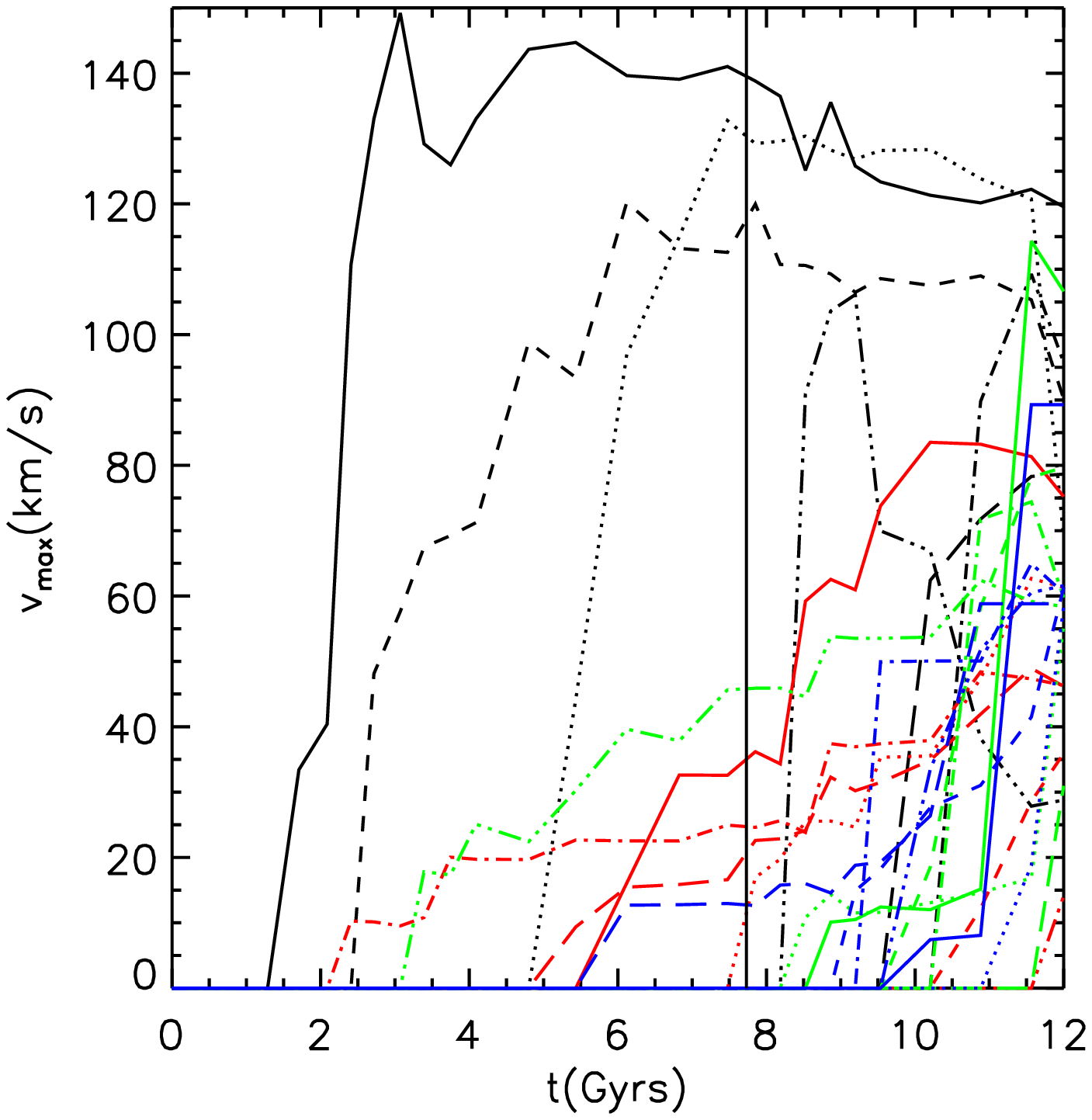}
\includegraphics[height=0.32\textwidth]{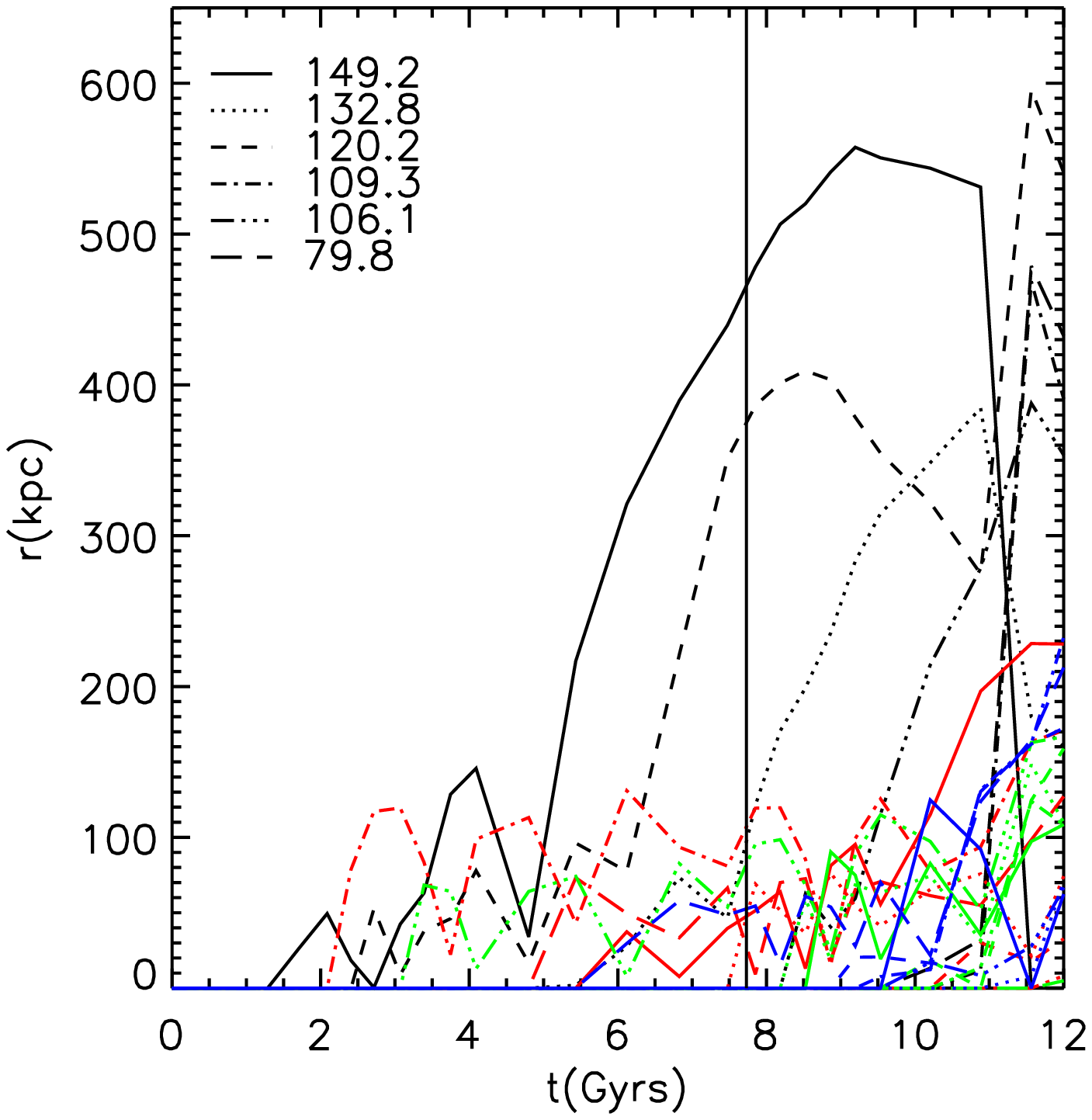}
\caption{{\bf Left:} cumulative maximum circular speed $\vmaxd$ functions for accreted (solid lines) and surviving (dashed lines) subhalos. We include only halos with greater than 50 particles. For the accreted halos, we consider only halos that at $z=0$ have 0.1 of their peak $\vmaxd$ considered over all times, and which pass within $r_\mathrm{merge}=50$\,kpc of the disc after $z=4.35$. Four host halos are shown, taken from the concordance cosmology simulation described in \S \ref{sec:cosmology} (1 black, 2 red, 3 green, 4 blue); their $\vmaxdhost$ are marked in km/s. The thick solid line shows a power-law fit to the surviving subhalo $\vmaxd$ function taken from \citet{2004MNRAS.355..819G} and \citet{2008arXiv0801.1127L}. A fit to the accreted subhalos is also shown (thick dotted line; see text for details). The vertical dotted lines mark $\vmaxd=40,60,80$\,km/s assuming $\vmaxdhost = 200$\,km/s. {\bf Middle \& Right:} evolutionary tracks in $\vmaxd$ and radius $r$ for the six most massive subhalos with $\vmaxd>40$\,km/s that are accreted by $z=0$ (solid, dotted, short dashed, dot dashed, triple-dot-dashed and long dashed lines). The different colours correspond to the same four host halos as in the left panel. For halo 1 (black lines) the $\vmaxd$ are marked on the right panel in km/s. Redshift $z=1$ ($\sim 8$\,Gyrs) is marked by the vertical solid lines. Note that the pericentres are likely over-estimated due to the limited number of simulation outputs.}
\label{fig:cosmoresults}
\end{center}
\end{figure*}

\subsection{Description of the simulation}\label{sec:cosmodescription}

We use the cosmological $\Lambda$CDM simulation already presented in \citet{2005MNRAS.364..367D}. The simulation was run using PkdGRAV \citep{2001PhDT........21S}, with cosmological parameters: $(\Omega_{\rm m},\Omega_{\Lambda},\sigma_8,h)$=(0.268,0.732,0.7,0.71), and a box of size $L_{\rm box}=90$ Mpc, with $300^3$ particles. The initial conditions were generated with GRAFIC2 \citep{2001ApJS..137....1B}. From the simulation volume, we extracted four Milky Way sized halos at a mass resolution of $m_p=5.7\times 10^5 M_\odot$. 

The subhalos inside each `Milky Way' and at each redshift output were identified using the {\it AHF}\footnote{{\tt http://www.aip.de/People/AKnebe/AMIGA/}} algorithm \citep{2004MNRAS.351..399G}. First, all particles are grouped together within isodensity regions. Second, unbound particles are iteratively removed until a bound structure is converged upon. We then assign particles to the smallest structure they appear in so that each particle is counted only once. We considered all (sub)halos with $>50$ particles. 

Once a subhalo was identified, we found its position by centring on it with the `shrinking sphere' method described in \citet{2006MNRAS.tmp..153R}; we used the peak of the rotation curve\footnote{Also called the `circular speed curve'.} of all bound particles about its mass centre as a proxy for mass. Each subhalo was then traced backwards in time, its progenitor being the subhalo at the previous redshift that contains the majority of its particles. To avoid ambiguities in this halo tracking, we ordered subhalos by mass so that subhalos were linked to their most massive progenitor not already claimed by a larger subhalo. A final complication can occur if two subhalos are about to merge. In this situation AHF sometimes over-estimates the mass of the smaller of the two. We dealt with this problem by searching for sudden spikes in mass at pericentre and removing these by assigning instead the mass found at the previous output time. 

We are interested in finding subhalos that merge with the disc at a given redshift. However, standard merger trees will not suffice since these define a merger as being when a subhalo {\it enters} the main halo. By this definition the LMC has already merged with the Milky Way! Instead, we define a subhalo as merged if it: (i) has less than a tenth of its peak circular speed considered over all times; and (ii) has passed within $r_\mathrm{merge}$ of the main halo after $z=4.35$. Our results are not sensitive to changes in the peak circular speed fraction or the redshift cut, but are sensitive to $r_\mathrm{merge}$, particularly for subhalos with peak circular speeds $\vmaxd\simlt 40$\,km/s. For this reason, we present results for a range of $r_\mathrm{merge} = 25,50,75$\,kpc. 

\subsection{The likelihood of disc plane mergers}

Figure \ref{fig:cosmoresults}, left panel, shows the cumulative maximum circular speed function of surviving (dashed) and accreted (solid) subhalos for the four Milky Way-sized halos. There is some cosmic variance amongst the halos -- in particular halo 1 is distinctly different from the other three. It has three very massive mergers with $\vmaxd > 90$\,km/s that occur at $z<1$. As we will show in section \S\ref{sec:simulations}, these are very damaging to a disc, while their late accretion leaves little time to rebuild a thin disc. As such, it seems likely that halos 2, 3  and 4 are more representative of the Milky Way. 

Our results for the surviving subhalos agree well with previous studies (\bcite{2004MNRAS.355..819G}; \bcite{2008arXiv0801.1127L}) and have cumulative $\vmaxd$ functions well fit by their fitting function:
\begin{equation}
N(>\vmaxd) = A \left(\frac{\vmaxd}{\vmaxdhost}\right)^{-\alpha}
\end{equation}
where $\vmaxdhost$ is the peak circular speed of the host galaxy; $\alpha = 2.559$; and $A = 0.06$. (Note that this underestimates slightly the low mass end. We find $\alpha=3$ gives a better formal fit in agreement with \citet{2004MNRAS.352..535D}.) Our accreted subhalos are well fit by a similar function with an exponential cut-off: 
\begin{equation}
N(>\vmaxd) = A \left(\frac{\vmaxd}{\vmaxdhost\,v_0}\right)^{-\alpha}\exp\left(-\left(\frac{\vmaxd}{\vmaxdhost\,v_0}\right)^{\beta}\right)
\end{equation}
The fitted values depend on our choice for the subhalos' point of closest approach, $r_\mathrm{merge}$. Figure \ref{fig:cosmoresults}, shows results for $r_\mathrm{merge}=50$\,kpc, for which we obtain best fit values: $A = 1.4$, $v_0 = 0.63$, $\alpha = 2$, and $\beta = 6$. However, decreasing $r_\mathrm{merge}$ decreases the number of lower mass subhalo mergers. For $r_\mathrm{merge} = 25$\,kpc we find: $A = 4.5$, $v_0 = 0.48$, $\alpha = 1.36$, and $\beta = 2.77$, while $r_\mathrm{merge} = 75$\,kpc gives: $A = 1.3$, $v_0 = 0.64$, $\alpha = 2.1$, and $\beta = 6.6$. We likely over-estimate the pericentres because of our limited number of simulation outputs, and because, if we were to include stars and gas in these simulations, the subhalos would be more resilient to tides (see e.g. \bcite{2006MNRAS.366.1529M}). However, our results are well converged for $\vmaxd > 40$\,km/s and $r_\mathrm{merge} < 75$\,kpc. Since it is these most massive halos that are of most interest for either damaging the disc or depositing accreted material, our choice of $r_\mathrm{merge}$ is not critical. We assume $r_\mathrm{merge} = 50$\,kpc from here on. 

From our above fits, assuming $\vmaxdhost=200$\,km/s, we find: 3 mergers with $\vmaxd > 80$\,km/s; 7 with $\vmaxd > 60$\,km/s; and 15 with $\vmaxd > 40$\,km/s. Nearly all mergers at $\vmaxd>60$\,km/s are complete by, or shortly after, $z\sim1$ (see the middle panel of Figure \ref{fig:cosmoresults}). Most of the massive accreted subhalos were found to be on highly eccentric orbits, with $e>0.8$ (see right-most panel of Figure \ref{fig:cosmoresults}). This agrees well with \citet{2007arXiv0708.1949K} and \citet{2004MNRAS.352..535D} who find that the accreted subhalos show a more eccentric orbit distribution than the survivors. It is important to stress that the {\it accreted subhalos are systematically more massive than the survivors}. As pointed out recently by \citet{2007arXiv0708.1949K}, it is essential to consider the accreted satellite distribution when studying satellite-disc interactions, otherwise the effects will be underestimated (see also \bcite{2007arXiv0704.1770S}). Here we add the additional point that it is important also to consider only the accreted halos that actually interact with the disc. Similar results have been derived previously from semi-analytic models \citep{2001ApJ...559..716T,2004MNRAS.348..811T,2005MNRAS.364..515T, 2003ApJ...598...49Z, 2005ApJ...624..505Z}.

Assuming isotropy, the probability that mergers will lie within an angle $\theta$ of the disc is given by $P=\sin\theta$. Hence, roughly a third of the mergers will occur with $\theta < 19.5^\mathrm{o}$, a third with $19.5^\mathrm{o} < \theta < 41.5^\mathrm{o}$, and a third with $41.5^\mathrm{o} < \theta < 90^\mathrm{o}$. Several recent studies -- that focus on the surviving satellites -- have found anisotropic satellite distributions, but some find polar alignment (the Holmberg effect; \bcite{1974ArA.....5..305H}; \bcite{2007arXiv0706.1350B}), others find planar alignment (the anti-Holmberg effect; \bcite{2005ApJ...628L.101B}; \bcite{2007arXiv0706.2009S}; \bcite{2007ApJ...662L..71F}), and some find no statistically significant alignment at all \citep{2006ApJ...645..228A}. From our four sample halos, we find two that have a very anisotropic merger history; indeed infall along filaments is expected to give rise to some anisotropy \citep{2004ApJ...603....7K,2005ApJ...629..219Z,2005MNRAS.363..146L}. However, we cannot address the issue directly with our simulations, since we do not know how the disc should align with the halo (see also \bcite{2006ApJ...650..550A}). In this work we will assume isotropy.

Putting all of the above together, we find that a typical Milky Way sized halo will have 1 subhalo merge near the disc plane ($\theta<20^\mathrm{o}$) with $\vmaxd > 80$\,km/s; $2-3$ with $\vmaxd > 60$; and 5 with $\vmaxd > 40$\,km/s. Away from the disc plane there will be twice as may mergers at the same mass. 

Our results agree well with a recent study by \citet{2007arXiv0711.5027S}. They use cosmological simulations to study a much larger sample of Milky Way halos than we do (17,000), but at lower resolution. They conclude that 95\% of all Milky Way halos have more than one merger with a subhalo $m > 5\times 10^{10}$\,M$_\odot$ ($\vmaxd \sim 75$\,km/s), while 70\% have a merger with a subhalo $m > 10^{11}$\,M$_\odot$ ($\vmaxd \sim 90$\,km/s). (They define a merger as being anything that enters the virial radius, which corresponds to the sum of our dashed and solid lines in Figure \ref{fig:cosmoresults}, left panel.) They suggest that these latter mergers could prove problematic for thin disc survival. In \S \ref{sec:simulations} we will show that such mergers do significantly heat the thin disc, but they do not destroy it; indeed, these mergers (and those at slightly lower mass) are likely to be essential for forming thick discs as massive as that seen in the Milky Way.

\begin{figure*}
\begin{center}
\includegraphics[width=\textwidth]{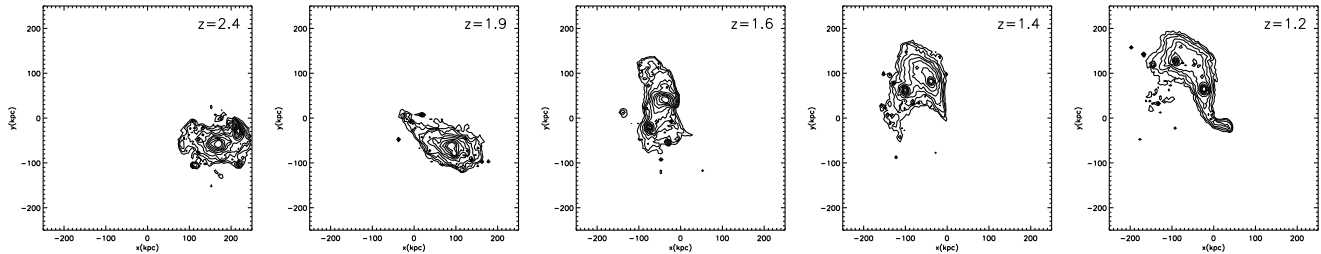}
\caption{Subhalos fall in inside loosely bound larger groups. Shown here are logarithmic density contours of a spherical region of radius 100\,kpc selected around the most massive subhalo in halo 2 at $z=2.4$, before this subhalo falls into the host halo. These same particles are tagged and plotted at later times, $z=1.9$, $z=1.6$, $z=1.4$, and $z=1.2$. The peak circular velocity of this subhalo is 75\,km/s, while its spherical overdensity mass and radius at $z=2.4$ are $M_\mathrm{SO}=2.55\times 10^{10}$M$_\odot$ and $r_\mathrm{SO} = 15.8$\,kpc, respectively. The much larger region extending to 100\,kpc and enclosing $10^{11}$\,M$_\odot$ (and some other smaller subhalos) largely co-moves with this subhalo.}
\label{fig:tag}
\end{center}
\end{figure*}

As a final point, we remark that it is surprising that subhalos with $\vmaxd \sim 60-70$\,km/s manage to merge in significantly less than a Hubble time. \citet{2004MNRAS.351..891Z} showed recently that, once mass loss due to tides is included, dynamical friction times for such subhalos in the Milky Way are greater than 10\,Gyrs. Indeed, we will verify that this is the case in \S\ref{sec:orbits} (\citet{2006ApJ...653.1180G} find similar results in their simulations). Yet such subhalos clearly do merge -- and rapidly -- in the cosmological simulations. The solution is presented in Figure \ref{fig:tag}, where we show logarithmic density contours of a spherical region of radius 100\,kpc selected around the most massive subhalo in halo 2 at $z=2.4$. These same particles are tagged and plotted at later times, $z=1.9$, $z=1.6$, $z=1.4$, and $z=1.2$. Notice that, although this subhalo has $\vmaxd = 75$\,km/s ($m=2.55\times 10^{10}$M$_\odot$), a much larger region around the subhalo enclosing $10^{11}$\,M$_\odot$ (and some other smaller subhalos) largely co-moves with it -- these subhalos fall in inside a larger loosely bound group environment and are effectively deposited at low apocentre. Similar results have been presented recently by \citet{2007arXiv0711.2429L}, who suggest that this might explain the apparent alignment of many Local Group galaxies in planes on the sky (see also \bcite{2008arXiv0802.0001L} and \bcite{2008arXiv0802.0302D}).

To approximate the effect of the larger group environment, for most of the simulations that follow in \S\ref{sec:simulations}, we will use a small initial apocentre of $r_a = 30$\,kpc. We justify this further in \S\ref{sec:orbits} where we show the effect of a larger apocentre orbit\footnote{\citet{1997ApJ...480..503H} argue, quite reasonably, for starting satellites at larger apocentre since this should be more cosmologically consistent. Here we suggest that it is actually less so because of this larger group environment.}. 

The above is particularly interesting in the context of the Sagittarius dwarf which is currently merging with the Milky Way. \citet{2004MNRAS.351..891Z} pointed out that it would have to have been uncomfortably massive in the past in order to arrive at its present location. This tension is resolved if Sagittarius accreted within a larger group environment. Modelling the effect of this in detail could be essential for a proper understanding of the Sagittarius stream and the orbits of the Large and Small Magellanic Clouds\footnote{\citet{2008arXiv0804.0426B} have recently studied isolated versus group mergers in clusters and find that 70\% of cluster galaxies are accreted in isolation. However, their study focuses on the {\it surviving} subhalos rather than the accreted ones, and on cluster environments, rather than the galaxy environment we study here, so detailed comparisons are necessarily difficult. We will pursue this issue of group versus isolated mergers in more detail in future work.}  

\section{Controlled simulations: dry satellite mergers}\label{sec:simulations}

In \S \ref{sec:cosmology}, we demonstrated that Milky Way sized galaxies will have on average $2-3$ satellites merge with $\vmaxd > 60$\,km/s near the disc plane ($\theta < 20^\mathrm{o}$) at early times, and twice as many at higher inclinations. In this section, we use a suite of collisionless simulations to measure the morphology and kinematics of the material accreted in such mergers, and the effect on the Milky Way disc. Some movies of our reference simulation \LMCten can be seen at: {\tt http://justinread.net}. 

\subsection{Description of the simulations}\label{sec:drydescription}

\begin{figure}
\begin{center}
\includegraphics[width=0.5\textwidth]{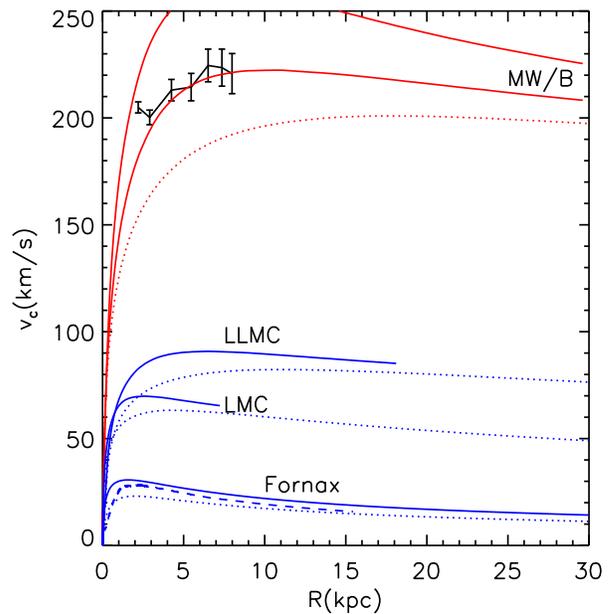}
\caption{Rotation curves for the different galaxy models. The upper red line is for MWB. The dotted lines show the dark matter halo contribution. The data points show the mean of
    HI measurements from
    \citet{1984ApJS...54..513B}, \citet{1974AAS...17..251W},
    \citet{1973AAS....8....1W}, \citet{1995ApJ...448..138M} and
    \citet{1986AAS...66..373K}. The blue dashed line shows the rotation curve from one of the subhalos taken from the cosmological simulations of \S \ref{sec:cosmology}.}
\label{fig:rotcurve}
\end{center} 
\end{figure}

We set up our Milky Way model (disc+halo system) by adiabatically growing a disc inside a
spherical halo. The initial spherical halos were generated from a distribution function using the method described in \citet{2004ApJ...601...37K} with the added refinement that the halo is composed of two mass species arranged on shells. The outer shell had particles $5\times$ more massive particles than the inner shell, in order to increase the effective resolution in the central parts. We then inserted a massless disc of particles with scale-length, $R_d = 3$ kpc. We slowly grew the mass of this disc over 1.5 Gyr to a final mass of $3\times 10^{10}$\,M$_\odot$ (MW) and $6\times 10^{10}$\,M$_\odot$ (MWB) while holding the disc particles fixed in place. After this time, we set the kinematics of the disc to give a constant Q = 1.5. Evolved in isolation, the MW system formed a small, weak bar after a Hubble time but otherwise did not change significantly; by contrast, the more massive MWB formed a strong bar. Recall that we do not know the total baryonic mass or gas fraction of the Milky Way disc at $z=1$ when most of these mergers took place, so it is useful to consider a range of disc masses. Our MW model has a peak circular velocity of $220$\,km/s and the dark matter halo provides 200\,km/s of this (see Figure \ref{fig:rotcurve}). Current mass models for the Milky Way favour lower halo contributions ($\sim 170$\,km/s; \bcite{2002ApJ...573..597K}), so our MW model corresponds to a minimum disc situation; MWB examines the effect of a more dominant disc. 

We chose three models for our satellite: Fornax, LMC and LLMC. Fornax, with {\it dark matter} contribution to the peak circular velocity $\vmaxd=25$\,km/s, models the satellite on the Fornax dwarf spheroidal galaxy which has a visible mass of $\sim 5\times 10^7$\,M$_\odot$, and a dynamical mass of $M(<1.5\,\mathrm{kpc}) \sim 7.5 \times 10^8$\,M$_\odot$ \citep{2005astro.ph.11465W}. It was initialised as a two component spherical galaxy, with star and dark matter velocities drawn from numerically calculated distribution functions as in \citet{2006MNRAS.tmp..153R}. LMC, with $\vmaxd=60$\,km/s, models the satellite on the Large Magellanic Cloud -- the largest surviving satellite of the Milky Way -- with visible mass of $M \sim 2.7 \times 10^9$\,$M_\odot$, and a dynamical mass of $M(<8.9\,\mathrm{kpc}) = 8.7\pm4.3\times10^9$\,M$_\odot$ \citep{2002AJ....124.2639V}. LLMC, with $\vmaxd=80$\,km/s, is a `large' LMC model that considers the effect of the most massive mergers found in \S\ref{sec:cosmology}. Both LMC and LLMC were set up as scaled versions of our MW Milky Way model. Our model satellite properties are given in Table \ref{tab:simulations} and Figure \ref{fig:rotcurve}. The total peak circular speeds $\vmaxt$ for Fornax, LMC and LLMC are $\vmaxt = 30,70,90$\,km/s, respectively. These are all larger than the quoted $\vmaxd$ above because those were for only the dark matter contribution to the peak circular speed. We quote $\vmaxd$ rather than $\vmaxt$ from here on for easier comparison with the results from \S \ref{sec:cosmology}. 

For the orbit parameters, we chose a wide range of initial inclination angles to the disc from $10-60^\mathrm{o}$, one retrograde orbit (Ret-10$^\mathrm{o}$), and range of pericentres and apocentres, as detailed in Table \ref{tab:simulations}. We chose a typical apocentre of 30\,kpc for reasons outlined in \S \ref{sec:cosmology}. We test the effect of a larger apocentre in one run: \LMCabig.  
 
The simulations were evolved using the collisionless tree-code, PkdGRAV \citep{2001PhDT........21S}. We used a maximum timestep of 0.005\,Gyrs with nine rungs of decreasing size in powers of two. The final evolved systems were mass and momentum centred using the `shrinking sphere' method described in \citet{2006MNRAS.tmp..153R}, and rotated into their moment of inertia eigenframe with the $z$ axis perpendicular to the disc. 

\begin{table*}
\begin{center}
\setlength{\arrayrulewidth}{0.3mm}
\begin{tabular}{@{\hspace{\myspace}}l@{\hspace{\myspace}}l@{\hspace{\myspace}}l@{\hspace{\myspace}}l@{\hspace{\myspace}}l@{\hspace{\myspace}}l@{\hspace{\myspace}}l@{\hspace{\myspace}}l@{\hspace{\myspace}}l@{\hspace{\myspace}}l}
{\it Galaxy model} & {\it Description} & $N_*$ & $N_\mathrm{DM}$ & $\epsilon_*$(kpc) & $\epsilon_\mathrm{DM}$(kpc) & $M_*(\mathrm{M}_\odot)$ & $r_*$(kpc) & $M_\mathrm{DM}(\mathrm{M}_\odot)$ & $r_\mathrm{DM}$(kpc) \\
\hline
MW & Milky Way; light disc  & $7.5\times 10^5$ & $2\times 10^6$ & 0.06 & 0.1; 1 & $3\times10^{10}$ & 3 & $10^{12}$ & 25 \\
MWB & Milky Way  & $7.5\times 10^5$ & $2\times 10^6$ & 0.06 & 0.1; 1 & $6\times10^{10}$ & 3 & $10^{12}$ & 25 \\
LLMC & Large-LMC  & $7.5\times 10^5$ & $2\times 10^6$ & 0.015 & 0.024; 0.234 & $3\times10^{9}$ & 1.78 & $1\times10^{11}$ & 15\\
LMC & LMC-like  & $7.5\times 10^5$ & $2\times 10^6$ & 0.015 & 0.024; 0.234 & $7\times10^{8}$ & 0.7 & $2.4\times10^{10}$ & 6\\
Fornax & Fornax-like  & $10^6$ & $10^6$ & 0.005 & 0.02 & $5.6\times10^{7}$ & 0.1 & $10^{9}$ & 2\\
\vspace{2.5mm}\\
{\it Simulation} & $(x,y,z)$\,(kpc) & $(v_x,v_y,v_z)$\,(km/s) & $i$ & $e$ & {\it Rotation} & {$T_\mathrm{out}$(Gyrs)} \\ 
\hline
\LMCten & (29.5,0.27,-5.2) & (-6.3,89.3,0.35) & 10$^\mathrm{o}$ & 0.8 & {\it prograde} & 6.5\\
\LMCtwe & (28.2,0.12,-10.3) & (-2.2,80.1,0.82)  & 20$^\mathrm{o}$ & 0.8 & {\it prograde} & 4.3\\
\LMCfor & (23,0.12,-19.3) & (-1.8,80.1,1.52)  & 40$^\mathrm{o}$ & 0.8 & {\it prograde} & 6.3\\
\LMCsix & (15,0.12,-26) & (-1.2,80.1,2.0)  & 60$^\mathrm{o}$ & 0.8 & {\it prograde} & 7\\
\LMCRet & (29.5,0.237,-5.2) & (-6.3,-89.3,0.35) & 10$^\mathrm{o}$ & 0.8 & {\it retrograde} & 5.3\\
\LMCeless & (29.5,0.27,-5.2) & (-6.3,143,0.35) & 10$^\mathrm{o}$ & 0.36 & {\it prograde} & 4.1\\
\LMCabig & (80,0.27,-15.2) & (-6.3,62.5,0.35) & 10$^\mathrm{o}$ & 0.74 & {\it prograde} & 6.8\\
\Fornax &  (29.5,0.27,-5.2) & (-6.3,89.3,0.35) & 10$^\mathrm{o}$ & 0.8 & {\it prograde} & 5\\
\LLMC & (29.5,0.27,-5.2) & (-6.3,89.3,0.35)  & 10$^\mathrm{o}$ & 0.8 & {\it prograde} & 5.5\\
\LLMCMWB & (29.5,0.27,-5.2) & (-6.3,89.3,0.35)  & 10$^\mathrm{o}$ & 0.8 & {\it prograde} & 4.5\\

\end{tabular}
\end{center}
\caption[]{Simulation labels and parameters. The top table shows the different galaxy models we use, two for the host galaxy (MW/MWB), and three different satellite galaxy models: LMC, LLMC and Fornax. The columns show from left to right: the number of star and dark matter particles, $N_*, N_\mathrm{DM}$; the force softenings, $\epsilon_*, \epsilon_\mathrm{DM}$; and the masses and scale lengths of each component, $M_*, r_*, M_\mathrm{DM}, r_\mathrm{DM}$. Where two force softenings are given, these are for the low mass inner and higher mass outer halo particles (see \S \ref{sec:drydescription} for details). The bottom table shows the different initial satellite orbits. The columns give from left to right: the initial phase space coordinates of the satellite, $x,y,z,v_x,v_y,v_z$; the initial inclination to the Milky Way disc, $i$; the eccentricity, $e$; the sense of rotation; and the simulation output time.}
\label{tab:simulations}
\end{table*}

\subsection{Some definitions and nomenclature}\label{sec:defines}

Some of the density profiles we obtain are not easily fit by a simple functional form. For this reason we use half mass scale lengths\footnotemark\ as this is then model independent. Discs are typically fit with an exponential profile, given by: 

\footnotetext{There can be some confusion here as `half mass scale length' depends on the choice of coordinates. We work in cylindrical polar coordinates $(R,\phi,z)$ so that $R_{1/2}$ means the radius containing half of the mass, having already integrated over $z$ and $\phi$. Similarly  $z_{1/2}$ means the $|z|$ that contains half of the mass, having already integrated over $R$ and $\phi$. In spherical polar coordinates $(r,\theta,\phi)$, the half mass radius $r_{1/2}$ is the radius containing half of the mass, having already integrated over $\theta$ and $\phi$. So $r_{1/2} \neq R_{1/2}$ even in the limit of spherical symmetry.}

\begin{equation}
\rho(R,z) = \frac{M_0}{4 \pi R_0^2 z_0} e^{-R/R_0} e^{-|z|/z_0}
\label{eqn:density}
\end{equation}
where $M_0$ is the disc mass, and $R_0, z_0$ are the scale length and height, respectively. For this distribution, the half mass scale lengths are related to the scale lengths as: $R_{1/2} = 1.68 R_0$, $z_{1/2} = 0.7 z_0$. Other common variants have $\rho(z) \propto \mathrm{sech}^2(z/z_s)$ and $\rho(z) \propto e^{-(z/z_e)^2}$, whose scale lengths interrelate as: $z_{1/2} = 0.55 z_s = 0.48 z_e = 0.7 z_0$. We will refer only to the half mass scale lengths from here on. 

\begin{figure*}
\begin{center}
\includegraphics[width=0.33\textwidth]{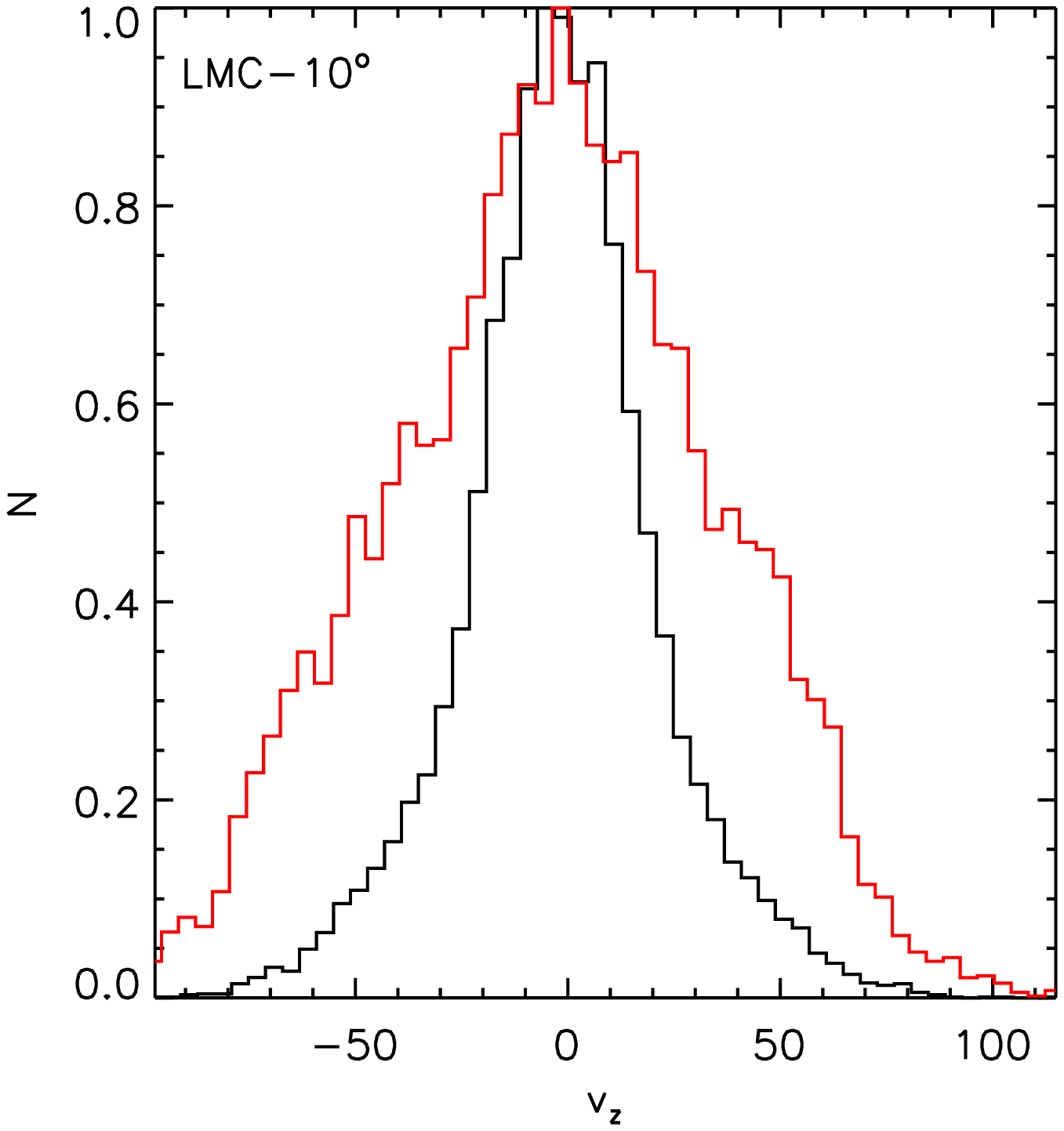}
\includegraphics[width=0.33\textwidth]{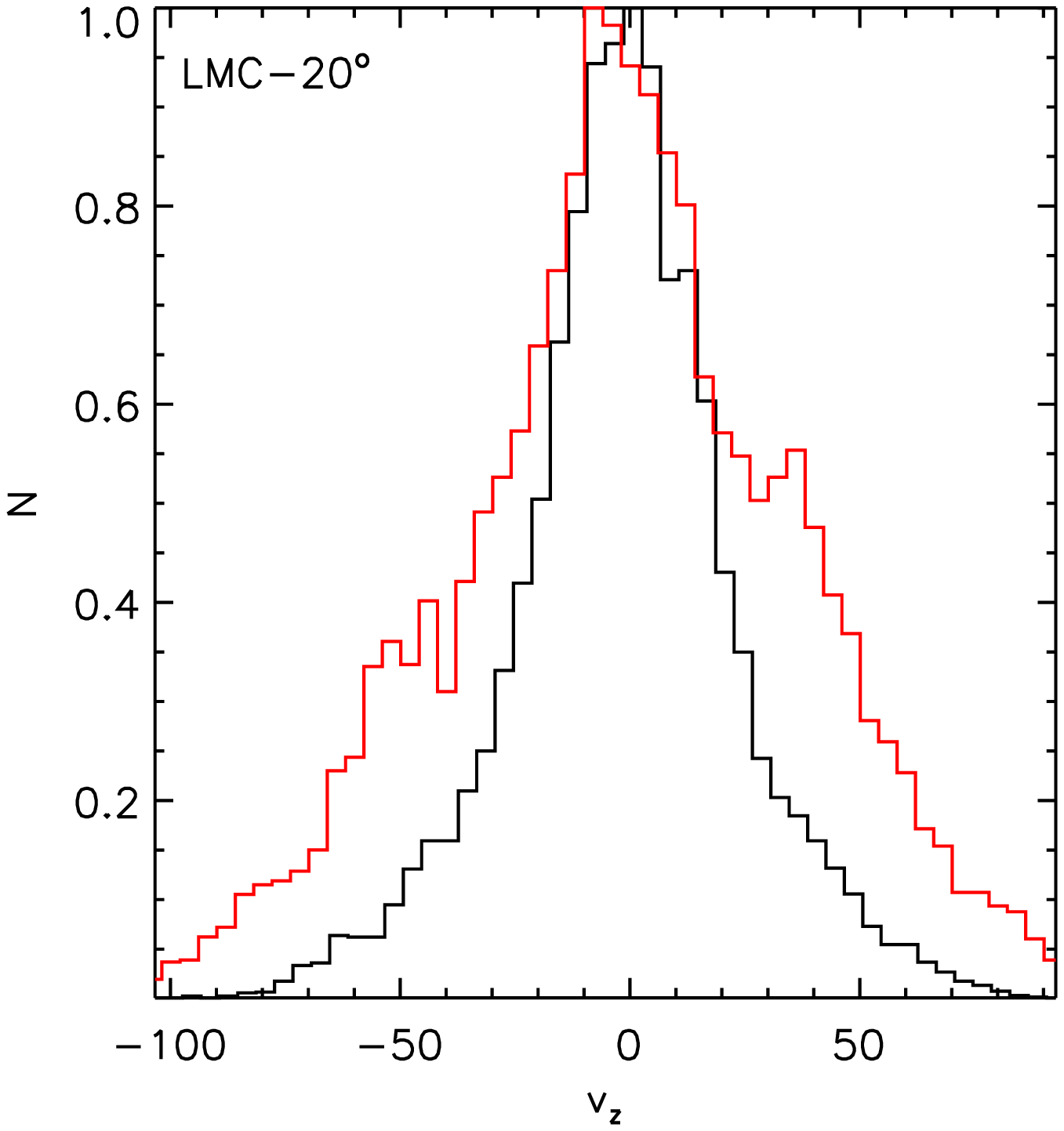}
\includegraphics[width=0.33\textwidth]{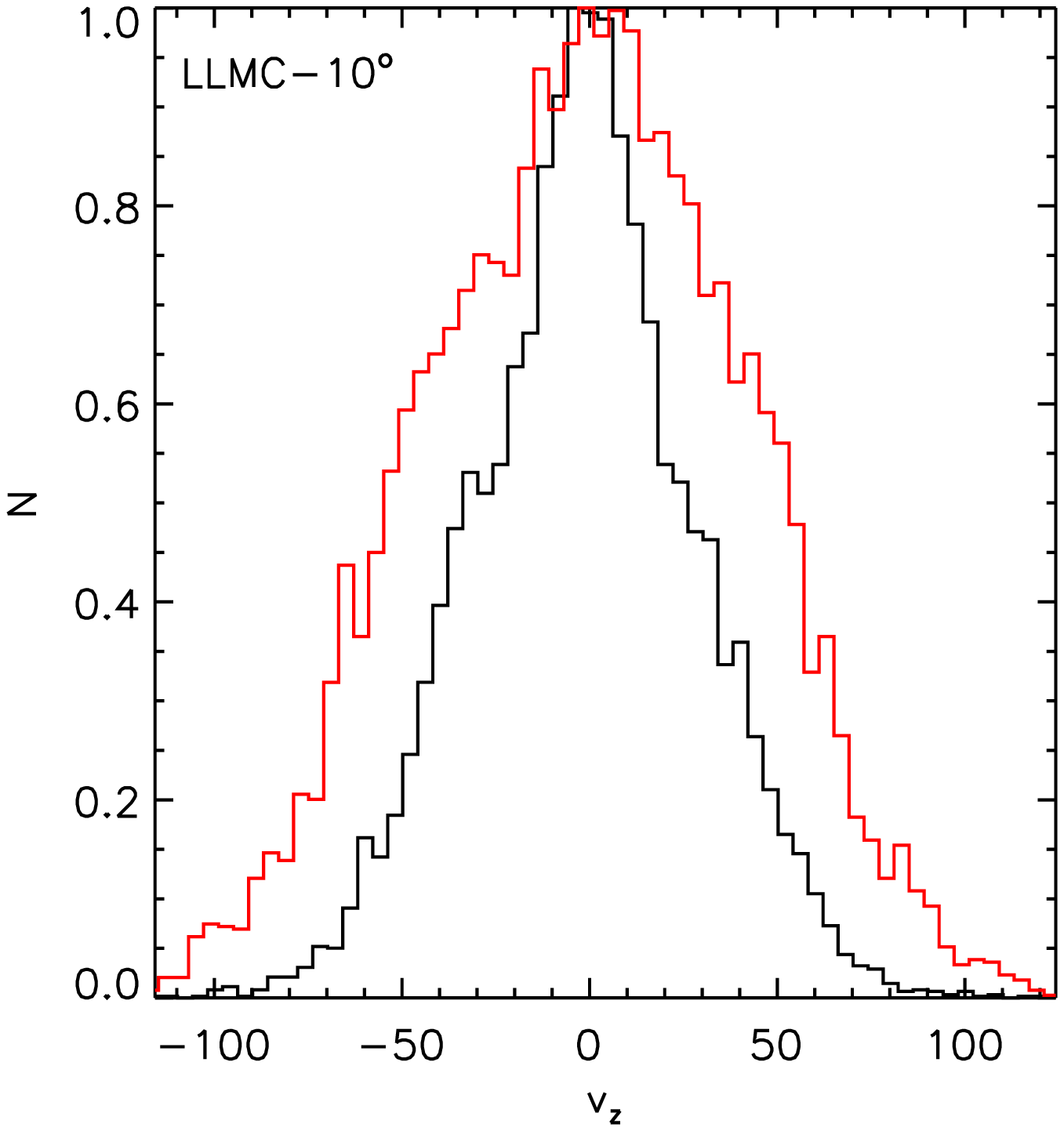}

\caption{The distribution of $v_z$ velocities at the solar neighbourhood, $8 < R < 9$\,kpc; $|z|<0.35$\,kpc, for \LMCten, \LMCtwe\ and \LLMC. The black line shows the Milky Way disc stars, the red the accreted satellite material. The distributions are normalised to peak at 1 and do not represent the mass in each component. Notice the prominent wings in the thin disc distribution for \LMCten\ and \LMCtwe: some of the stars originating in the thin disc show `thick disc' kinematics. For the more massive \LLMC merger, the wings are now very prominent: the heated thin disc distribution looks like the accreted distribution in \LMCten\ and little thin disc component remains.}
\label{fig:veldecomp}
\end{center}
\end{figure*}

We also refer frequently to the `solar neighbourhood'. By this we mean a cylindrical slice centred on $R_\odot = 8.5$\,kpc, 1\,kpc thick -- i.e. $8 < R < 9$\,kpc; a slice $7.5 < R < 8.5$\,kpc gives nearly indistinguishable results. 

\subsection{Results}\label{sec:results}

Our main results are shown in Figures \ref{fig:impact1}, \ref{fig:orbchange1} and \ref{fig:satchange1}. Figure \ref{fig:impact1} shows the effect of increasing the impact angle; Figure \ref{fig:orbchange1}, the effect of changing the satellite orbit; and Figure \ref{fig:satchange1}, the effect of changing the satellite and host galaxy properties. Some simulations were run for longer to ensure that the final state was in equilibrium. Typical times were $\sim 5$\,Gyrs (see Table \ref{tab:simulations}). 

Figure \ref{fig:veldecomp} shows the distribution of $v_z$ velocities of stars at the solar neighbourhood for \LMCten, \LMCtwe\ and \LLMC. For \LMCten\ and \LMCtwe, some of the stars originating in the Milky Way thin disc are scattered such that they have thick disc like kinematics (notice the prominent wings), but the majority of stars with thick disc kinematics are accreted. For the more massive satellite merger -- \LLMC\ -- these wings are now very prominent and little thin disc remains. The heated thin disc distribution now looks like the accreted distribution in \LMCten. 

In most of what follows, we will decompose the final distributions into Milky Way stars and dark matter, and those accreted from the infalling satellite. This decomposition makes sense, since we can expect the accreted satellite stars to be chemically distinct from the thin disc stars. Observations at the solar neighbourhood show that thick and thin disc stars do show different chemistry, while there are some thin disc stars with thick-disc like kinematics (\S \ref{sec:introduction} and see \bcite{2007ApJ...663L..13B}). In \S\ref{sec:thickdisc}, we investigate alternative decompositions where we include heated thin disc stars in the thick disc.  

\newlength{\myfspace}
\setlength{\myfspace}{-0.03\textwidth}
\begin{figure*}
\begin{center}

\includegraphics[width=0.22\textwidth]{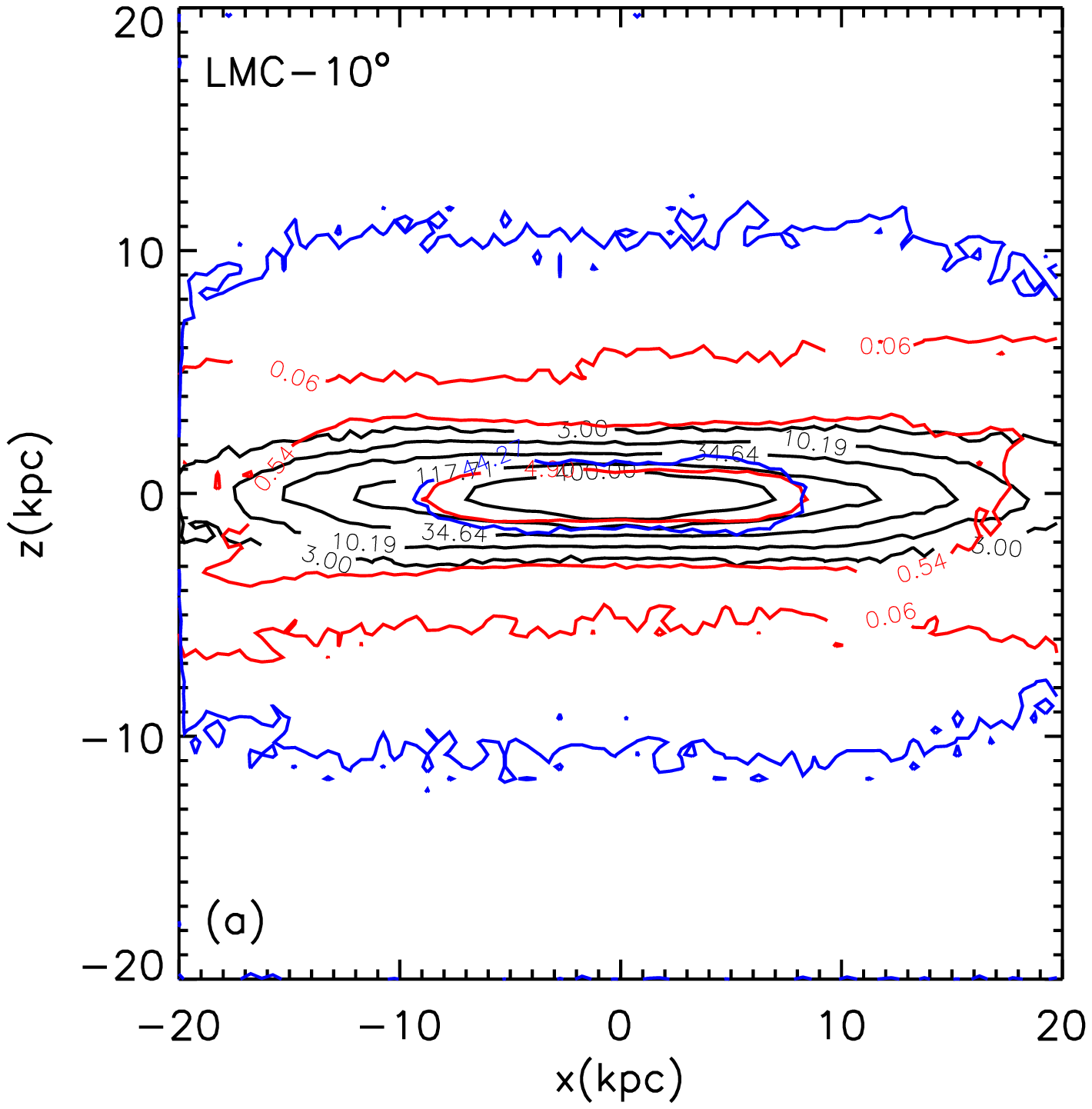}\hspace{\myfspace}
\includegraphics[width=0.22\textwidth]{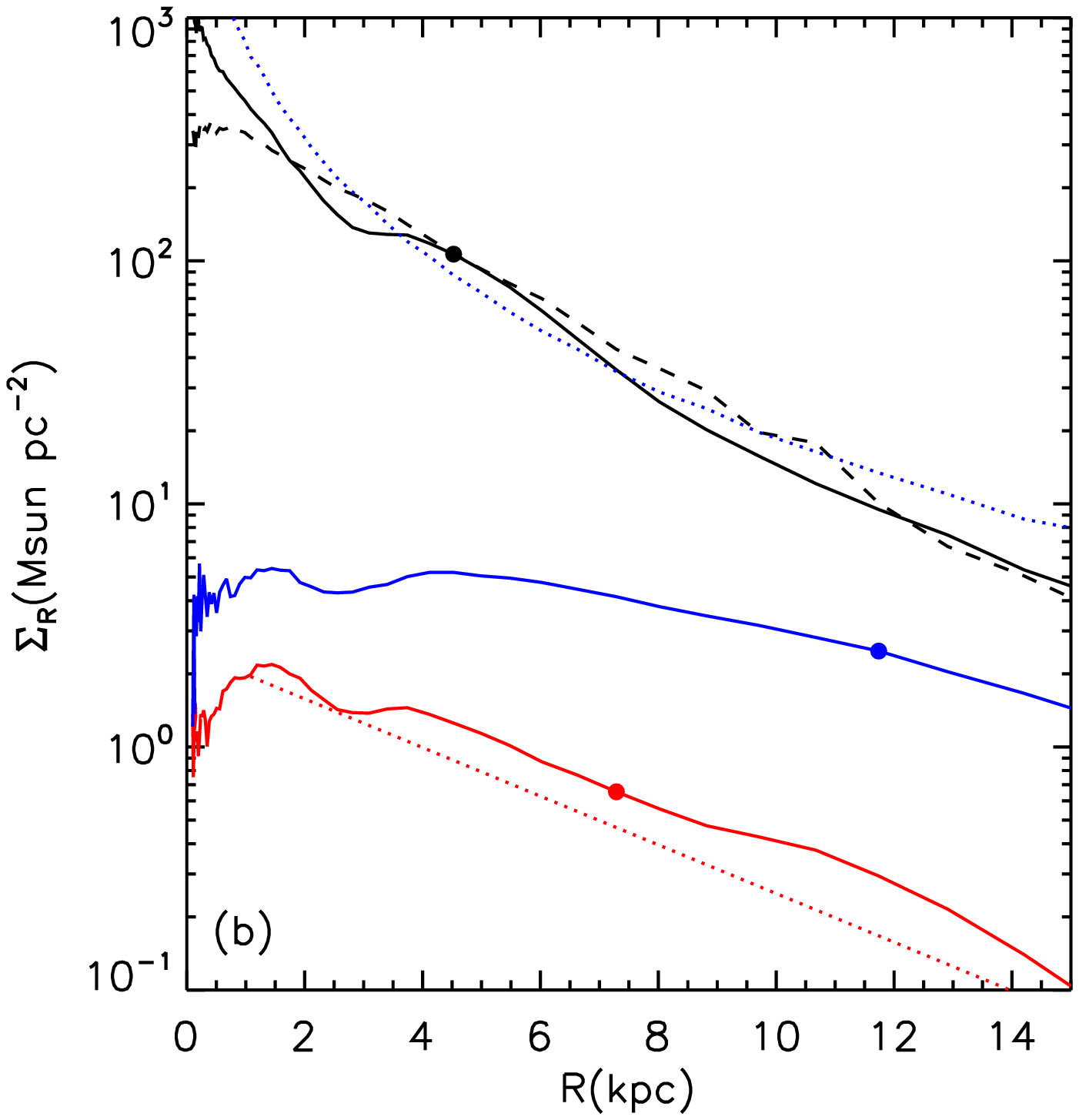}\hspace{\myfspace}
\includegraphics[width=0.22\textwidth]{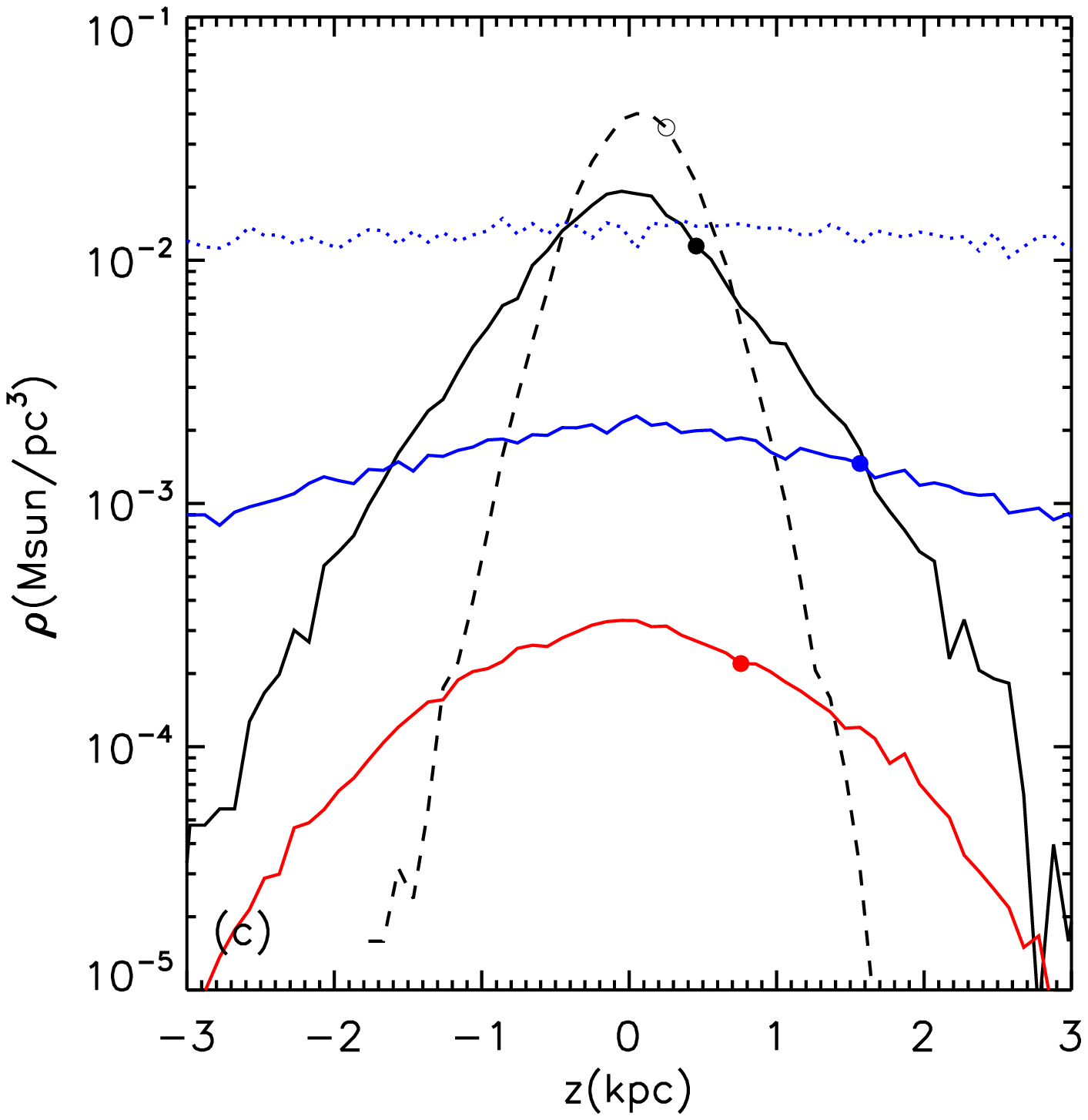}\hspace{\myfspace}
\includegraphics[width=0.22\textwidth]{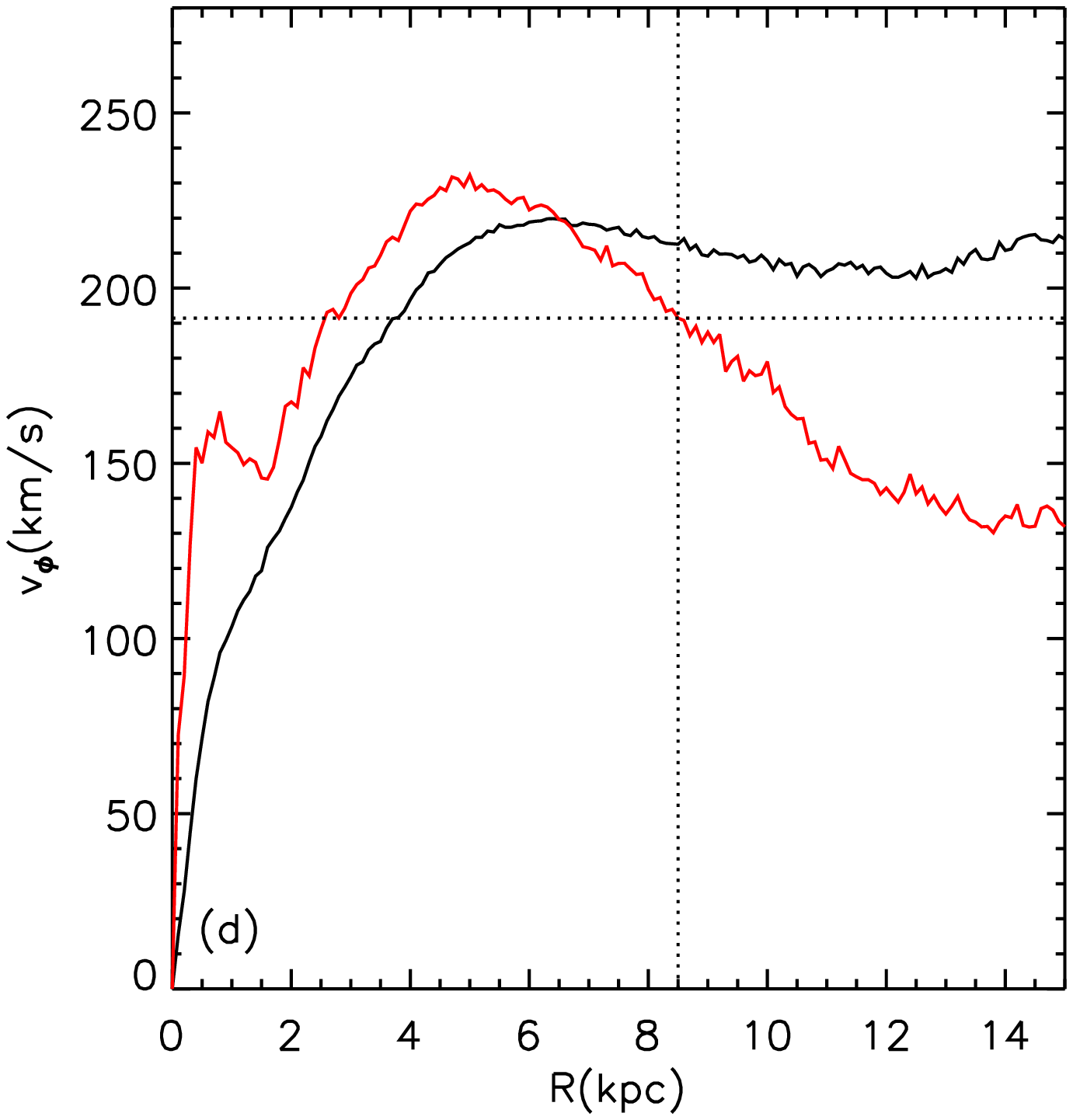}\hspace{\myfspace}
\includegraphics[width=0.22\textwidth]{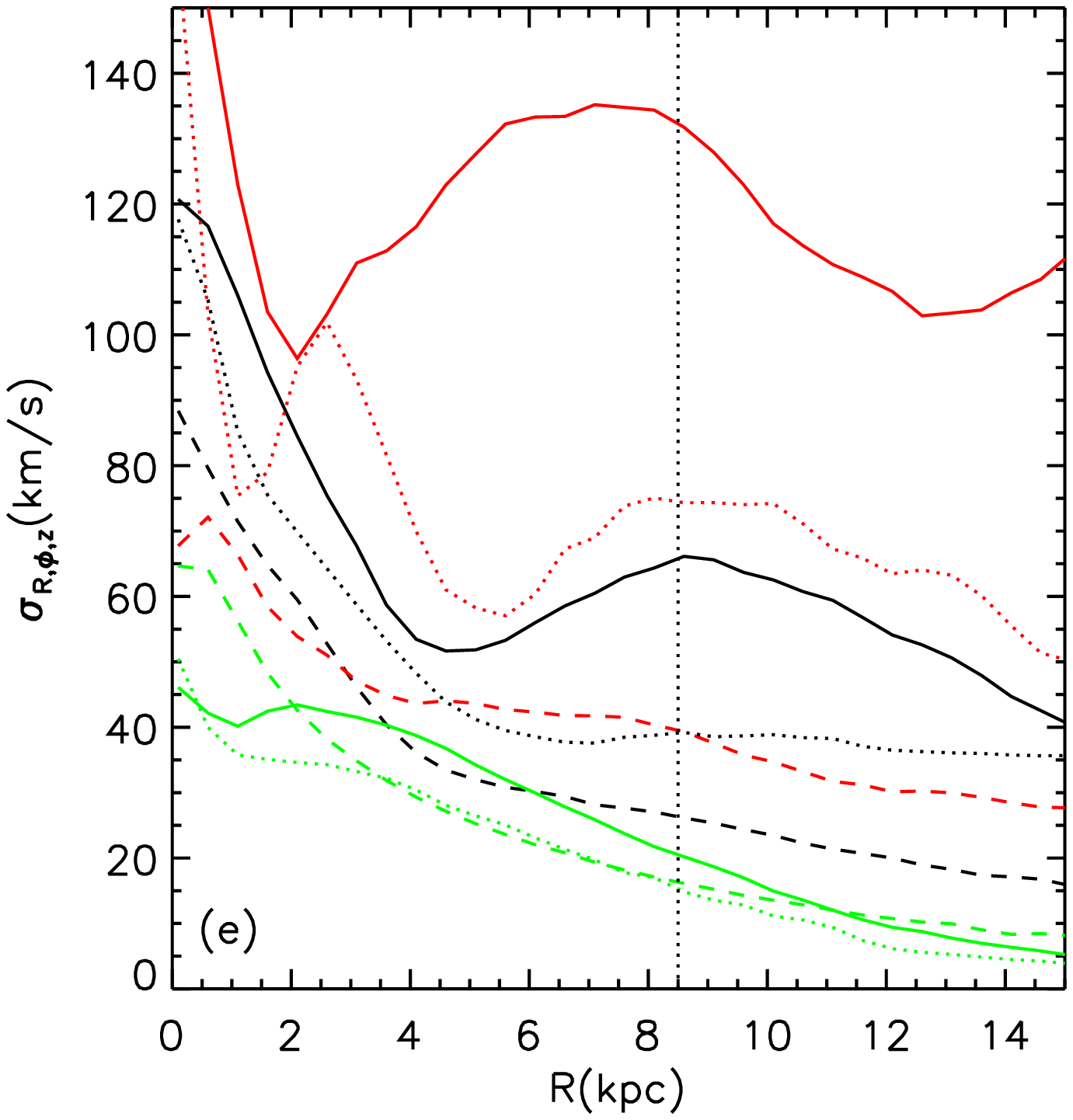}\hspace{\myfspace}\\

\includegraphics[width=0.22\textwidth]{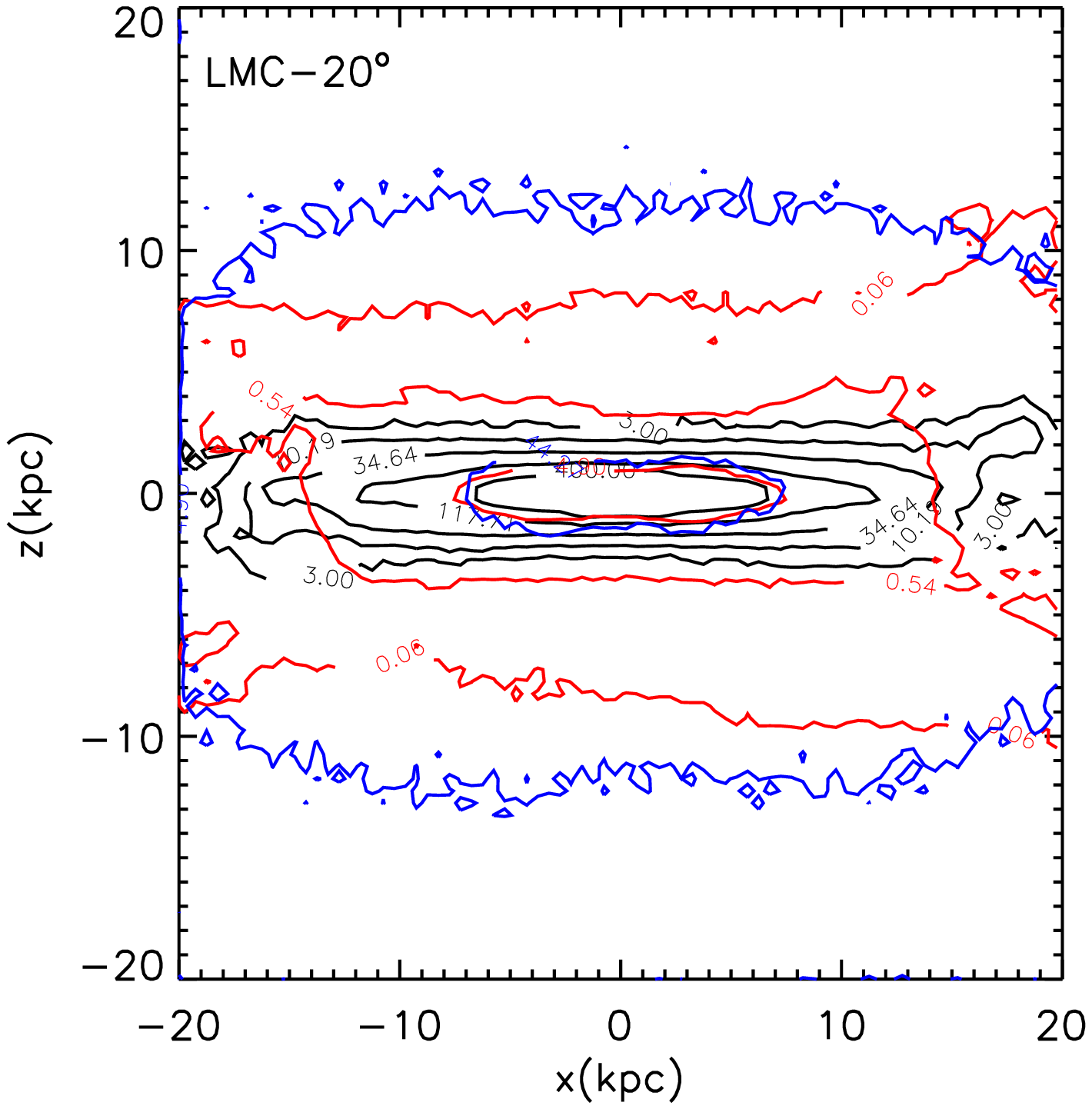}\hspace{\myfspace}
\includegraphics[width=0.22\textwidth]{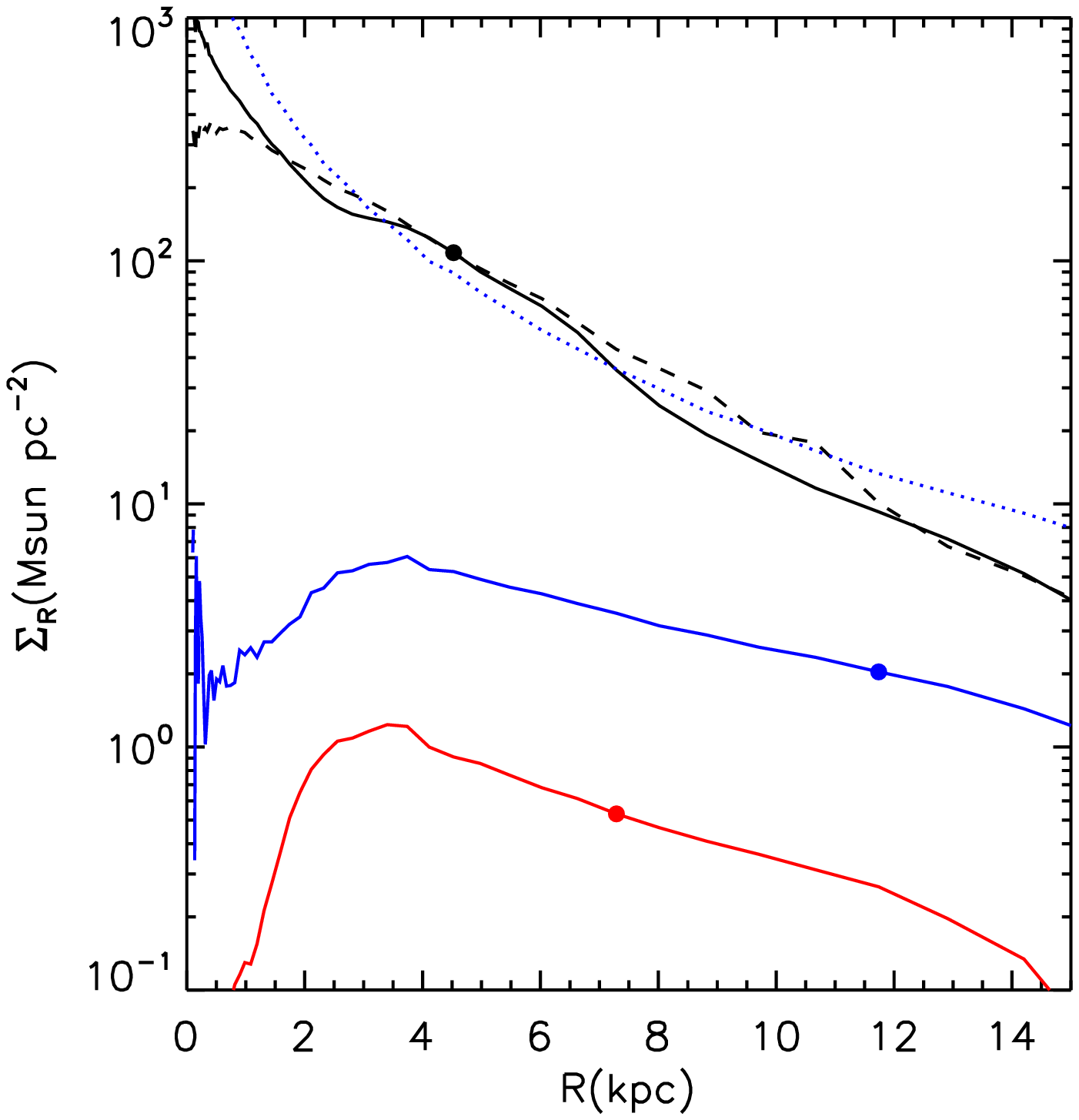}\hspace{\myfspace}
\includegraphics[width=0.22\textwidth]{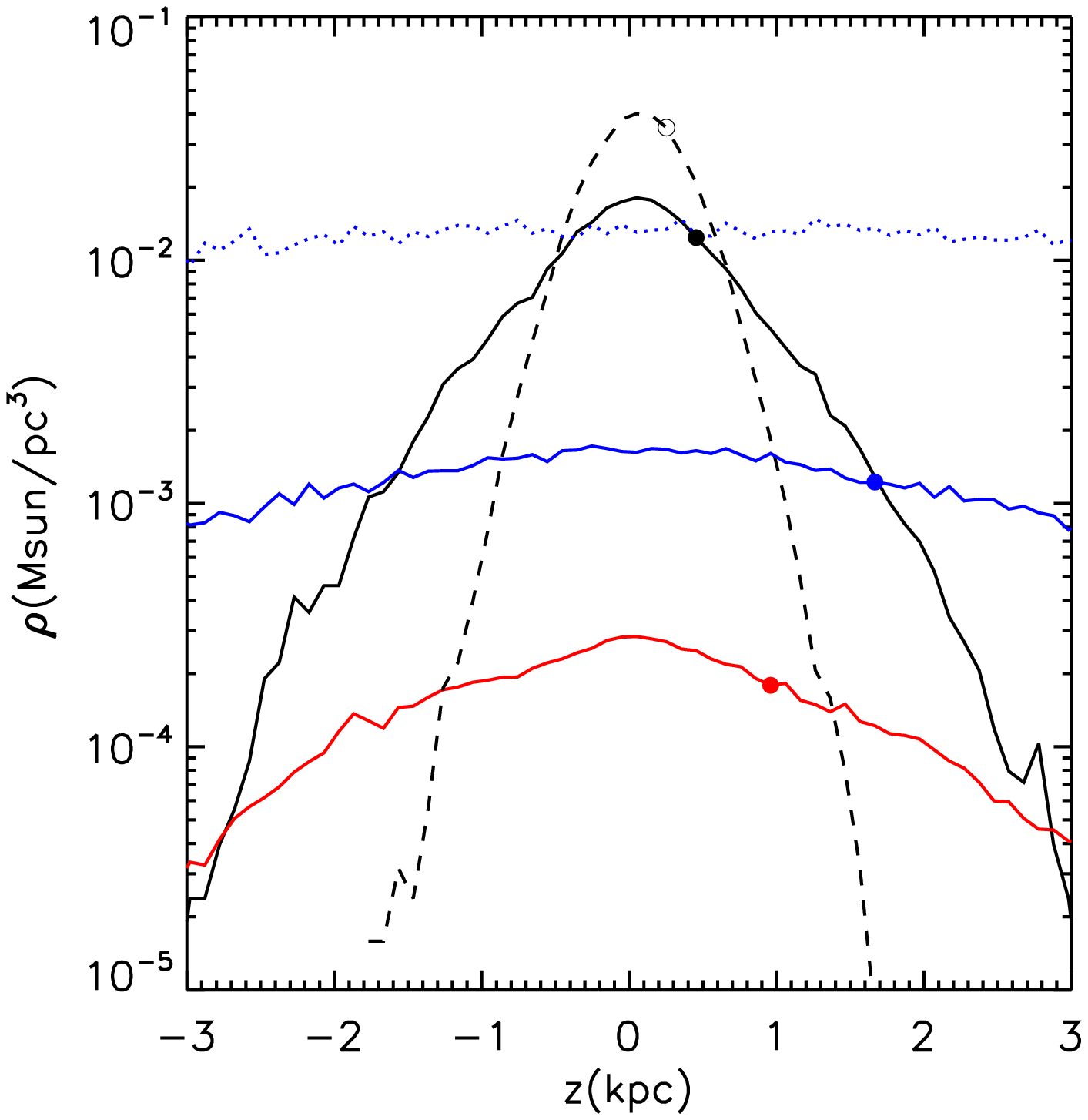}\hspace{\myfspace}
\includegraphics[width=0.22\textwidth]{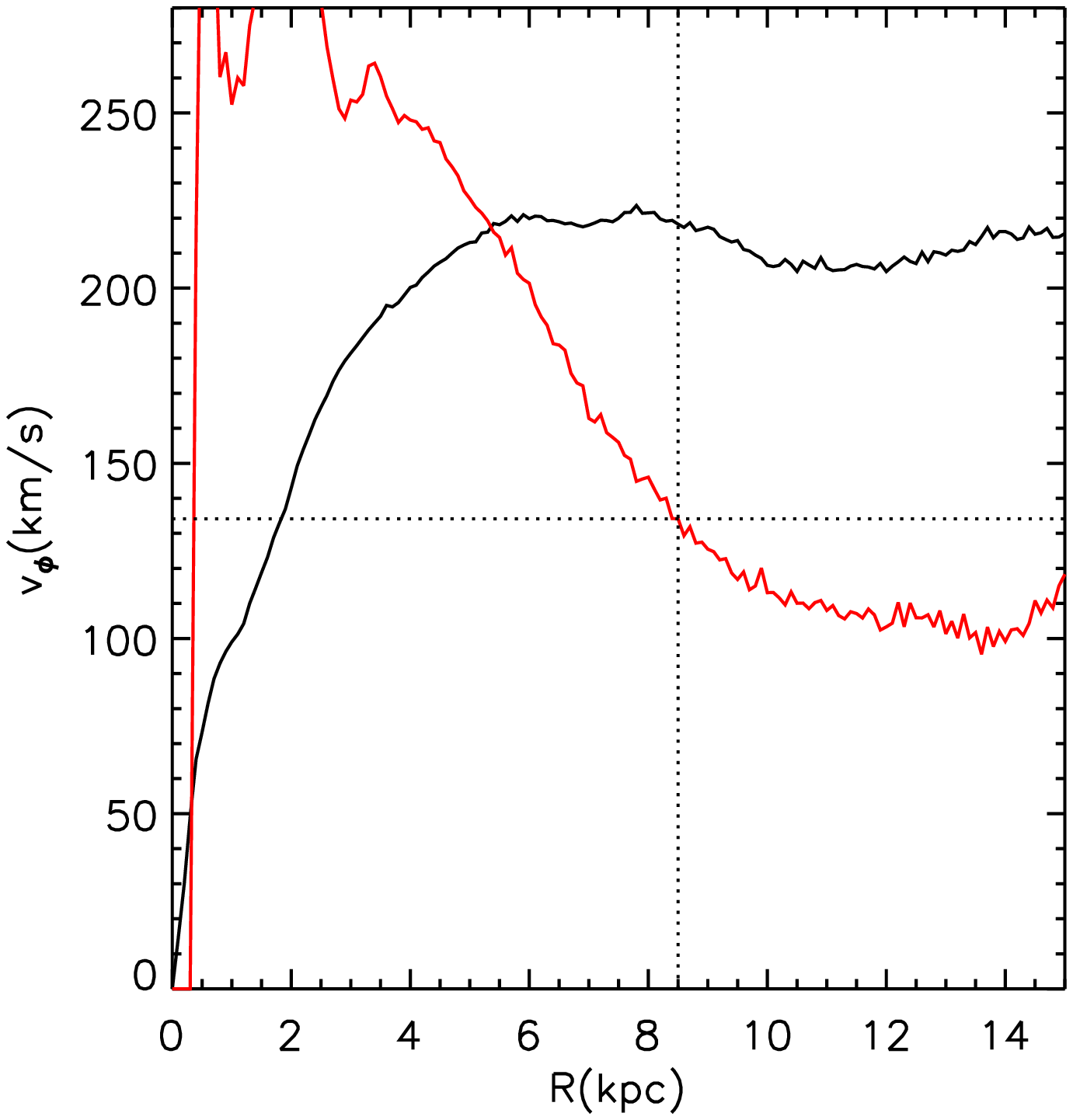}\hspace{\myfspace}
\includegraphics[width=0.22\textwidth]{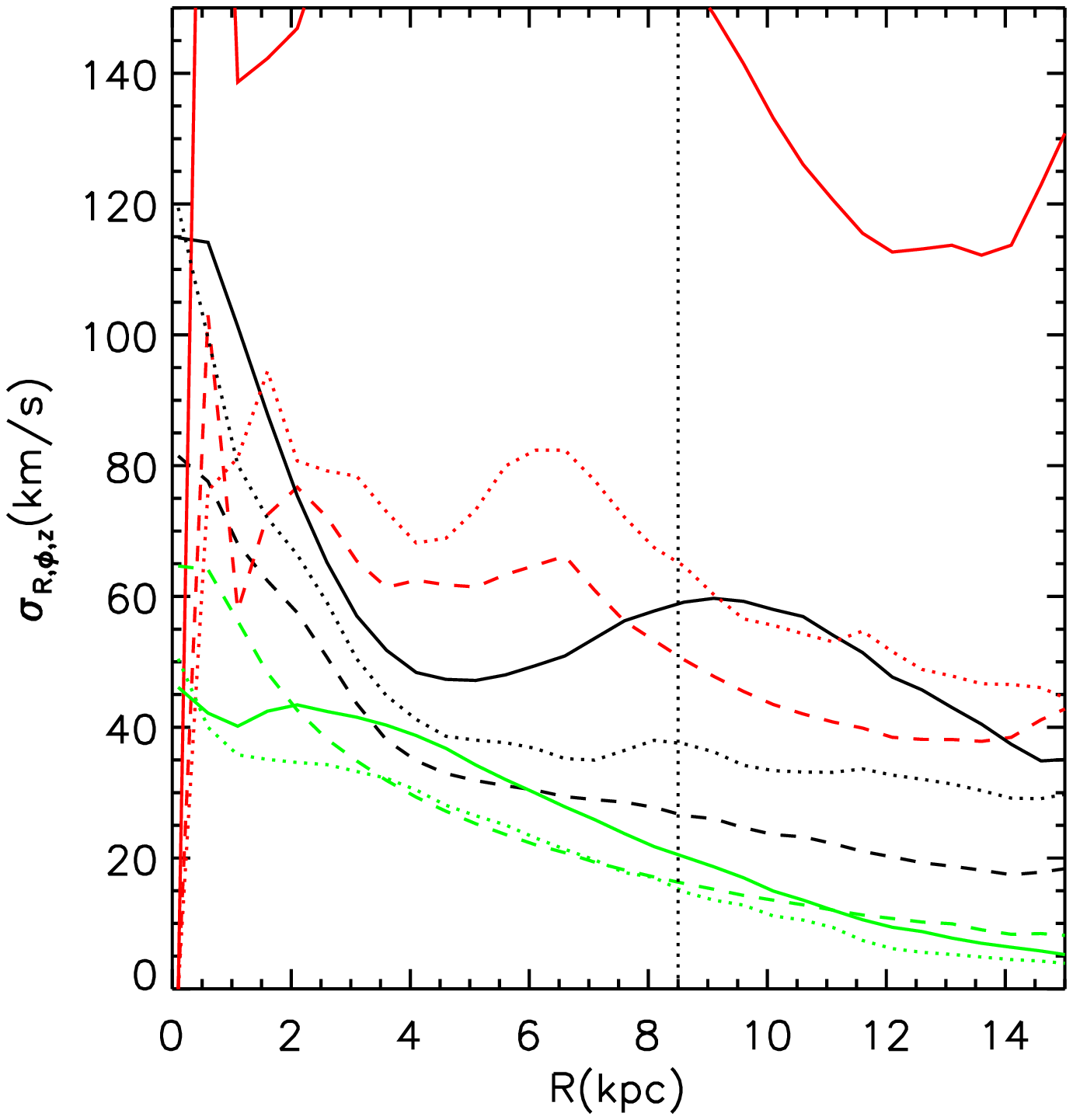}\hspace{\myfspace}\\

\includegraphics[width=0.22\textwidth]{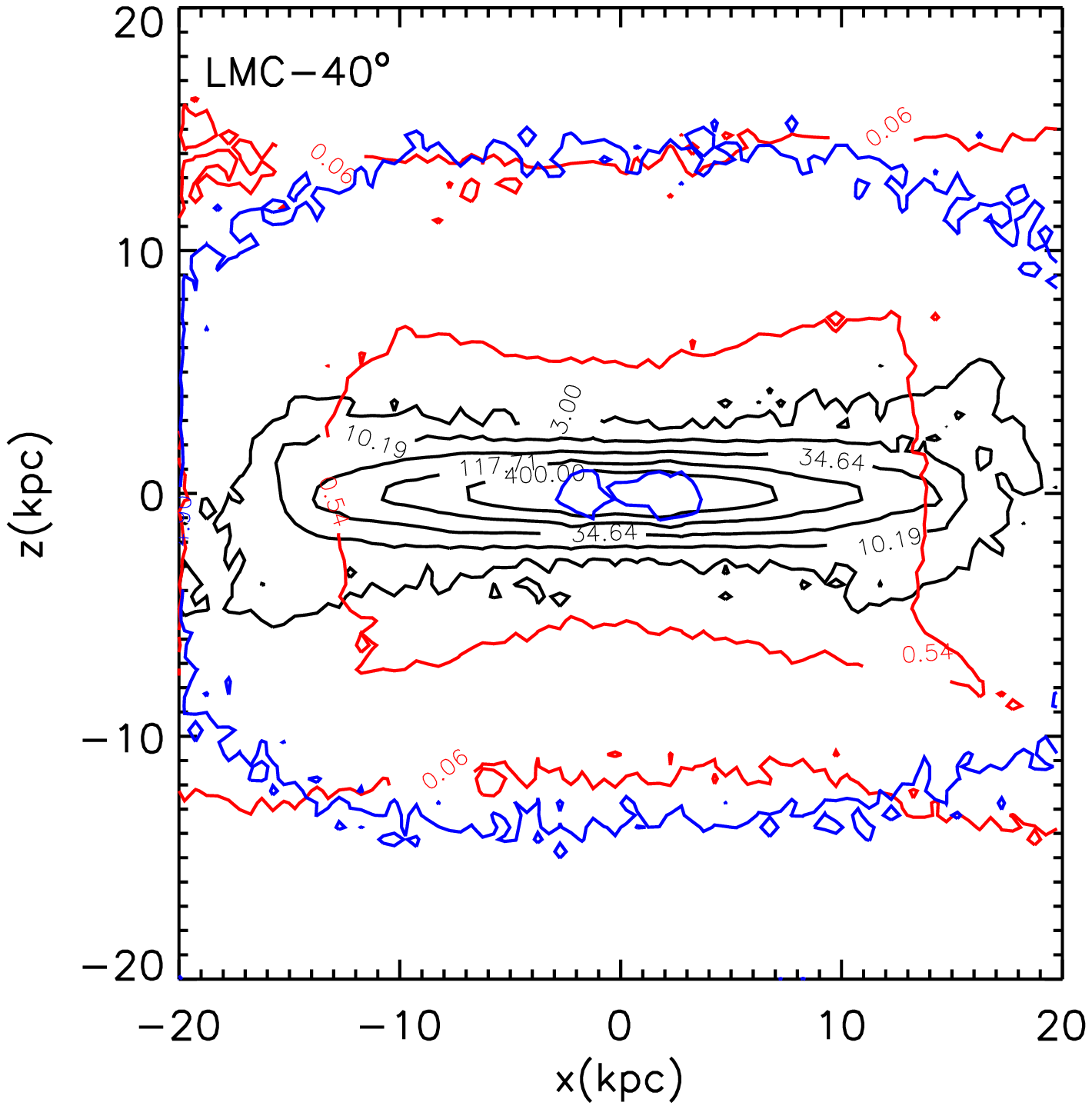}\hspace{\myfspace}
\includegraphics[width=0.22\textwidth]{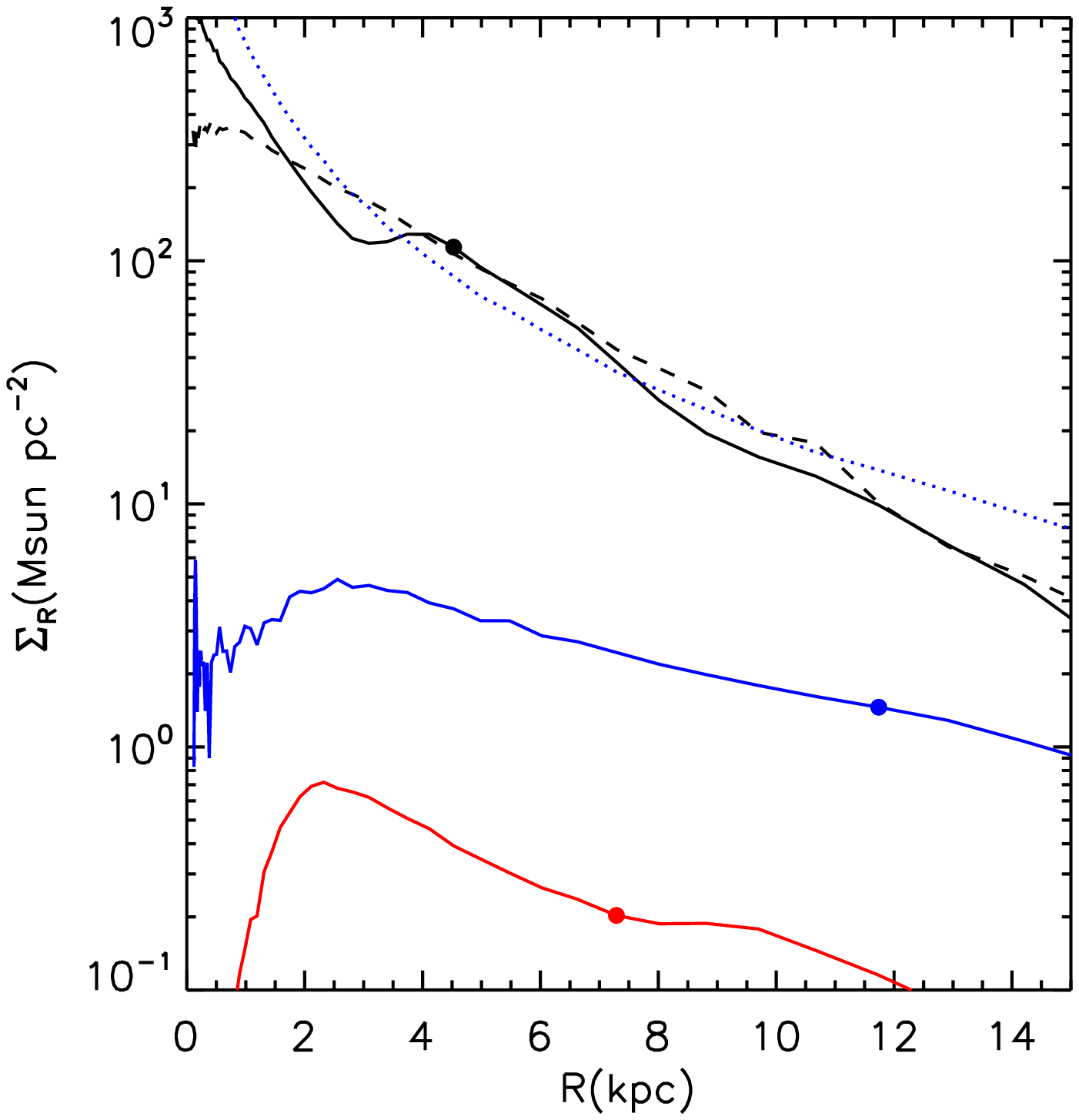}\hspace{\myfspace}
\includegraphics[width=0.22\textwidth]{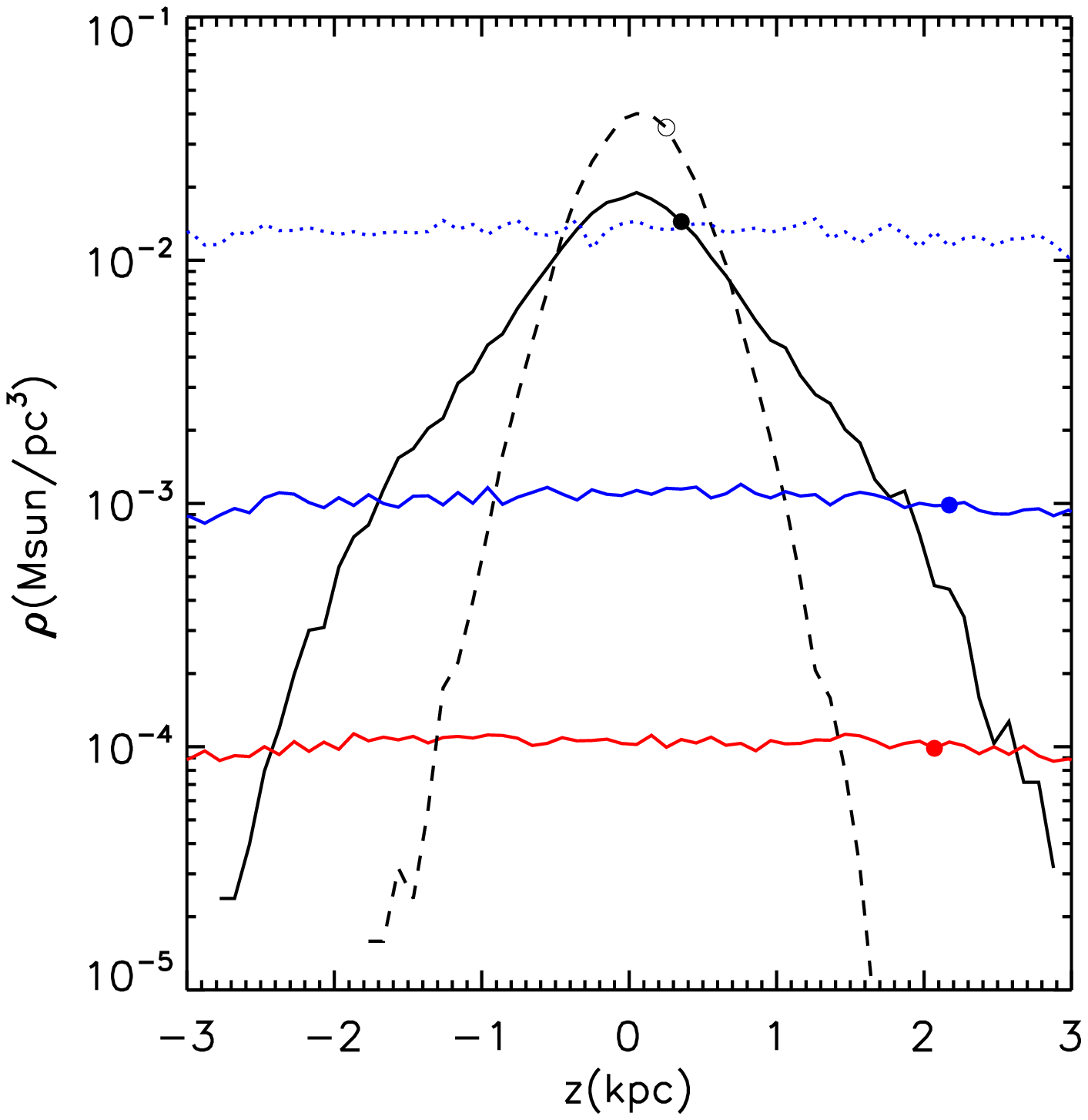}\hspace{\myfspace}
\includegraphics[width=0.22\textwidth]{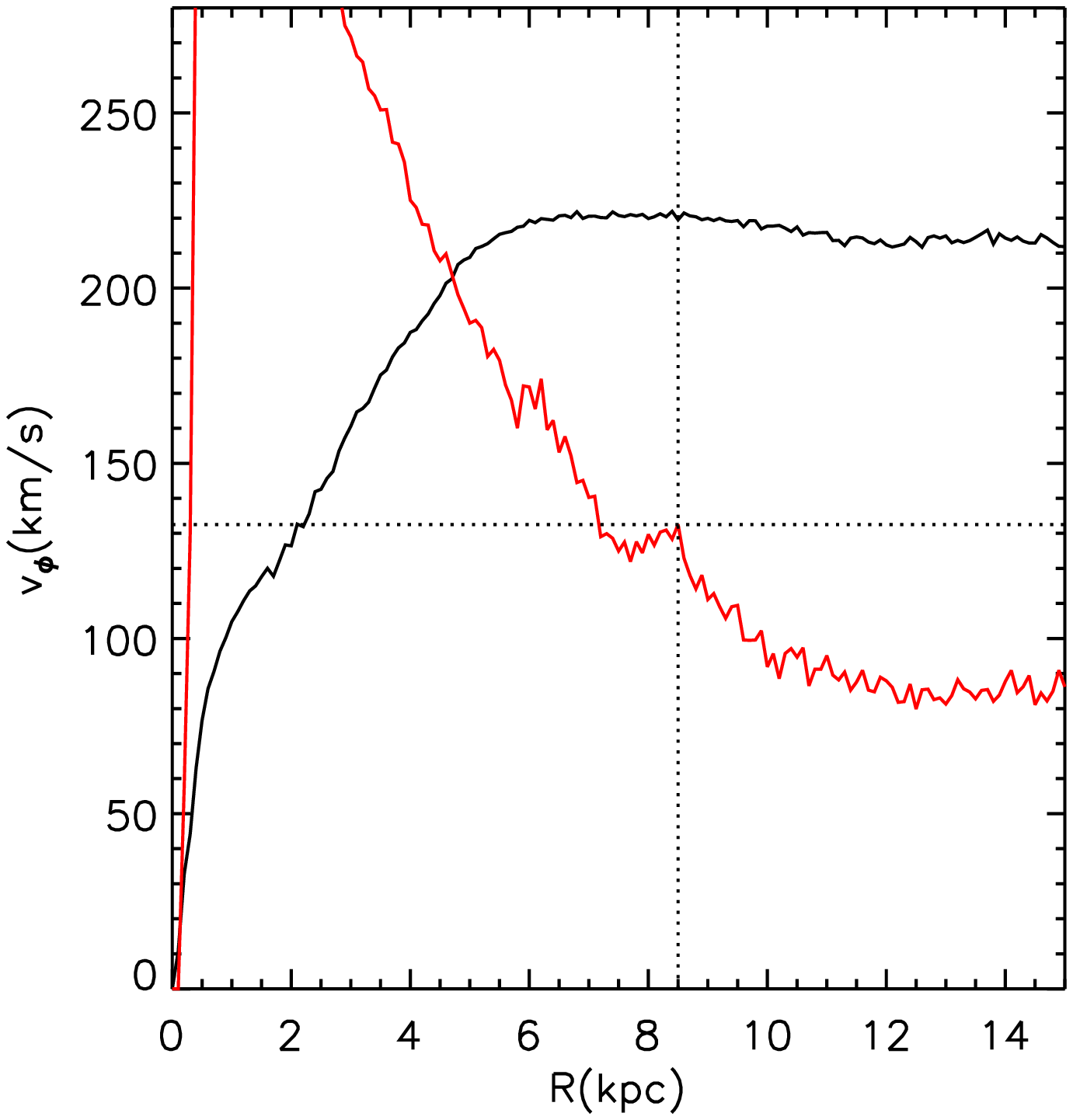}\hspace{\myfspace}
\includegraphics[width=0.22\textwidth]{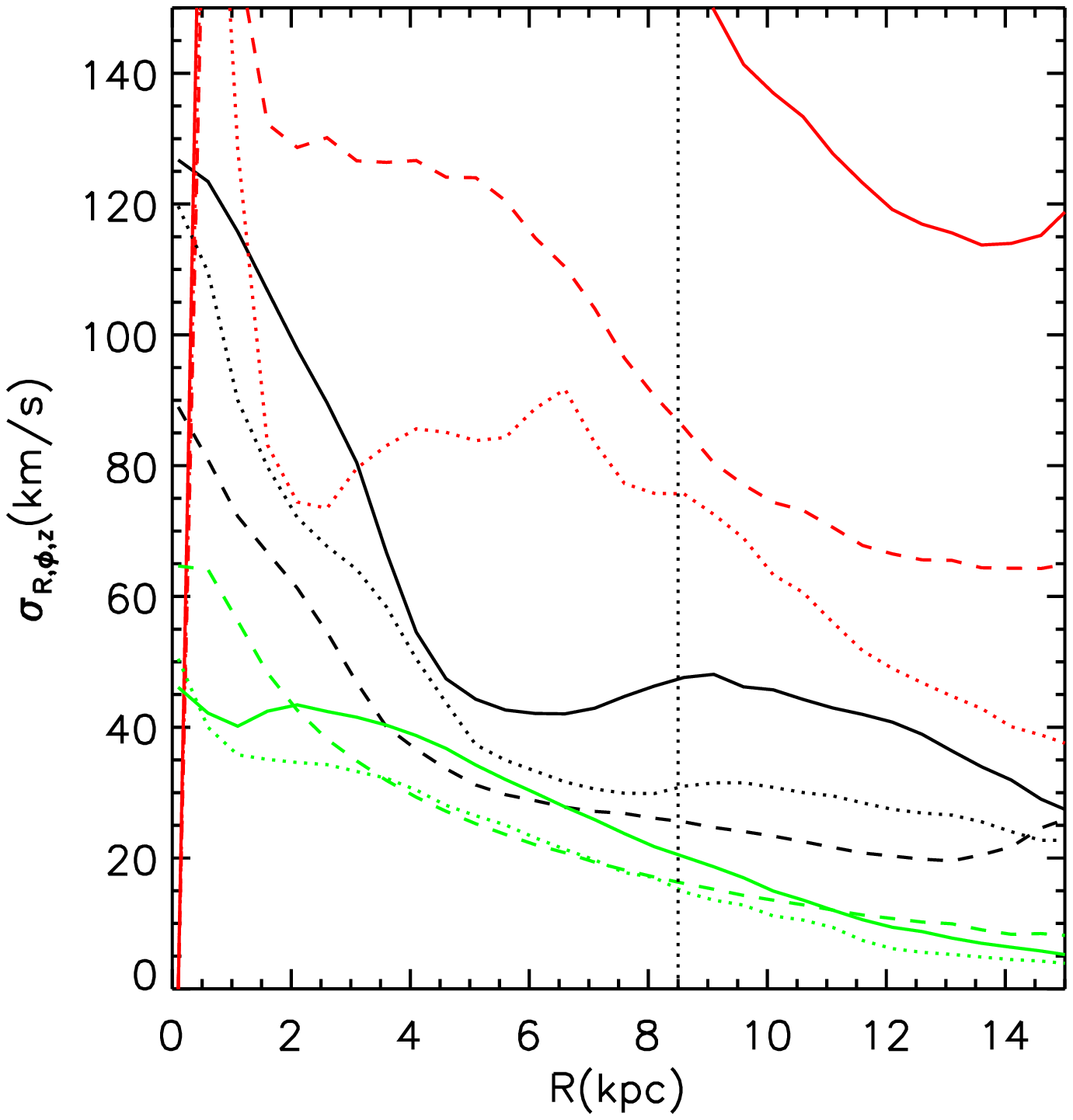}\hspace{\myfspace}\\

\includegraphics[width=0.22\textwidth]{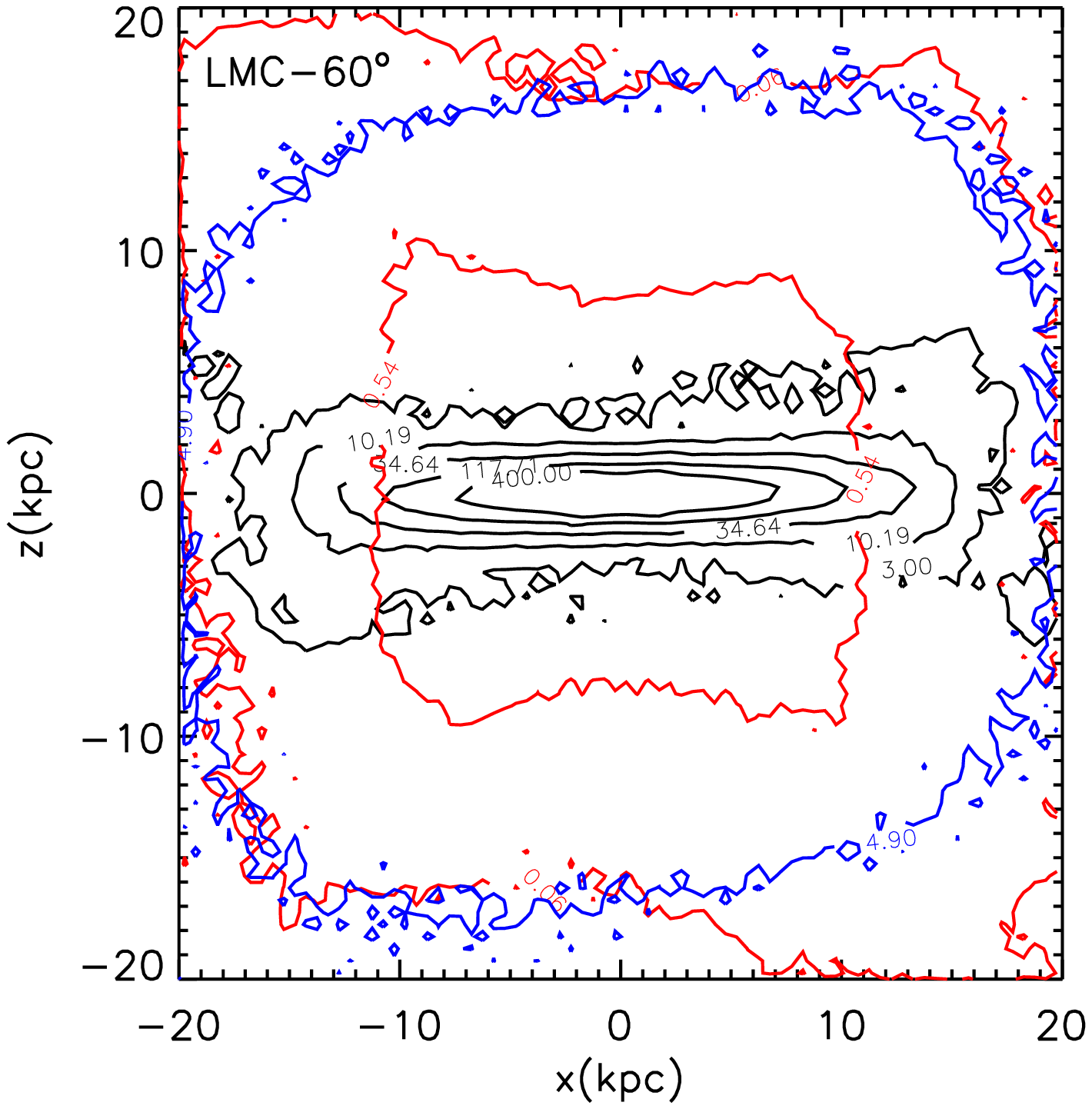}\hspace{\myfspace}
\includegraphics[width=0.22\textwidth]{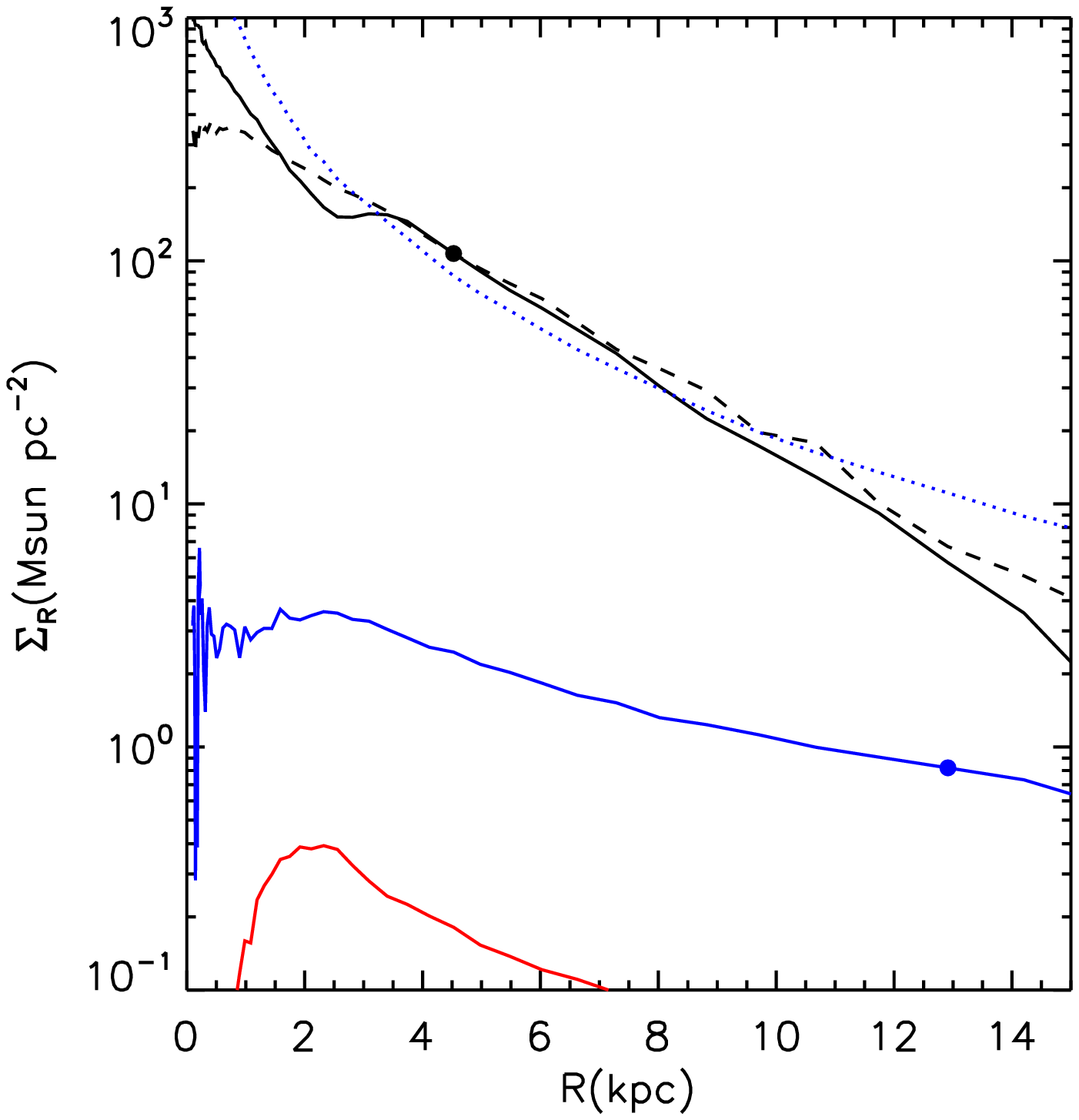}\hspace{\myfspace}
\includegraphics[width=0.22\textwidth]{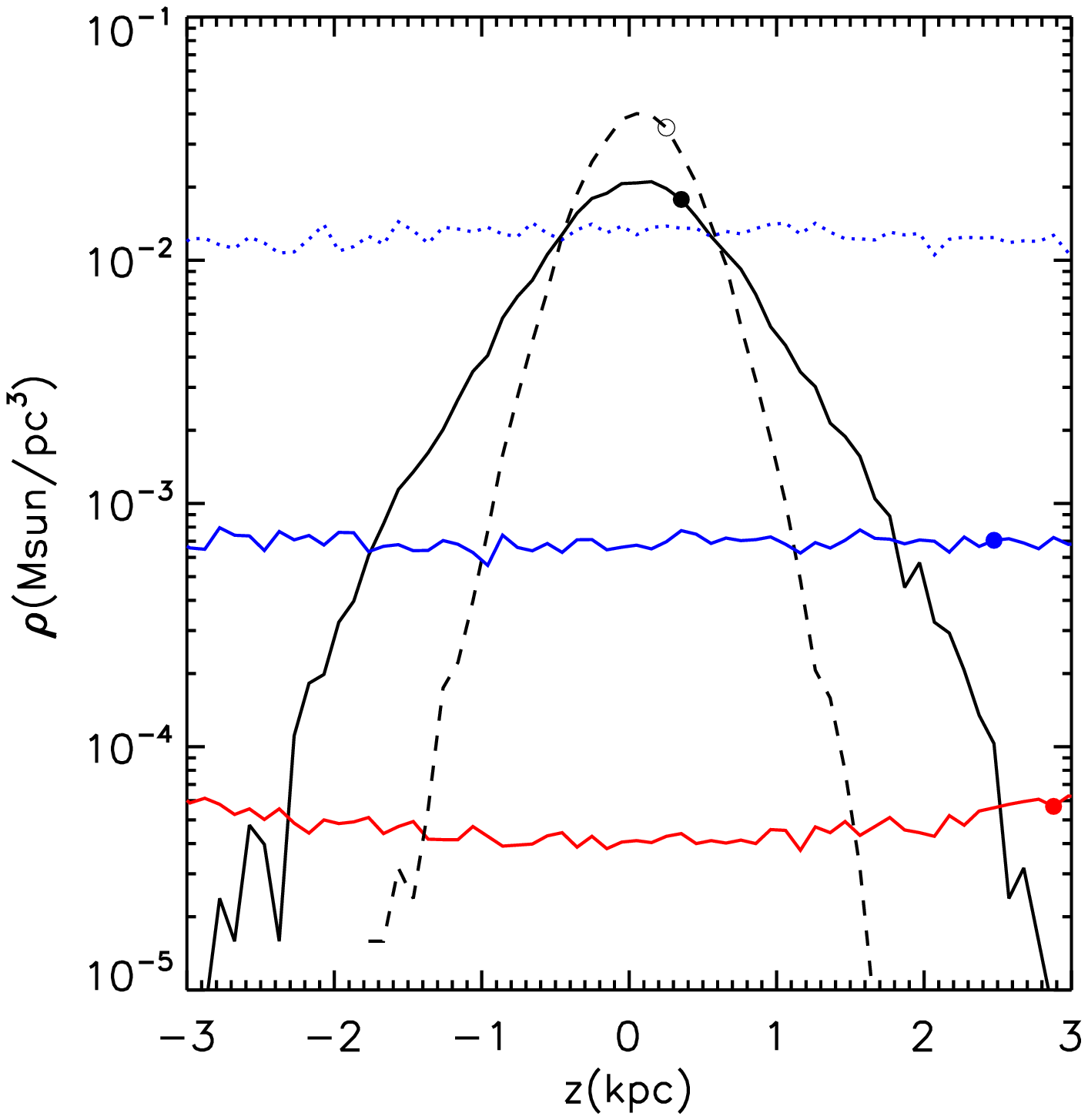}\hspace{\myfspace}
\includegraphics[width=0.22\textwidth]{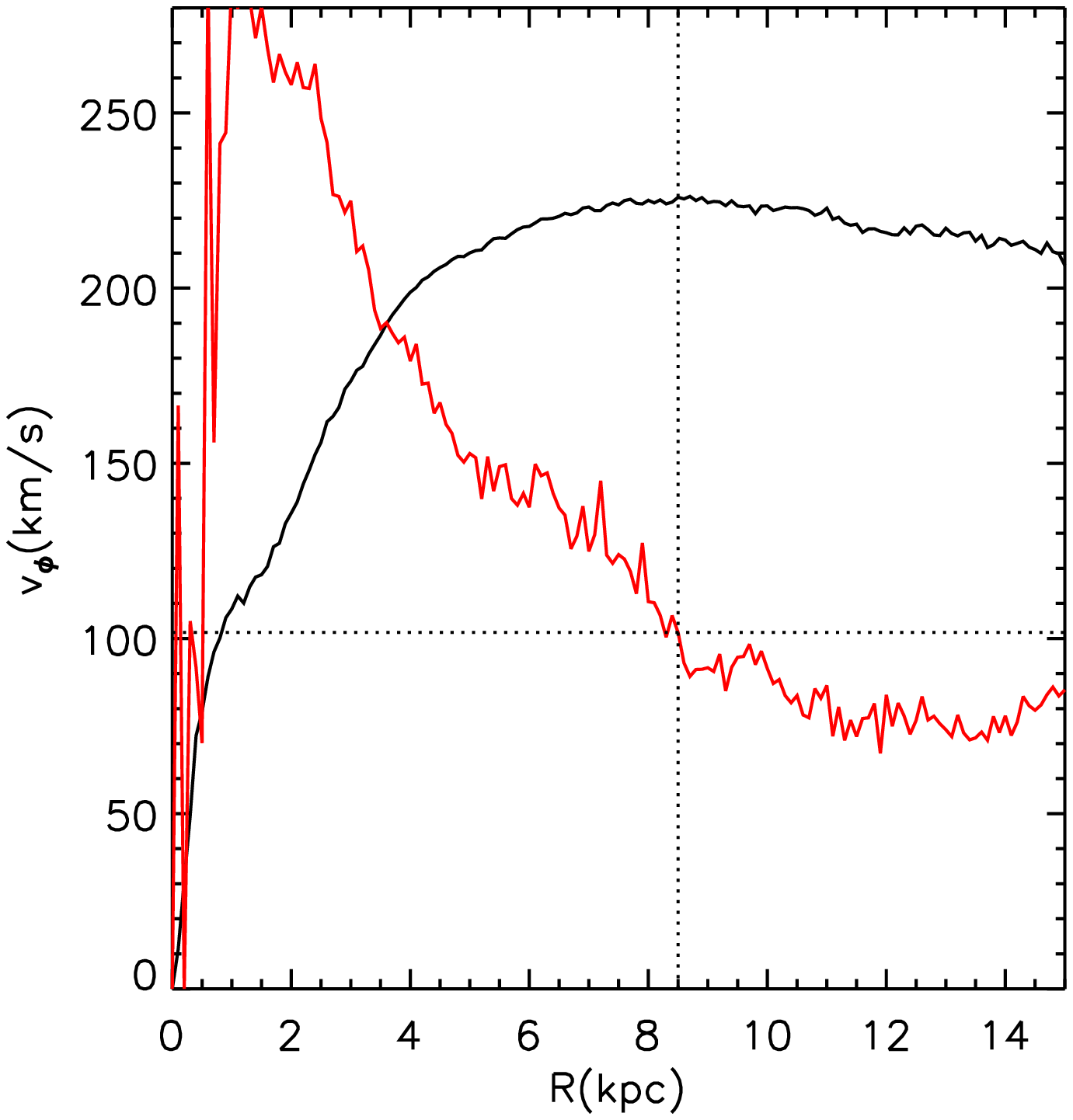}\hspace{\myfspace}
\includegraphics[width=0.22\textwidth]{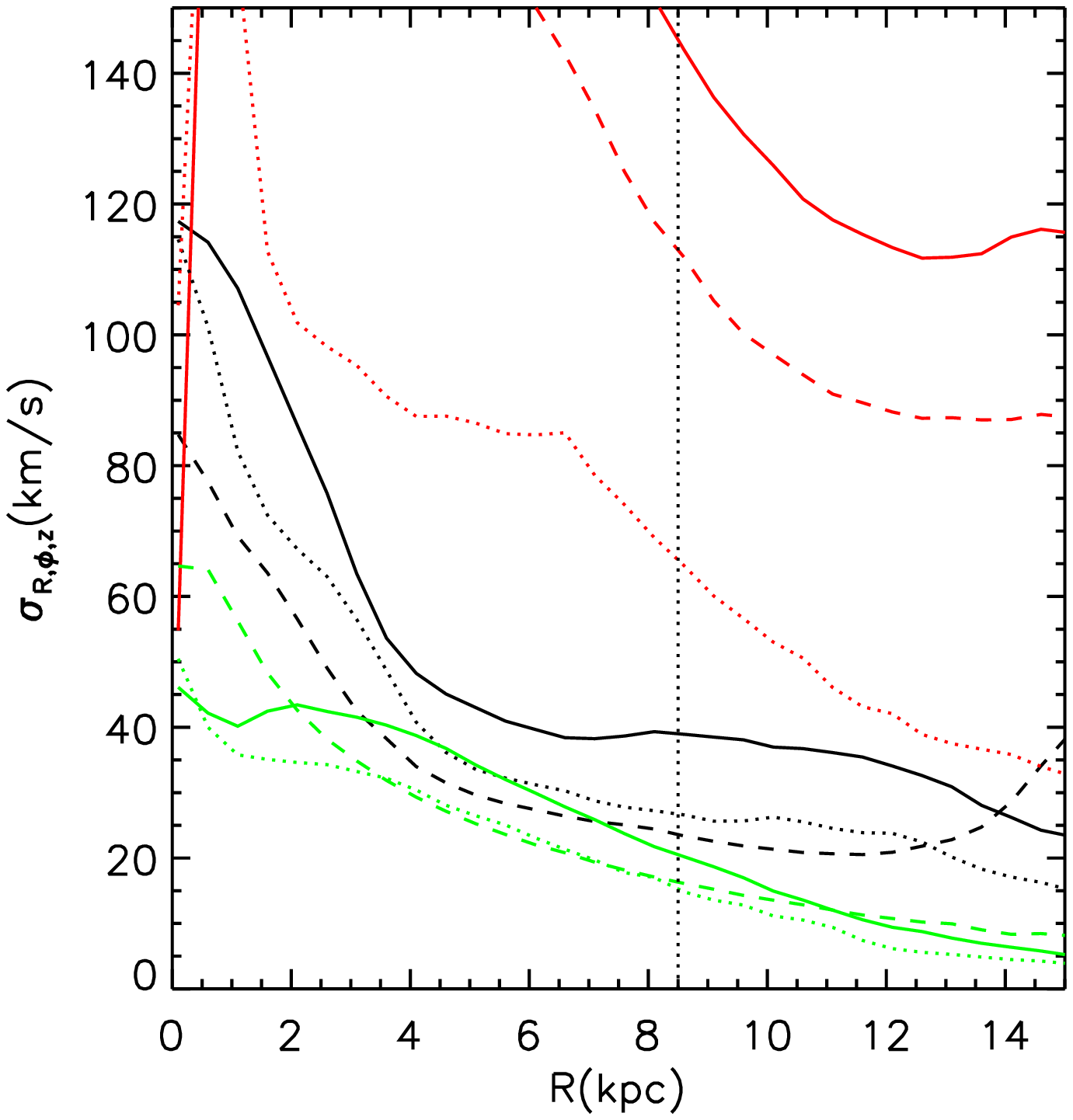}\hspace{\myfspace}\\

\caption{The effect of increasing impact angle: simulations: \LMCten, \LMCtwe, \LMCfor\ and \LMCsix. From left to right, the panels show: {\bf (a)} logarithmic density contours, viewed from side, in units of M\,$_\odot$\,pc$^{-2}$; {\bf (b)} surface density as a function of $R$ in a slice $|z|<1.1$\,kpc (the red dotted line [offset] for simulation LMC-10$^\mathrm{o}$ shows an exponential fit with scale length $R_0 = R_{1/2}/1.68 = 4.3$\,kpc); {\bf (c)} density as function of $z$ in a slice $8<R<9$\,kpc; {\bf (d)} the rotation curve ($v_\phi(R)$ for $|z|<0.35$\,kpc); and {\bf (e)} the $R$ (solid), $\phi$ (dotted) and $z$ (dashed) components of the stellar velocity dispersion as a function of projected radius. In all cases, we show the Milky Way stars (black), accreted satellite stars (red), Milky Way dark matter (blue dotted) and satellite accreted dark matter (blue). The black dashed lines (left three panels) and green lines (right panel) show the Milky Way disc initial conditions. The solid dots mark the half mass scale lengths for each component. The black dotted lines mark the solar position. Notice that a thick disc of stars forms for impact angles $\simlt 20^\mathrm{o}$, with a corresponding thick disc of accreted dark matter. Higher impact angles give more boxy stellar and dark matter distributions that rotate more slowly and are hotter. They also do increasing damage to the Milky Way thin disc, producing a flared outer disc that for \LMCfor \ and \LMCsix\ is better described as a warp.}
\label{fig:impact1}
\end{center}
\end{figure*}

\subsubsection{Varying the impact angle}\label{sec:impact}

The rows in Figure \ref{fig:impact1} show the effect of increasing the impact angle: simulations \LMCten, \LMCtwe, \LMCfor\ and \LMCsix. From left to right, the panels show: (a) logarithmic density contours, viewed from side, in units of M\,$_\odot$\,pc$^{-2}$; (b) surface density as a function of $R$ in a slice $|z|<1.1$\,kpc (the red dotted line for simulation \LMCten\ shows an exponential fit with scale length $R_0 = R_{1/2}/1.68 = 4.3$\,kpc); (c) density as function of $z$ in a slice $8<R<9$\,kpc; (d) the rotation curve ($v_\phi(R)$ for $|z|<0.35$\,kpc); and (e) the $R$ (solid), $\phi$ (dotted) and $z$ (dashed) components of the stellar velocity dispersion as a function of projected radius. In all cases, we show the Milky Way stars (black), accreted satellite stars (red), Milky Way dark matter (blue dotted) and satellite accreted dark matter (blue). The black dashed lines (left three panels) and green lines (right panel) show the Milky Way disc initial conditions. The solid dots mark the half mass scale lengths for each component. 

\begin{figure}
\begin{center}
\includegraphics[width=0.5\textwidth]{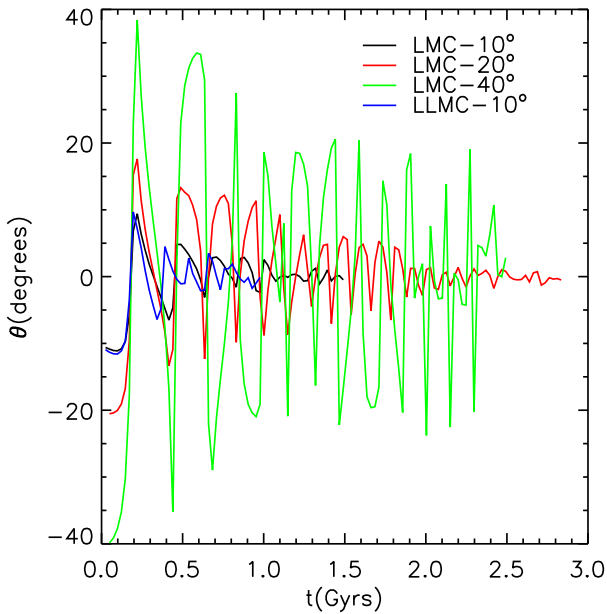}
\caption{The satellite-disc inclination angle $\theta$ as a function of time; lines are truncated when the satellite is fully accreted. For $\theta < 20^\mathrm{o}$ initially, the satellite is dragged into the disc plane; for $\theta=40^\mathrm{o}$ initially, the satellite remains out of the disc plane at all times. Increasing the satellite mass (\LLMC) leads to faster disc plane dragging.}
\label{fig:thetaplot}
\end{center}
\end{figure}

For $\theta < 20^\mathrm{o}$ we form a thick disc of stars and dark matter. At the solar neighbourhood ($8 < R < 9$\,kpc) these lag the thin disc rotation (by 20-85\,km/s); are hotter ($\sigma_z \sim 40-60$\,km/s); and are of longer scale length and larger scale height ($R_{1/2} \sim 7$\,kpc; $z_{1/2}\sim0.9$\,kpc). The rotational lag of the stellar thick disc and $\sigma_z$ both increase with the satellite inclination angle, $\theta$. Notice that the thick disc stars lead the rotation towards the Galactic centre. This is because the thick disc distributions have a hole in the centre (see Figure \ref{fig:impact1}(b)). At the solar neighbourhood, a large fraction of thick disc stars are near apocentre and have small tangential velocity. Towards the Galactic centre, this cannot be maintained because of the decreasing density; most thick disc stars at radii $\simlt 4$\,kpc are near pericentre and have high tangential velocity. 

For $\theta > 20^\mathrm{o}$ we no longer form a thick disc of stars or dark matter. Instead, we have boxy stellar and dark matter distributions that are much hotter ($\sigma_z > 100$\,km/s) and rotate more slowly ($v_c < 100$\,km/s). The boxy stellar material is a strong candidate for the new `inner halo' structure, recently discovered by \citet{2007arXiv0706.3005C} (see Table \ref{tab:milkyway}). A similar retrograde encounter could also explain some of the outer halo. Recall that these high inclination encounters ($\theta > 20^\mathrm{o}$) are twice as likely as the lower inclination ones (\S\ref{sec:cosmology}). 

The movement from disc-like to boxy structures is also clearly seen in Figure \ref{fig:impact1}(c). Notice that for \LMCten, there is a definite disc in the accreted stars (red line) and dark matter (blue line). For \LMCtwe, the discs expand in scale height and  length. For \LMCfor\ and \LMCsix, both distributions are constant in density over the range $-3 < z < 3$ and can no longer be described as discs. 

Figure \ref{fig:thetaplot} demonstrates that dynamical friction plane dragging is responsible for the formation of accreted thick discs. In all simulations, the inclination angle of the satellite to the disc decreases with time. However, only for $\theta < 20^\mathrm{o}$ initially, is the satellite completely dragged into the disc plane.

\subsubsection{Varying the orbit}\label{sec:orbits}

\begin{figure*}
\begin{center}
\includegraphics[width=0.22\textwidth]{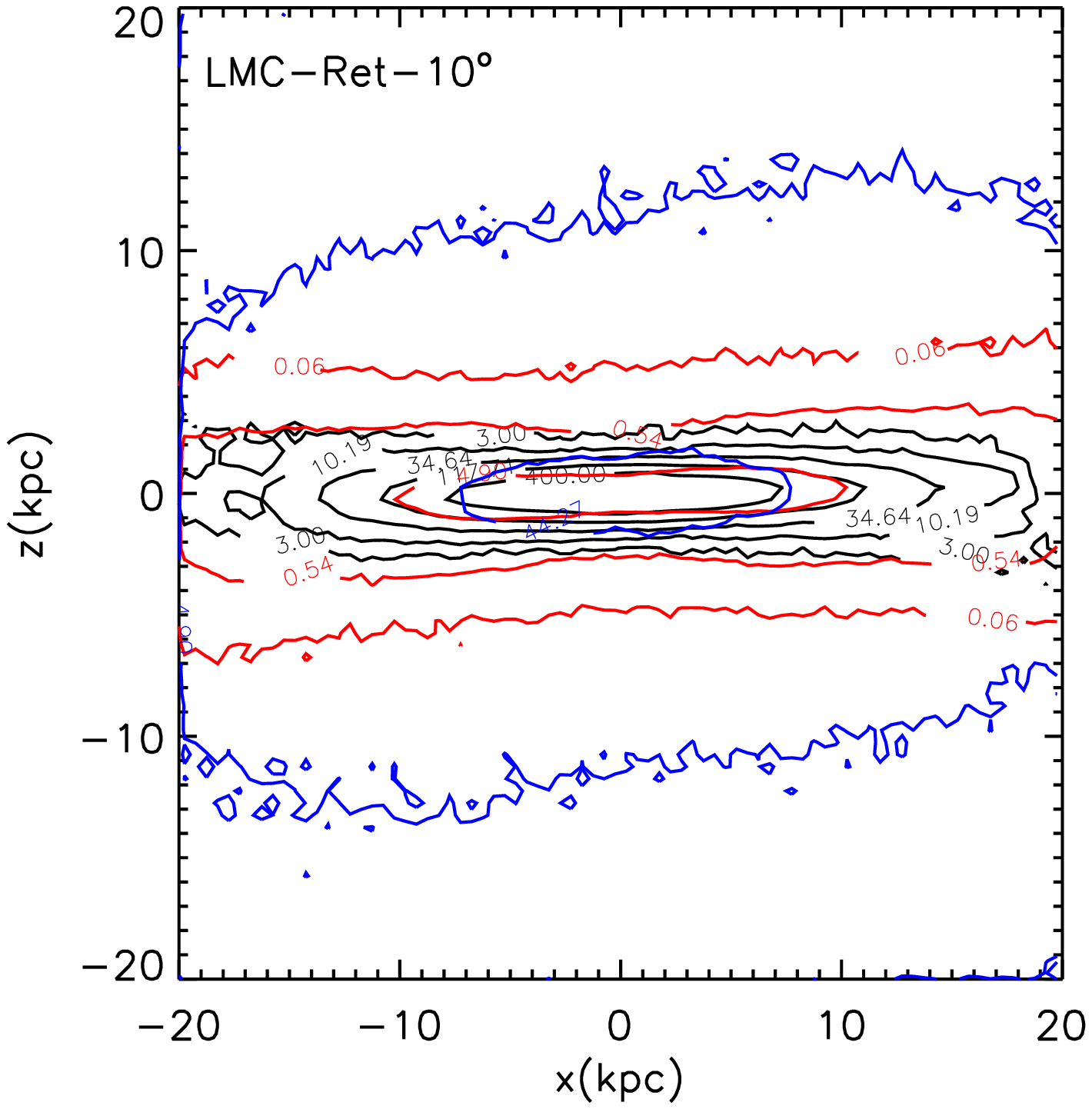}\hspace{\myfspace}
\includegraphics[width=0.22\textwidth]{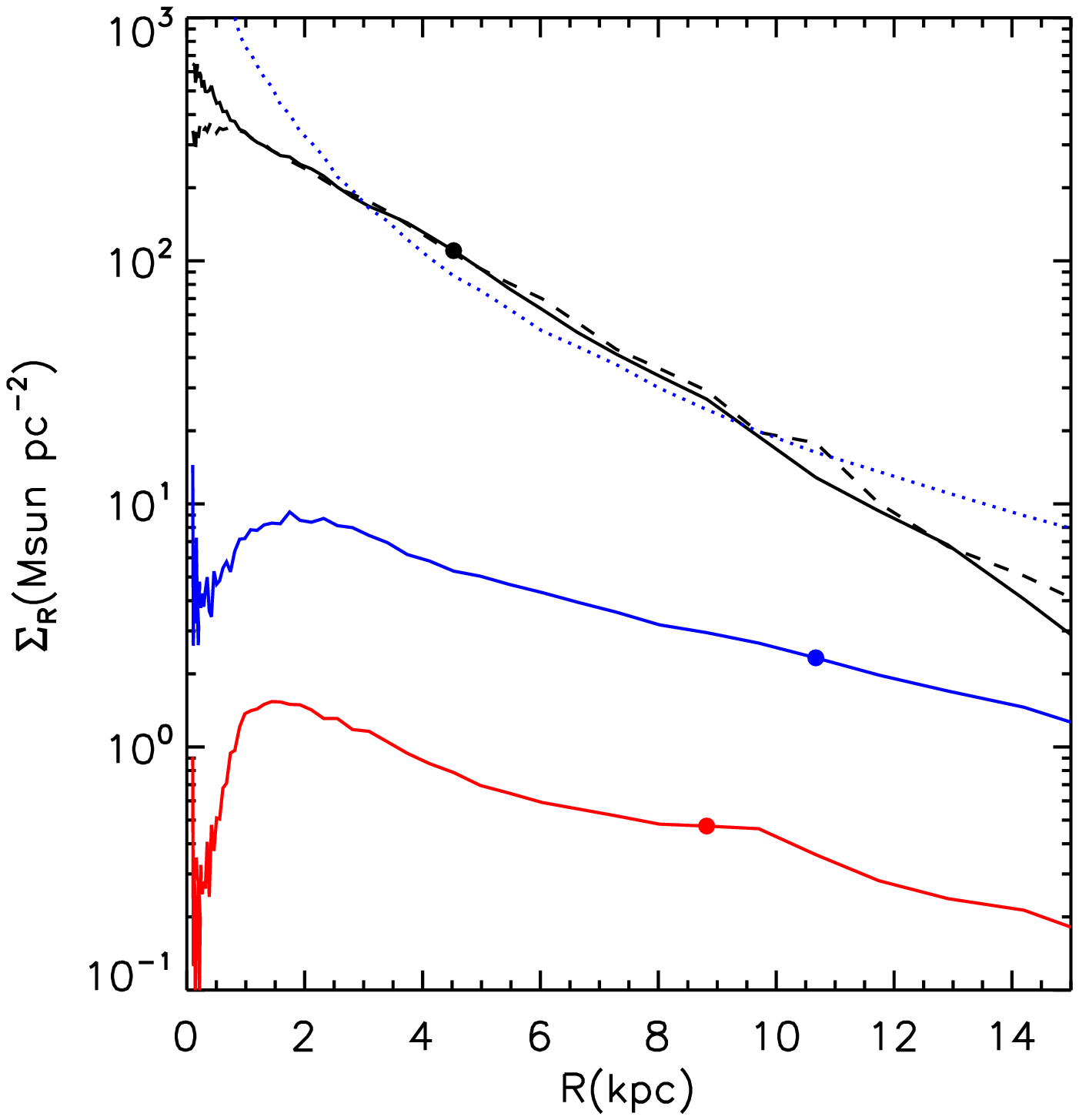}\hspace{\myfspace}
\includegraphics[width=0.22\textwidth]{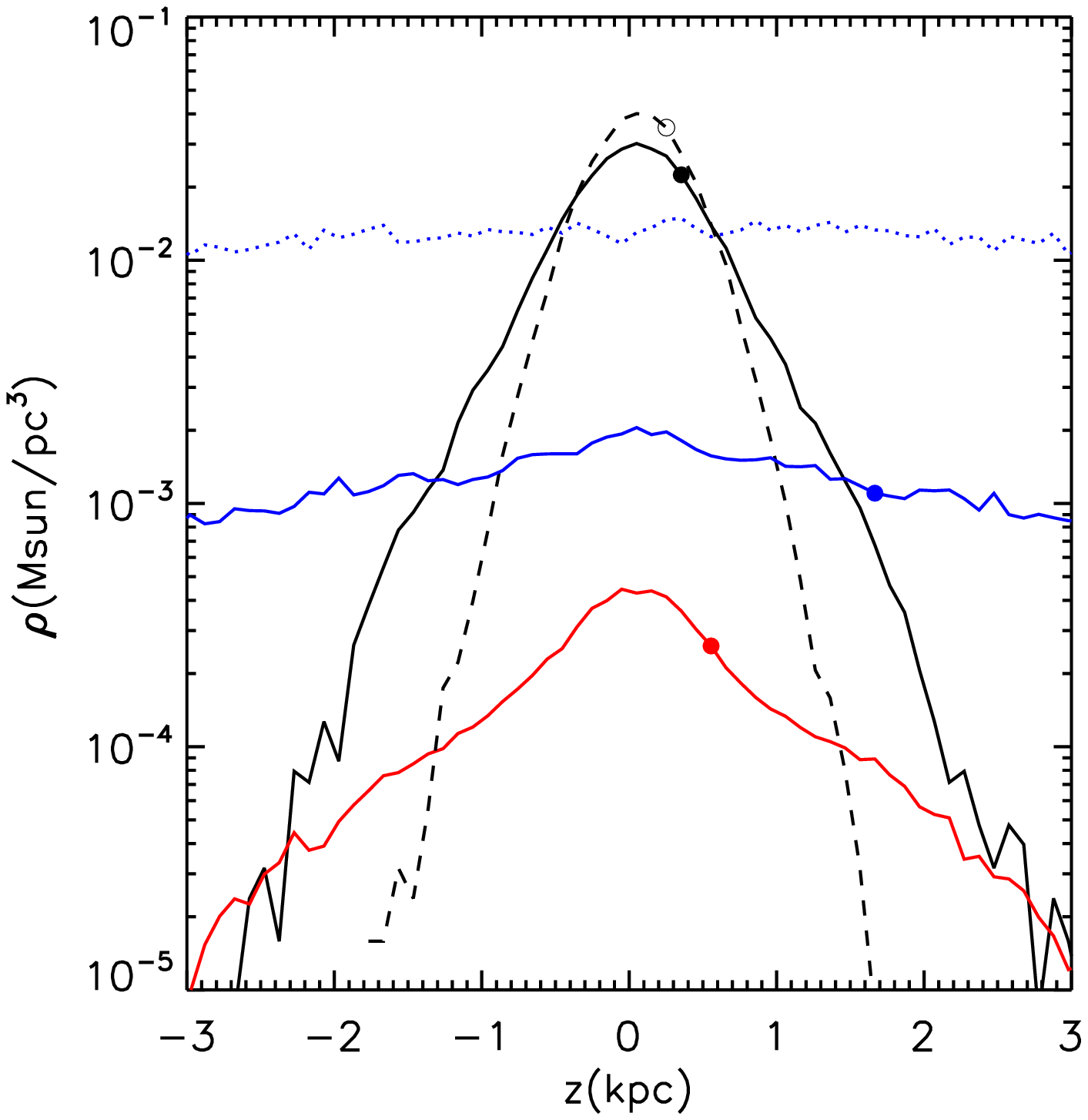}\hspace{\myfspace}
\includegraphics[width=0.22\textwidth]{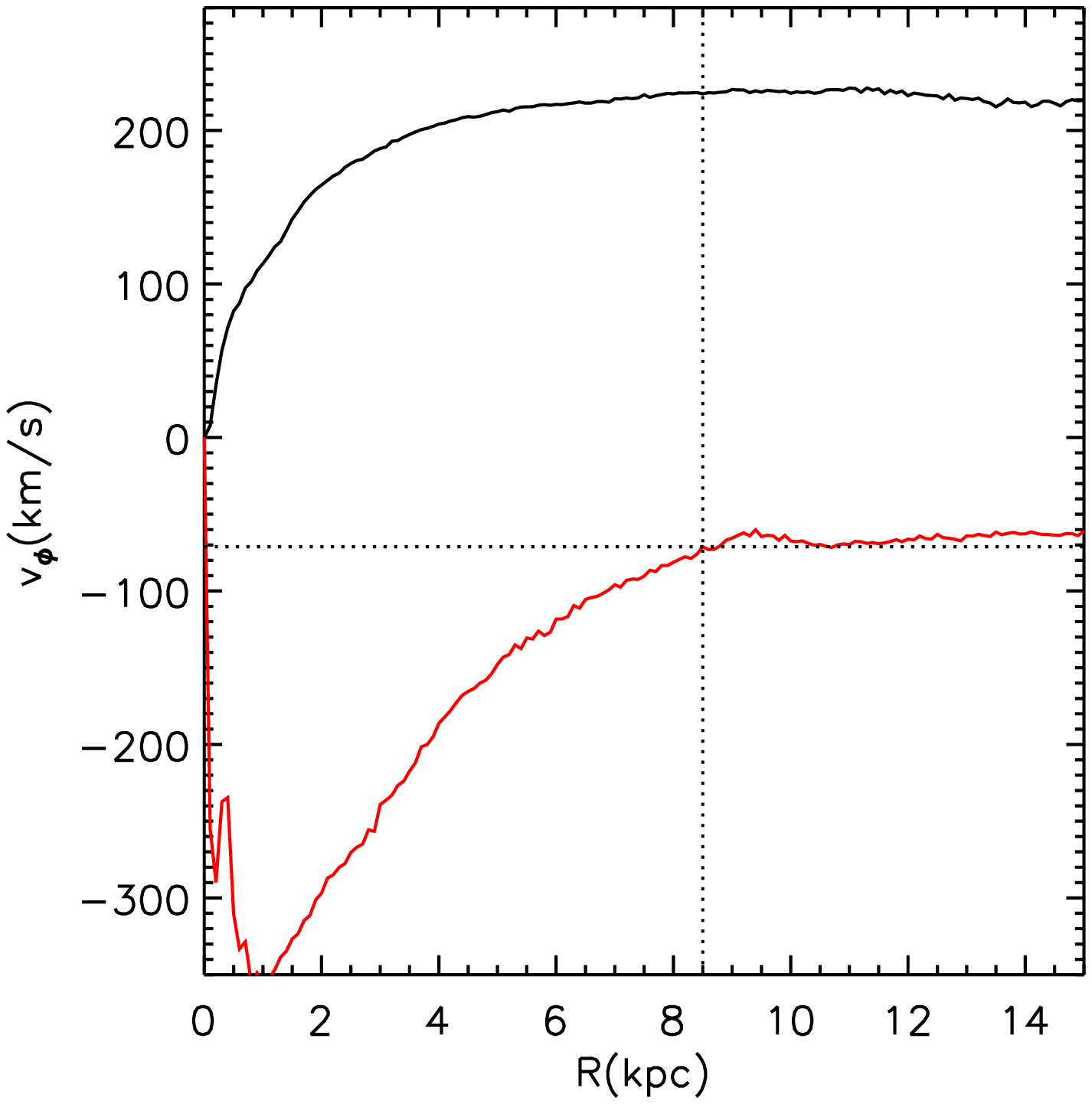}\hspace{\myfspace}
\includegraphics[width=0.22\textwidth]{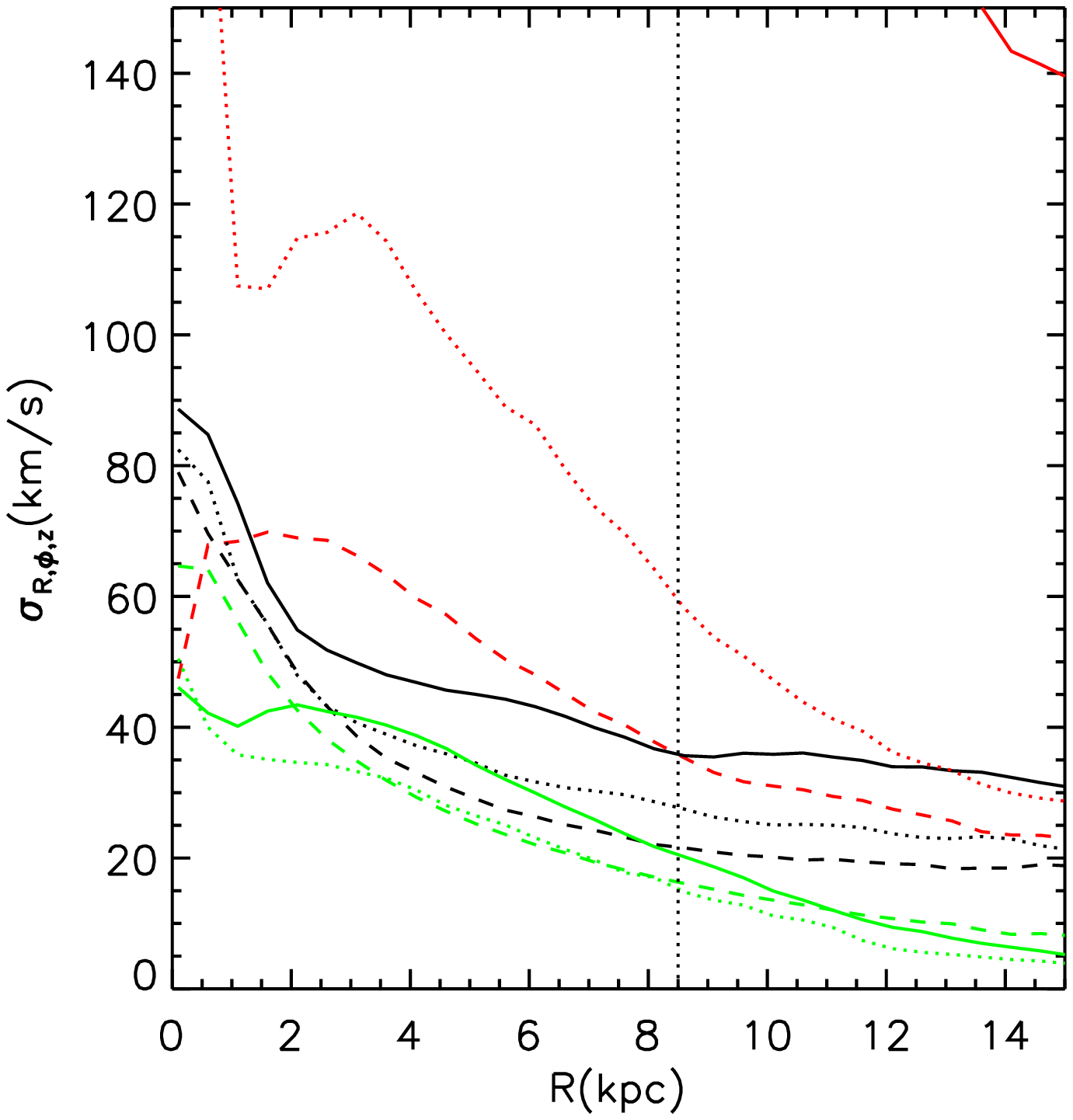}\hspace{\myfspace}\\

\includegraphics[width=0.22\textwidth]{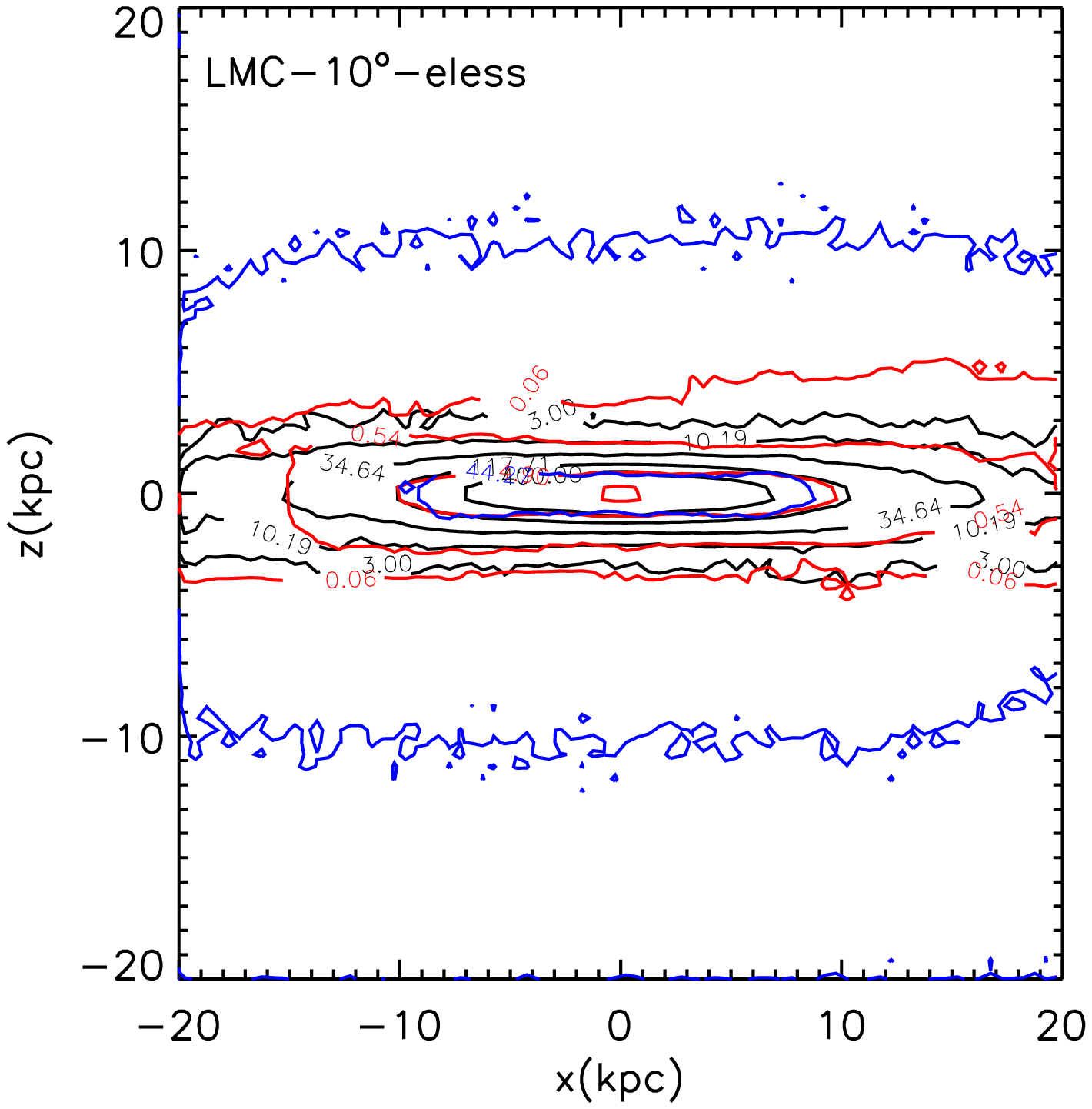}\hspace{\myfspace}
\includegraphics[width=0.22\textwidth]{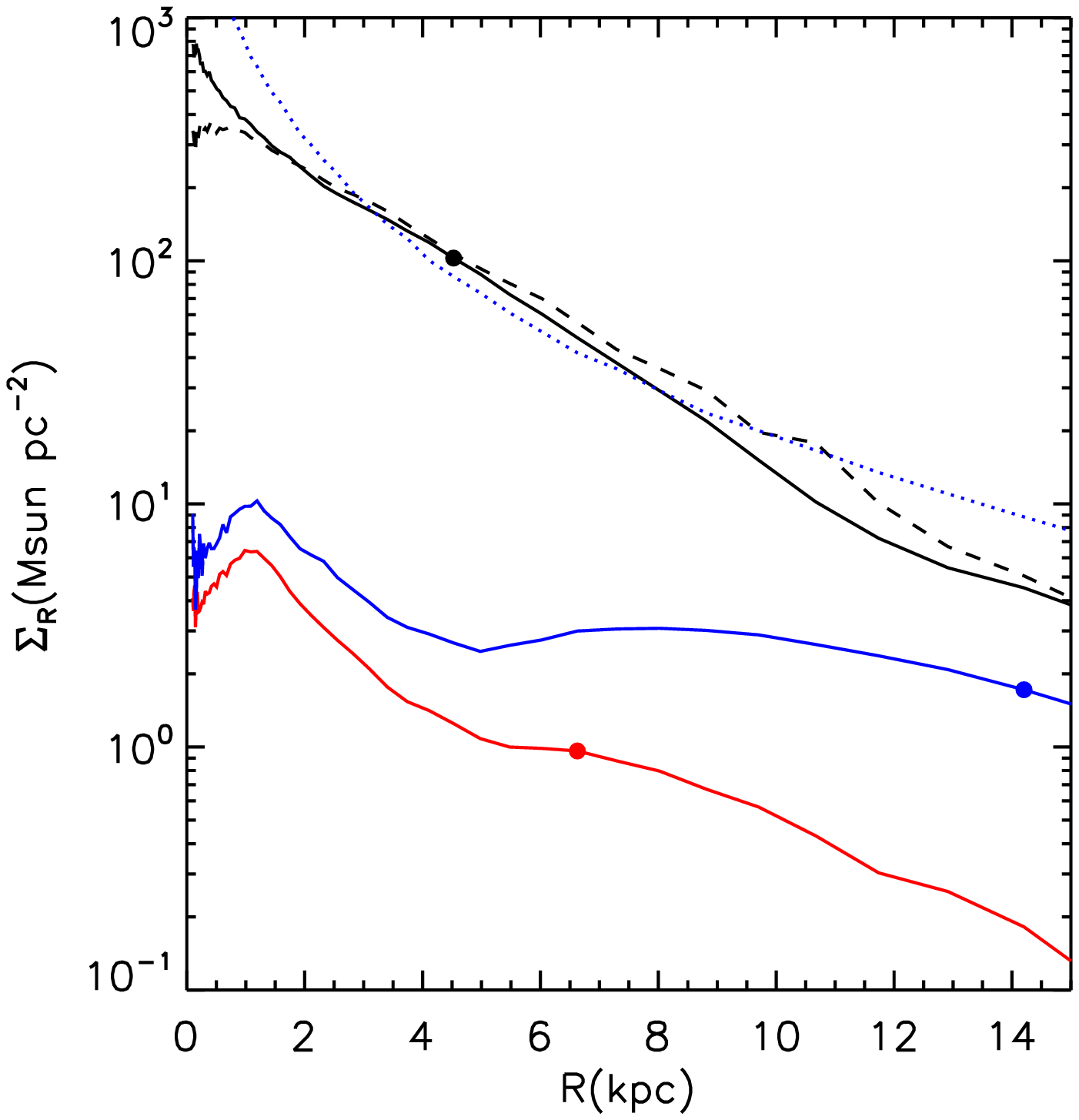}\hspace{\myfspace}
\includegraphics[width=0.22\textwidth]{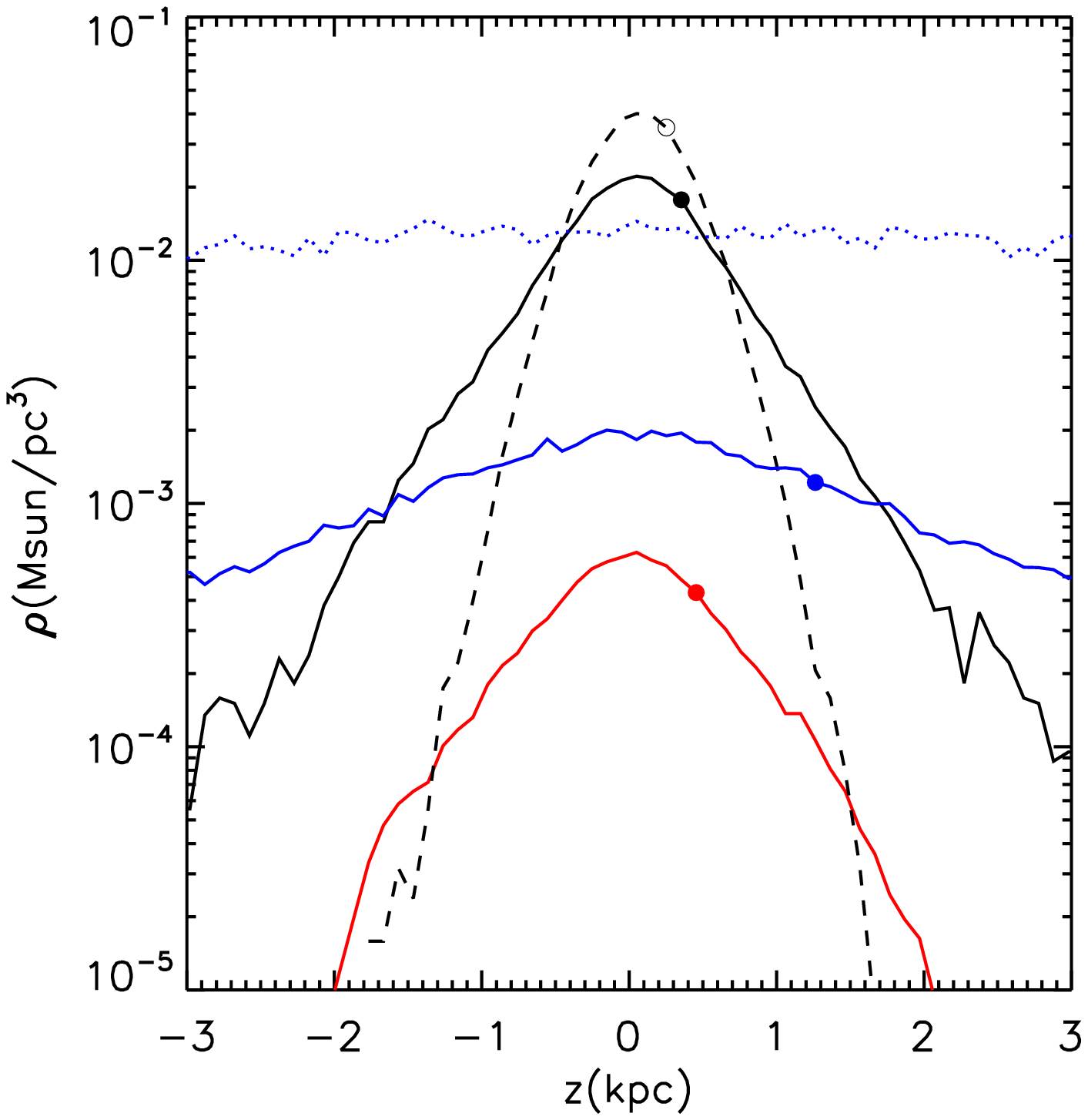}\hspace{\myfspace}
\includegraphics[width=0.22\textwidth]{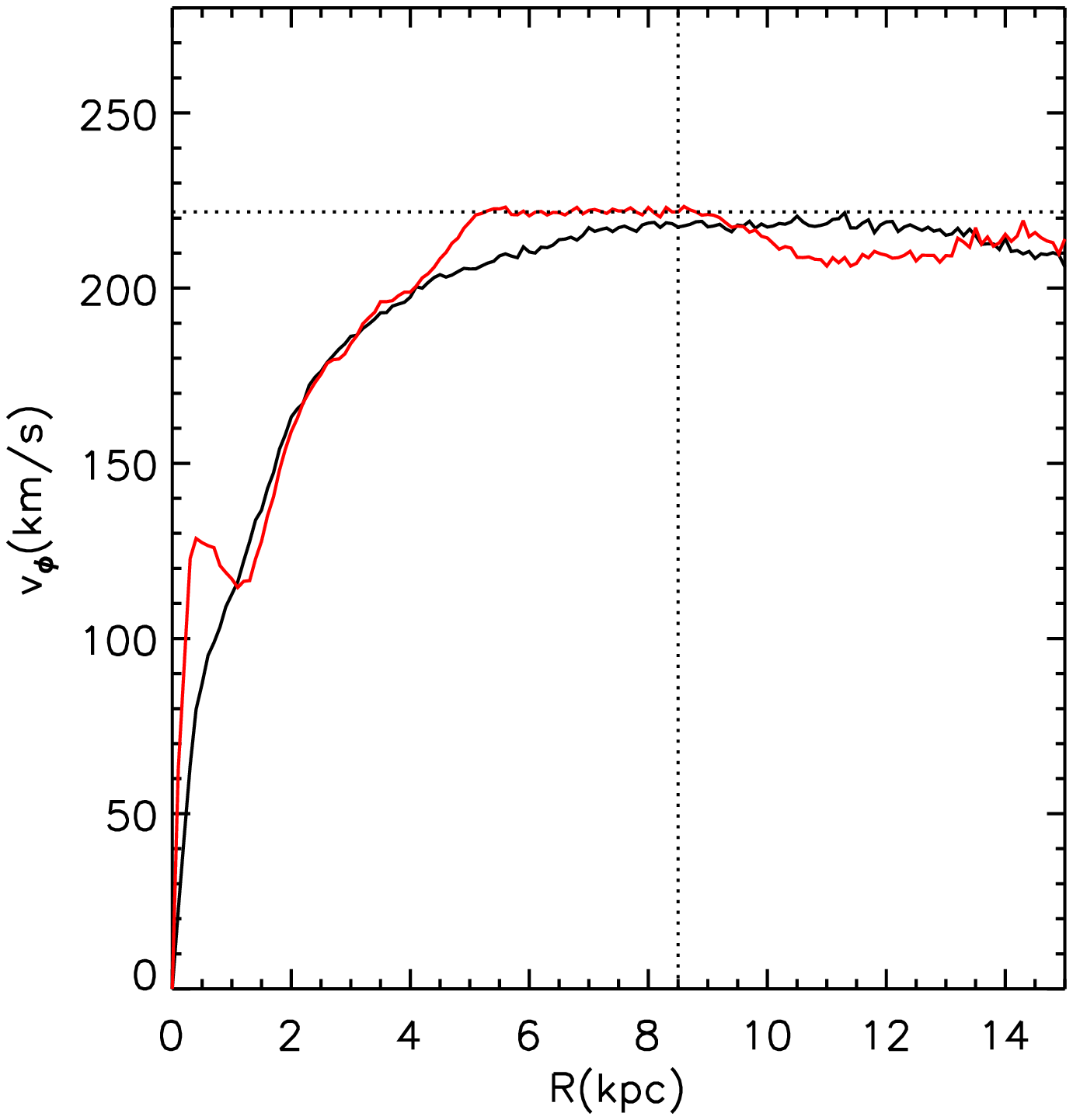}\hspace{\myfspace}
\includegraphics[width=0.22\textwidth]{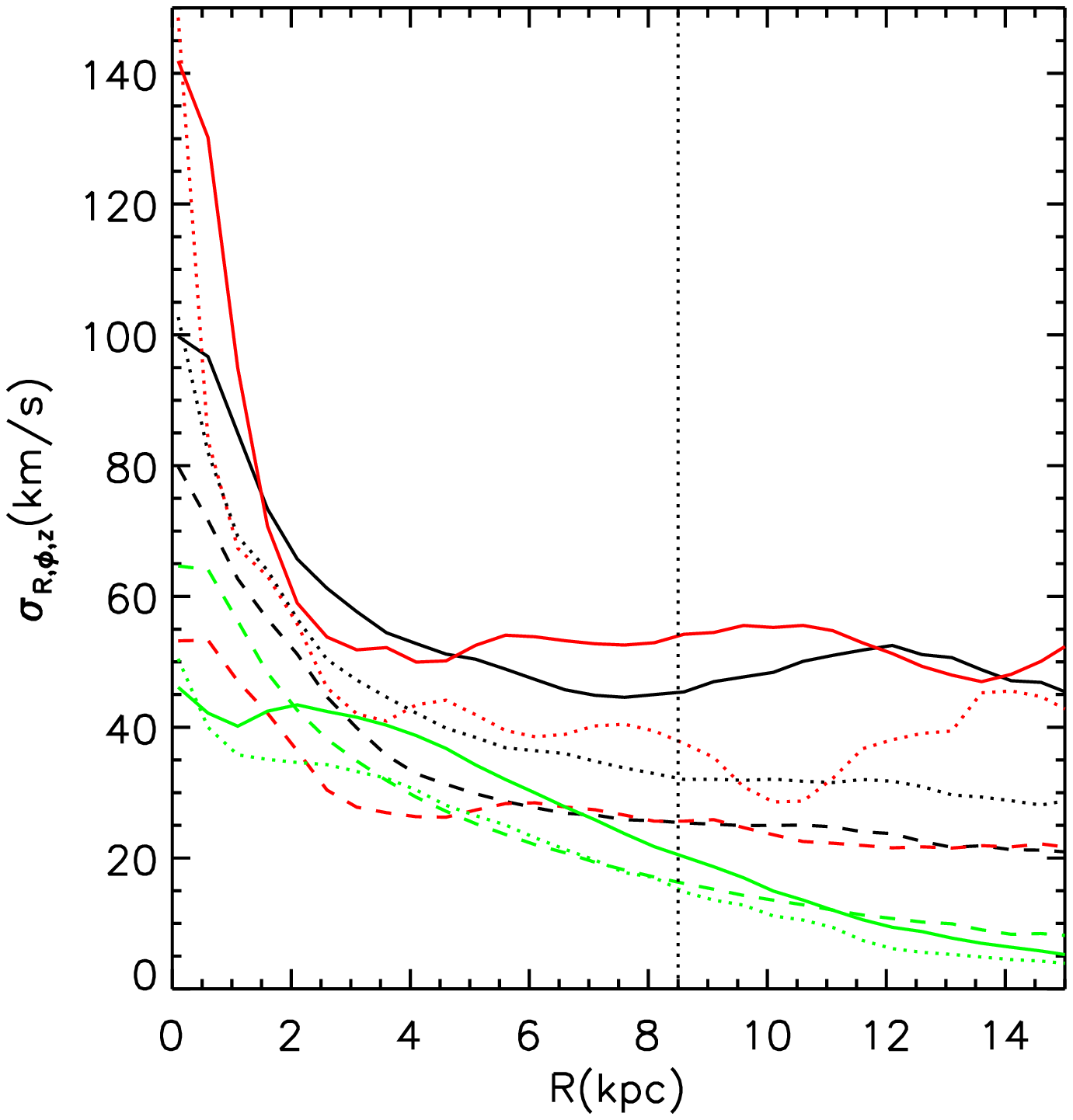}\hspace{\myfspace}\\

\includegraphics[width=0.22\textwidth]{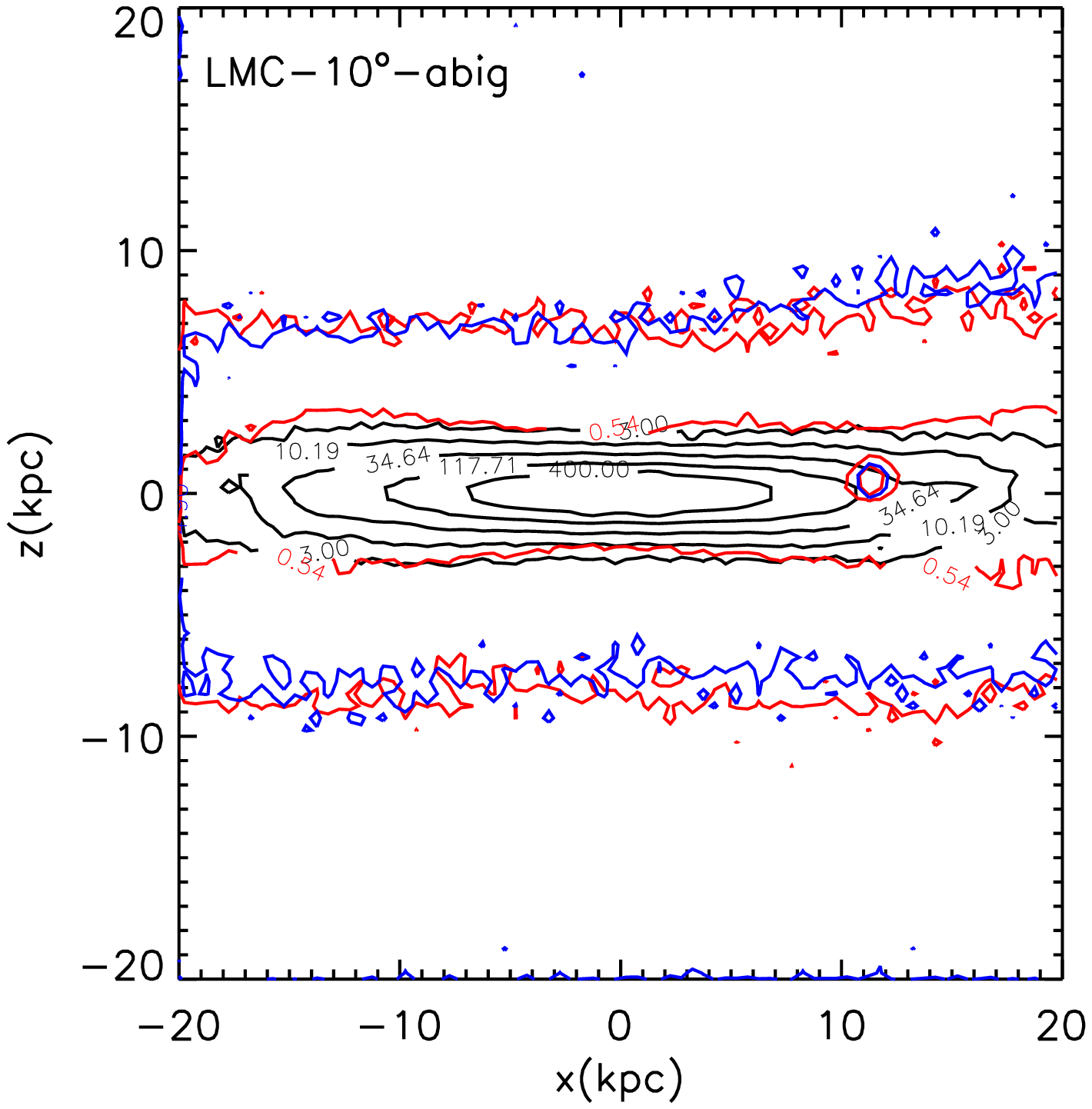}\hspace{\myfspace}
\includegraphics[width=0.22\textwidth]{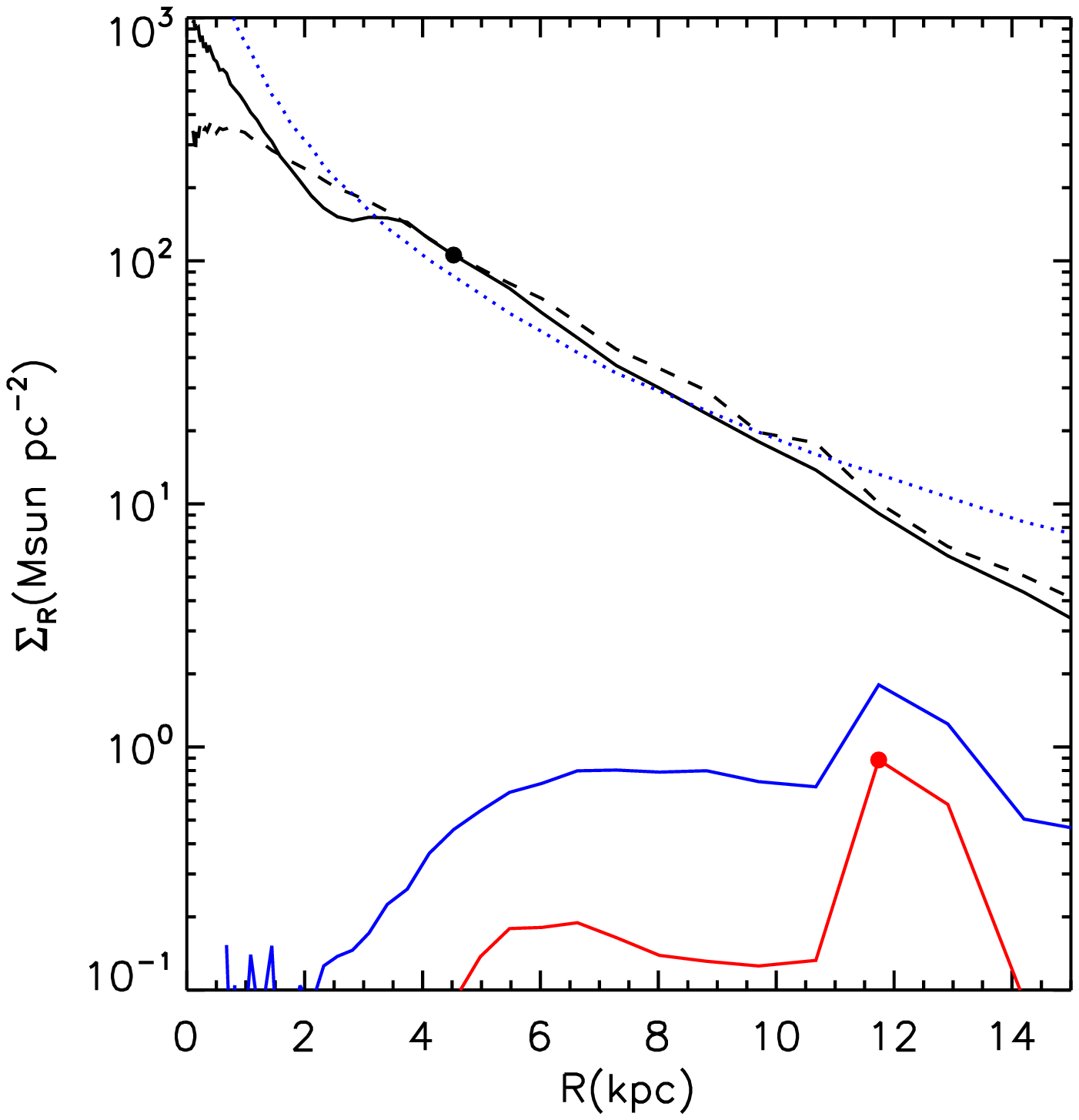}\hspace{\myfspace}
\includegraphics[width=0.22\textwidth]{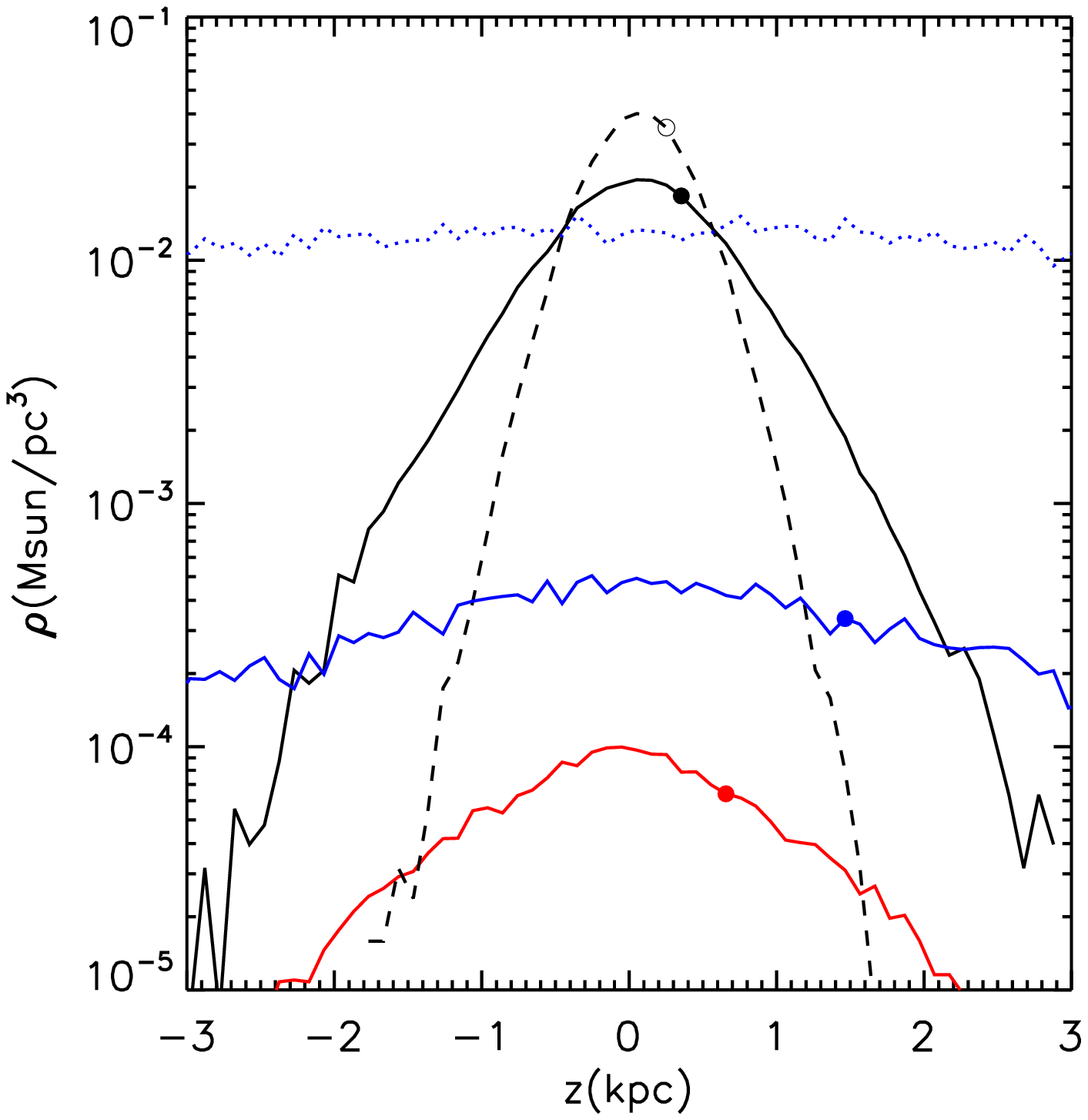}\hspace{\myfspace}
\includegraphics[width=0.22\textwidth]{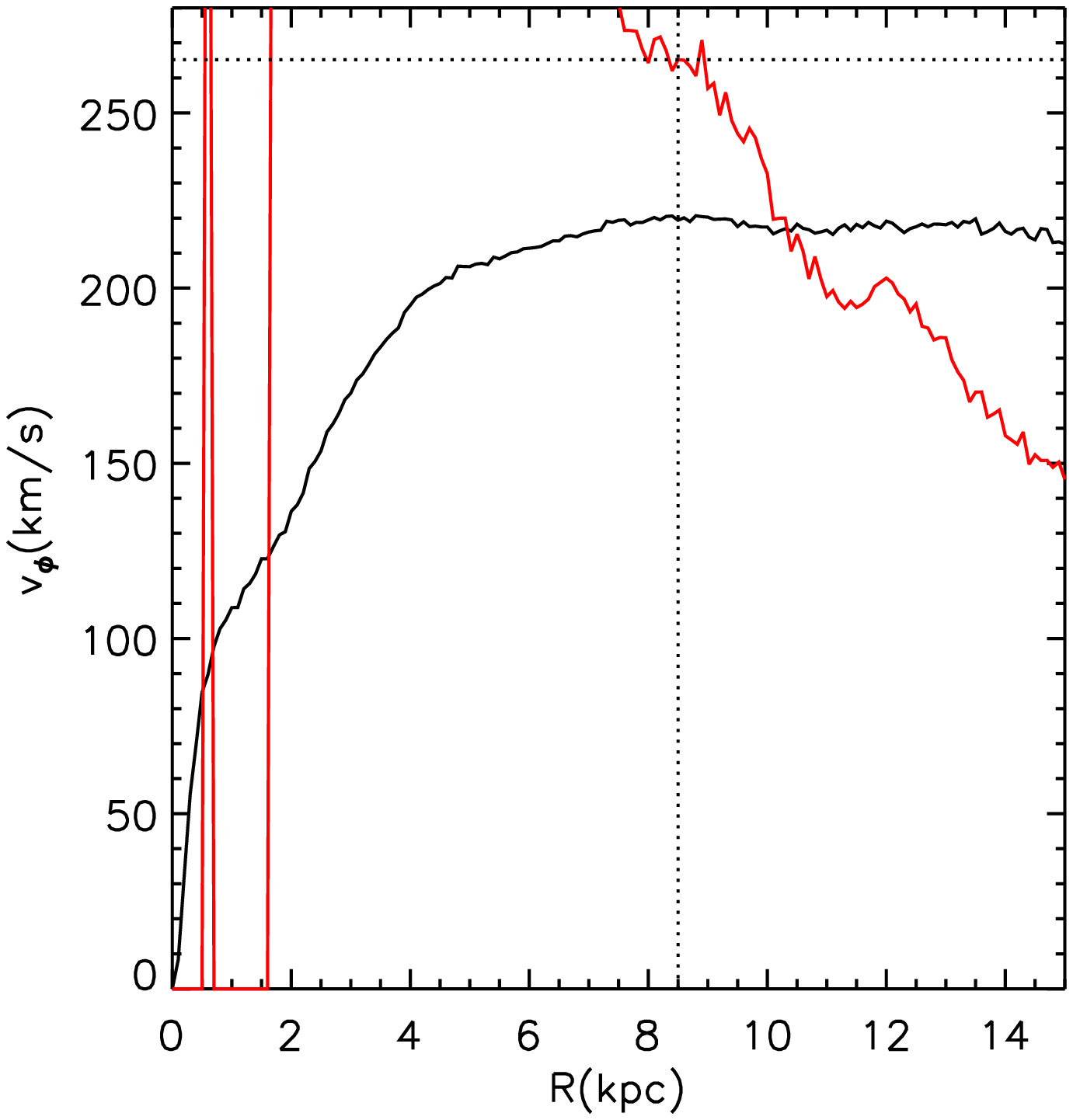}\hspace{\myfspace}
\includegraphics[width=0.22\textwidth]{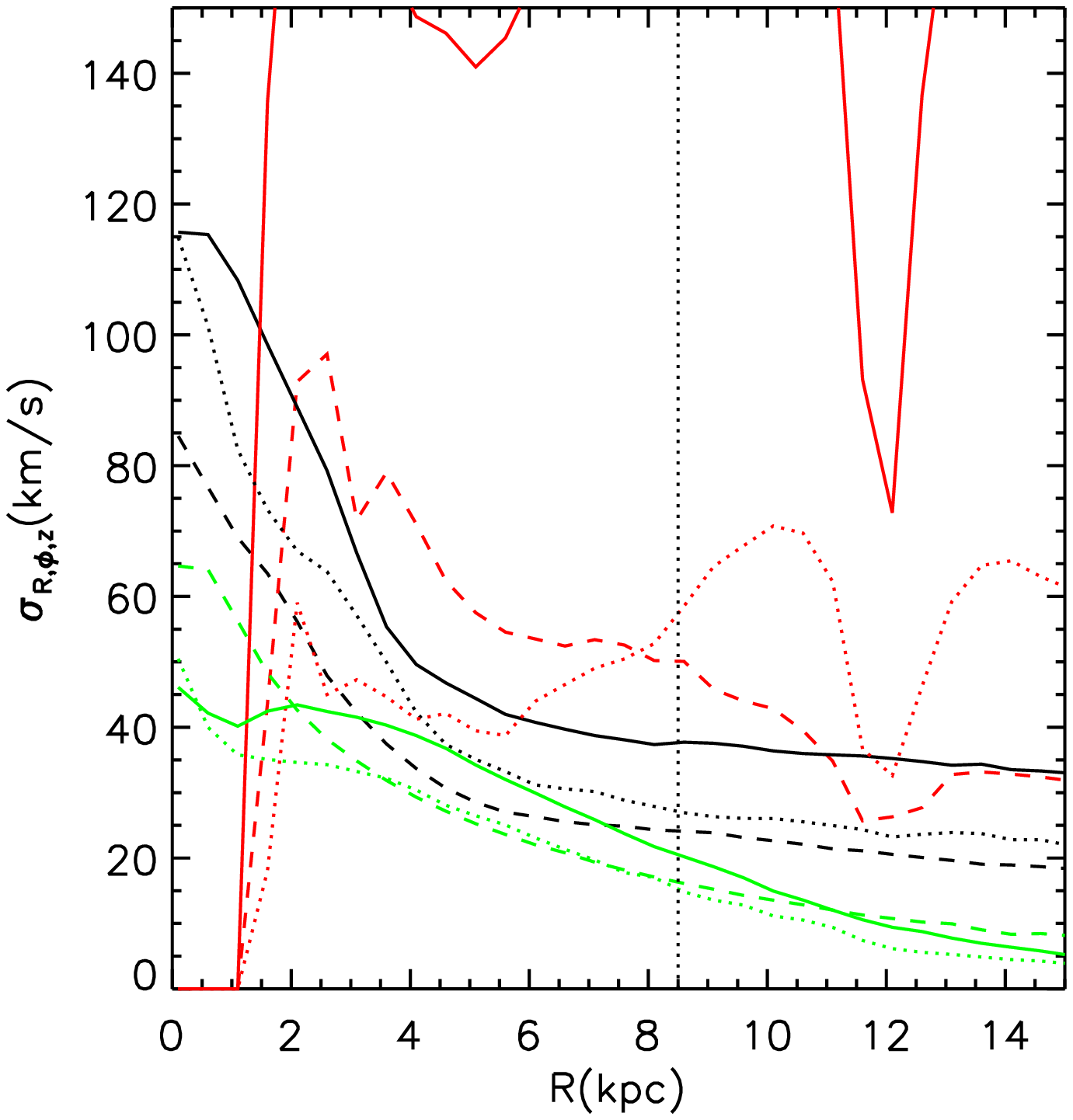}\hspace{\myfspace}\\

\caption{The effect of changing the orbit: simulations: \LMCRet, \LMCeless\ and \LMCabig. The lines and panels are as in Figure \ref{fig:impact1}. Notice that the retrograde merger, \LMCRet, forms a thick disc very similar to the equivalent prograde merger (\LMCten; Figure \ref{fig:impact1}), though it disrupts further from the centre. It has a quite different effect on the Milky Way disc: a bar does not form, and there is a less pronounced warp/flare. The low eccentricity merger, \LMCeless, produces a higher stellar density at the solar neighbourhood than our reference simulation \LMCten, but the dark disc density is very similar. The high apocentre merger, \LMCabig, fails to complete merging even over 7\,Gyrs (the remnant is visible in the left panel; see also Figure \ref{fig:thickorbits}) and a significant thick disc is not formed.}
\label{fig:orbchange1}
\end{center}
\end{figure*}

\begin{figure}
\begin{center}
\includegraphics[width=0.5\textwidth]{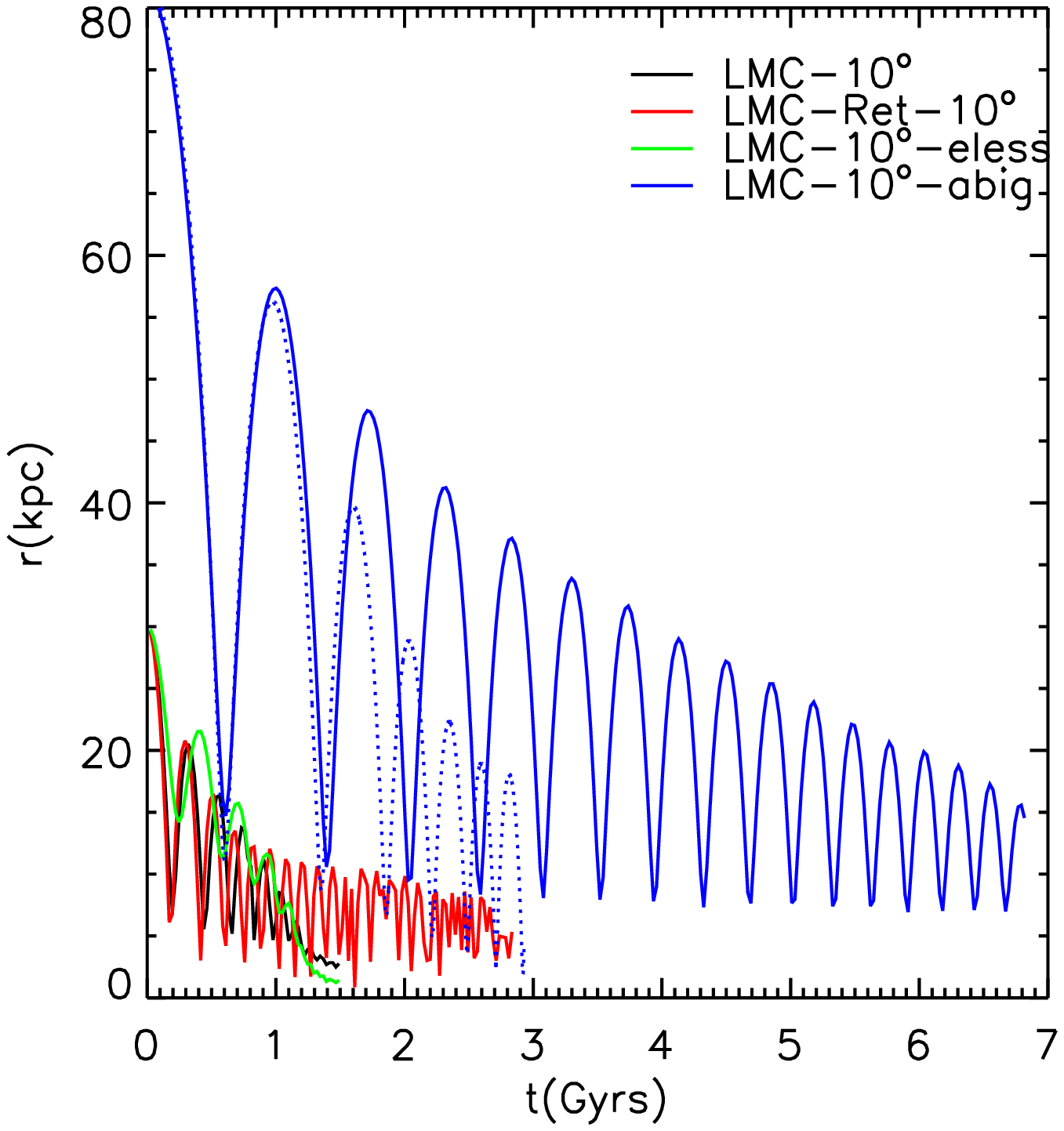}
\caption{Decay of the satellite orbit as a function of time for selected simulations. The dotted blue line shows the analytic decay rate expected for \LMCabig, assuming Chandrasekhar friction and that the satellite does not lose any mass.}
\label{fig:thickorbits}
\end{center}
\end{figure}

Figure \ref{fig:orbchange1} investigates the effect of changing the satellite orbit; lines and panels are as in Figure \ref{fig:impact1}. We consider a retrograde encounter with the disc (\LMCRet); an orbit with low eccentricity (\LMCeless; $e=0.36$), and an orbit with large apocentre (\LMCabig; $r_a = 81$\,kpc). In the cosmological simulations of \S\ref{sec:cosmology}, we find no strong bias towards prograde or retrograde orbits, suggesting that retrograde mergers should be common (other recent studies also find only weak net rotation for the satellites, see e.g. \bcite{2007arXiv0704.1770S}). From the observational point of view, it is interesting that, of the two extragalactic disc galaxies studied in detail to date, one has a counter-rotating thick disc \citep{2005ApJ...624..701Y}. The orbital decay of the satellite for each of these runs and the reference simulation, \LMCten, are shown in Figure \ref{fig:thickorbits}. 

As compared to the equivalent prograde encounter (\LMCten), the retrograde satellite disrupts further out leaving a larger hole behind. This can be seen in the radial surface density (Figure \ref{fig:orbchange1}(b)). This is expected because dynamical friction goes as $1/v^2$. The prograde satellite is nearly stationary with respect to the disc and falls in faster than the retrograde one \citep{1986ApJ...309..472Q}. The effect is small, however, because the dark halo dominates the mass as all radii in our MW model (this is not the case for MWB). The larger difference between the prograde and retrograde encounters is in the response of the Milky Way disc; we discuss this in \S\ref{sec:milkyway}.

The lower eccentricity run, \LMCeless\ ($e=0.36$), gives a lower $\sigma_R$ and a smaller lag in the rotation of the accreted material. The small rotation lag can be increased by increasing the impact angle, $\theta$ (see \S\ref{sec:impact}).

The high apocentre run, \LMCabig, is particularly interesting. Even over 7\,Gyrs it has not fully accreted and dynamical friction proceeds slowly, despite the satellite being initially of LMC mass. The orbit is shown in Figure \ref{fig:thickorbits}. The dotted blue line on this plot shows the expected orbit for \LMCabig\ assuming a static background potential and Chandrasekhar dynamical friction \citep{1987gady.book.....B}. We use a Coulomb logarithm that varies with radius as in \citet{2003ApJ...582..196H}. The simulated orbit decays much more slowly than the Chandrasekhar model because we have not accounted for satellite mass loss due to tides. The satellite, which starts out at $2.4\times 10^{10}$\,M$_\odot$, is just a few $10^{8}$\,M$_\odot$ at the end of the simulation. \citet{2004MNRAS.351..891Z} have recently highlighted this problem. They show that it is difficult to understand how mergers -- like the observed Sagittarius dwarf merger -- could proceed in less than a Hubble time without Sagittarius having been unrealistically massive in the past. A solution, as presented in \S \ref{sec:cosmology}, is that galaxies like Sagittarius `ride in' inside larger loosely bound groups and are effectively deposited at low apocentre. This is why in our other runs we assume an initial apocentre of $r_a = 30$\,kpc.  

\subsubsection{Varying the satellite and host galaxy properties}\label{sec:sat}

\begin{figure*}
\begin{center}
\includegraphics[width=0.22\textwidth]{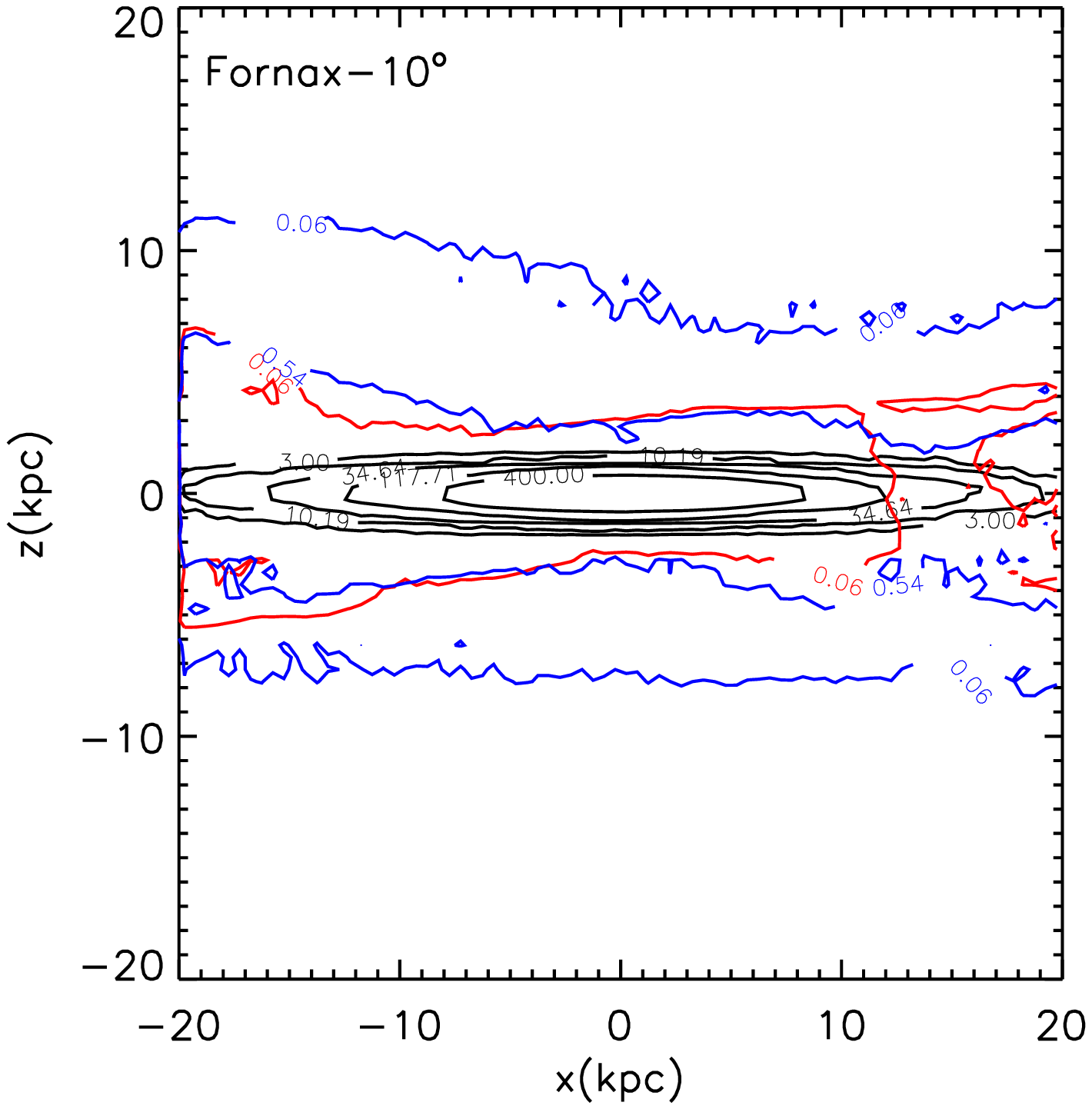}\hspace{\myfspace}
\includegraphics[width=0.22\textwidth]{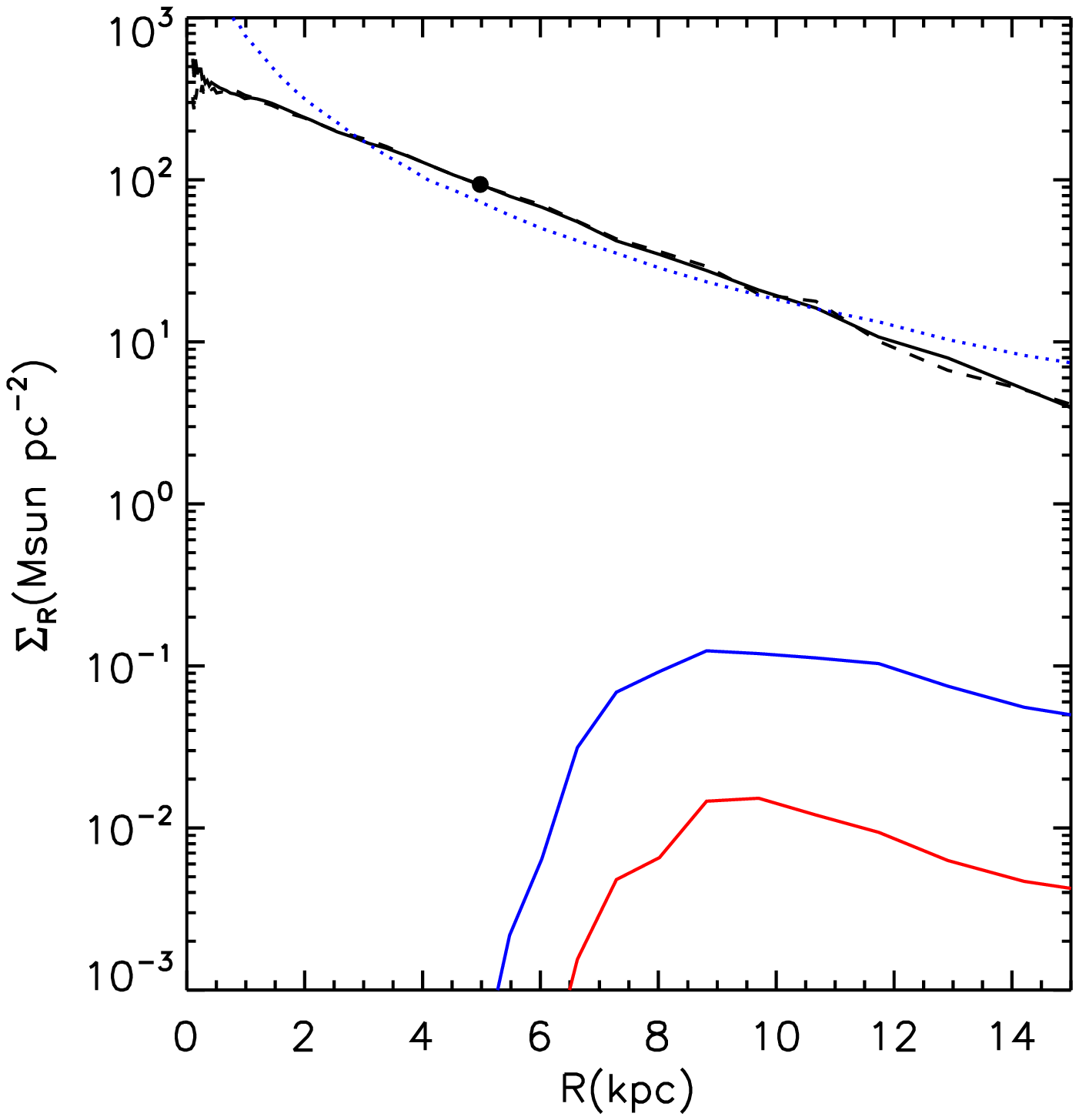}\hspace{\myfspace}
\includegraphics[width=0.22\textwidth]{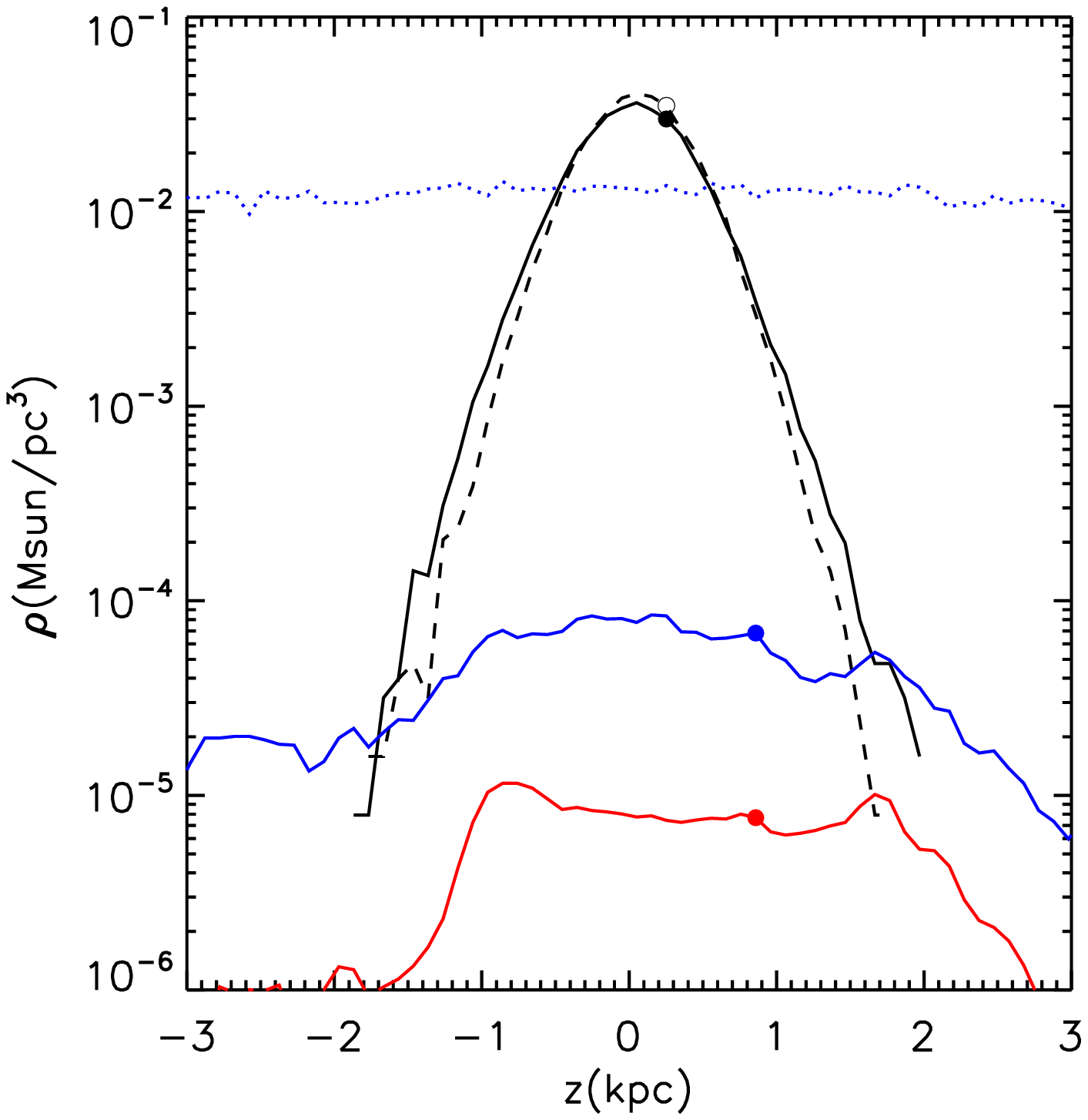}\hspace{\myfspace}
\includegraphics[width=0.22\textwidth]{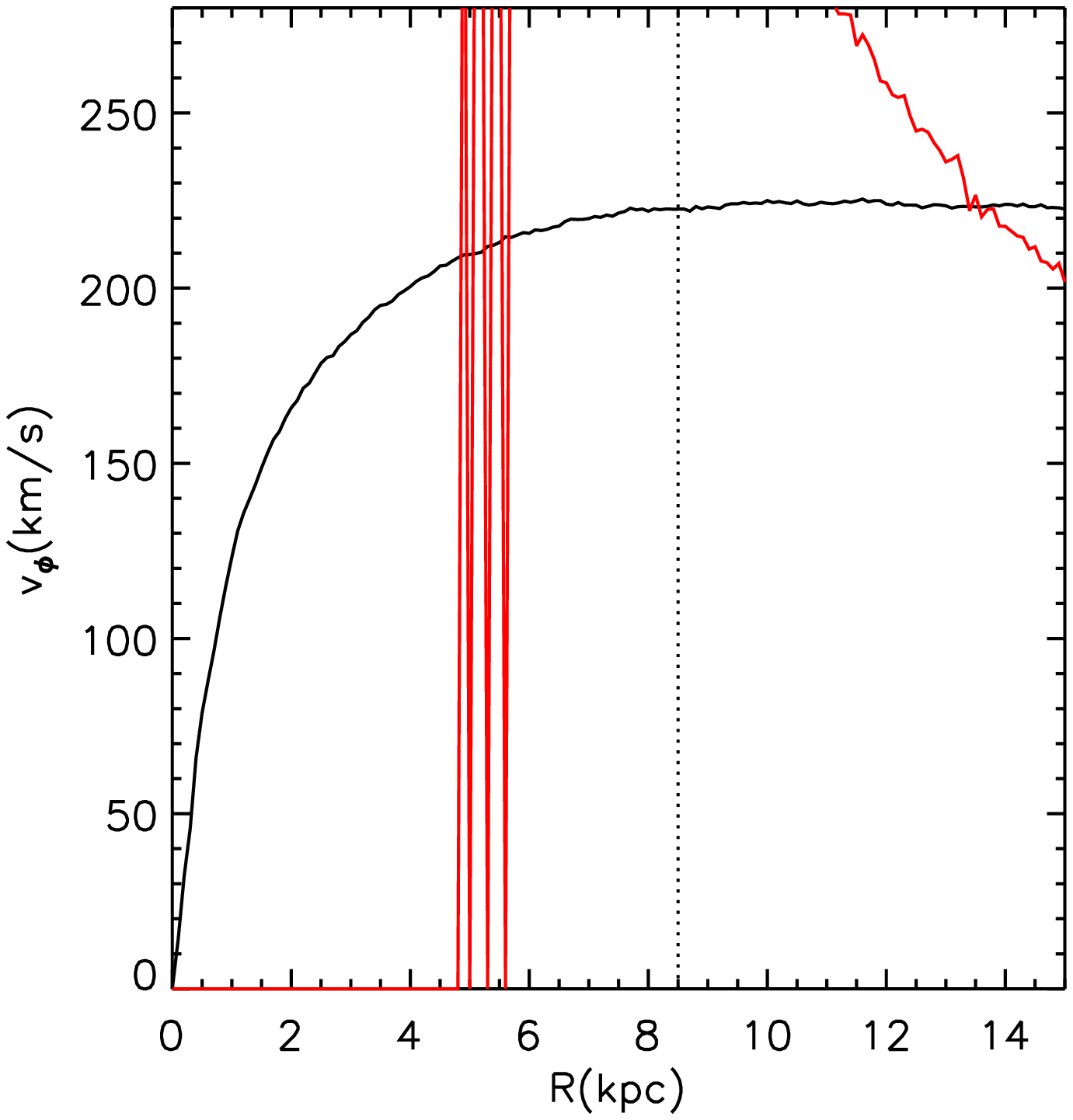}\hspace{\myfspace}
\includegraphics[width=0.22\textwidth]{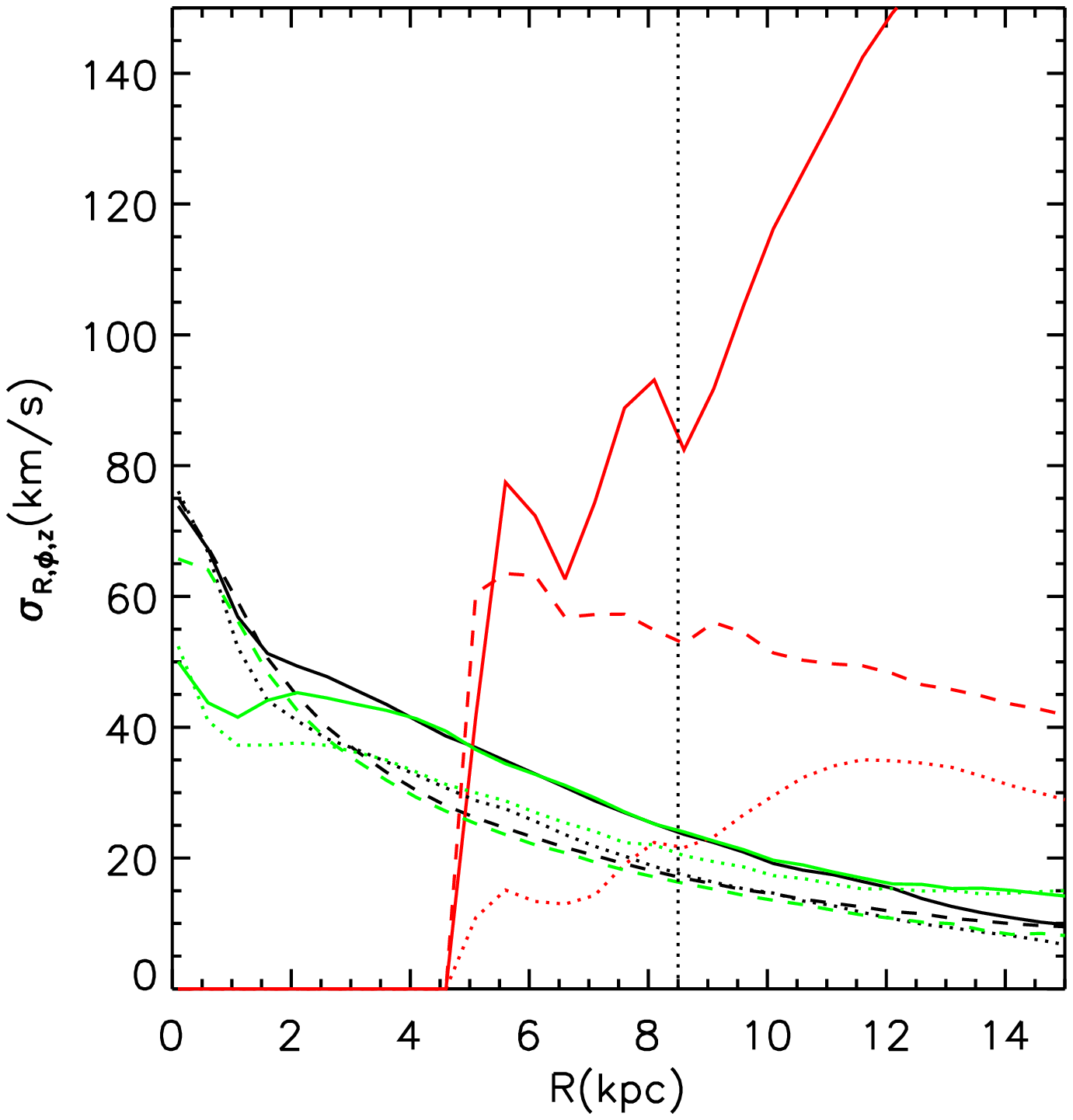}\hspace{\myfspace}\\

\includegraphics[width=0.22\textwidth]{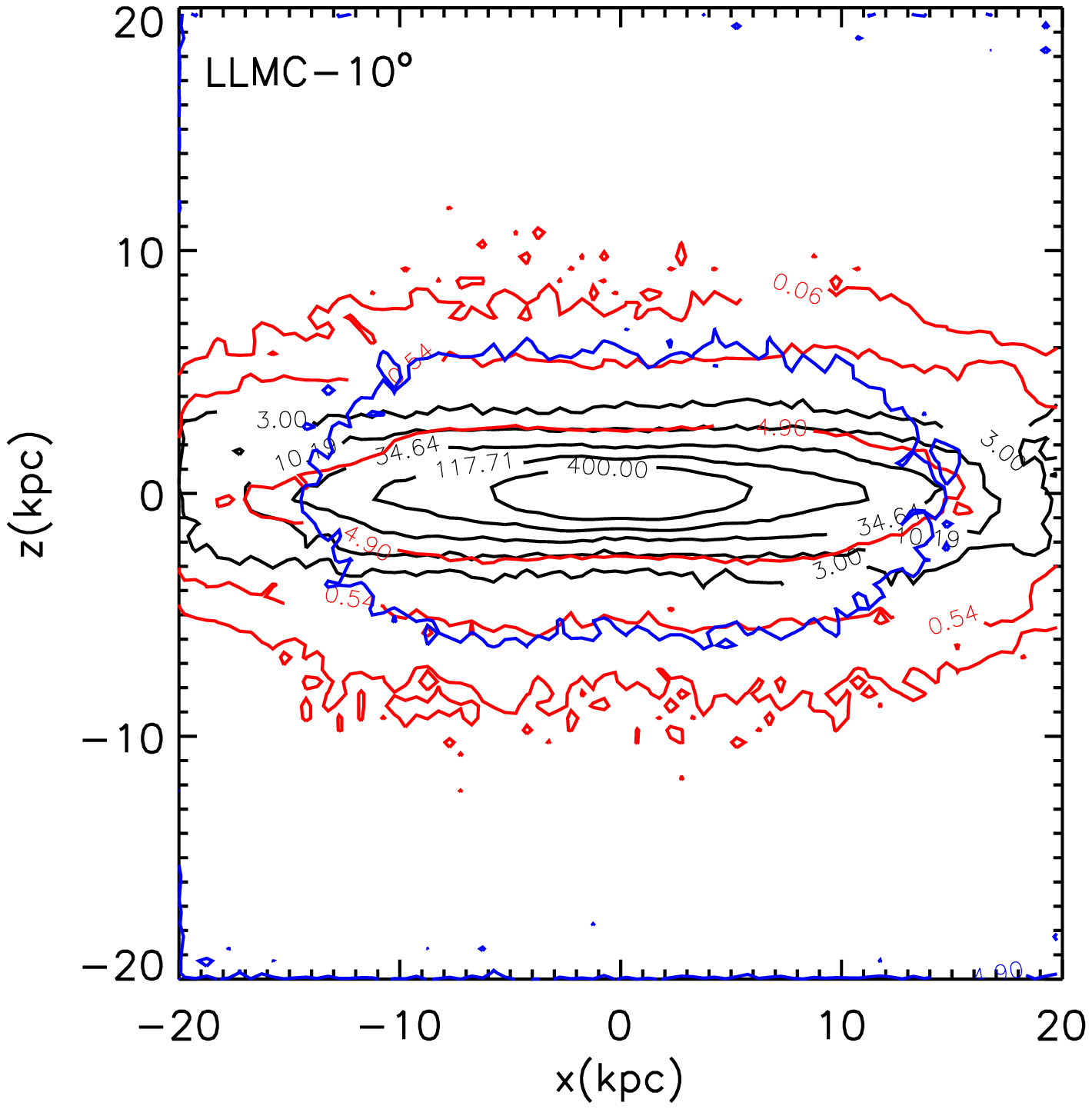}\hspace{\myfspace}
\includegraphics[width=0.22\textwidth]{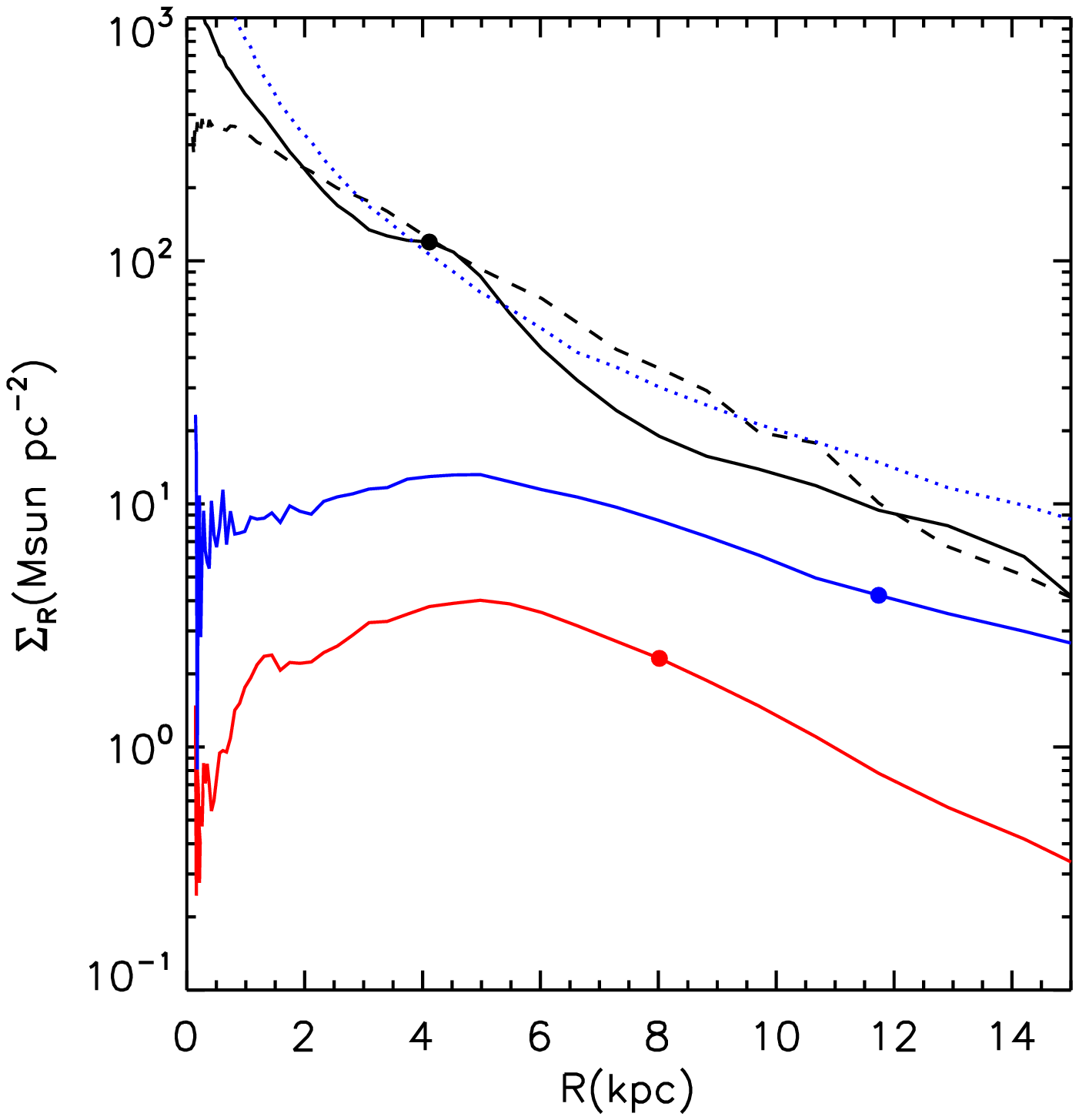}\hspace{\myfspace}
\includegraphics[width=0.22\textwidth]{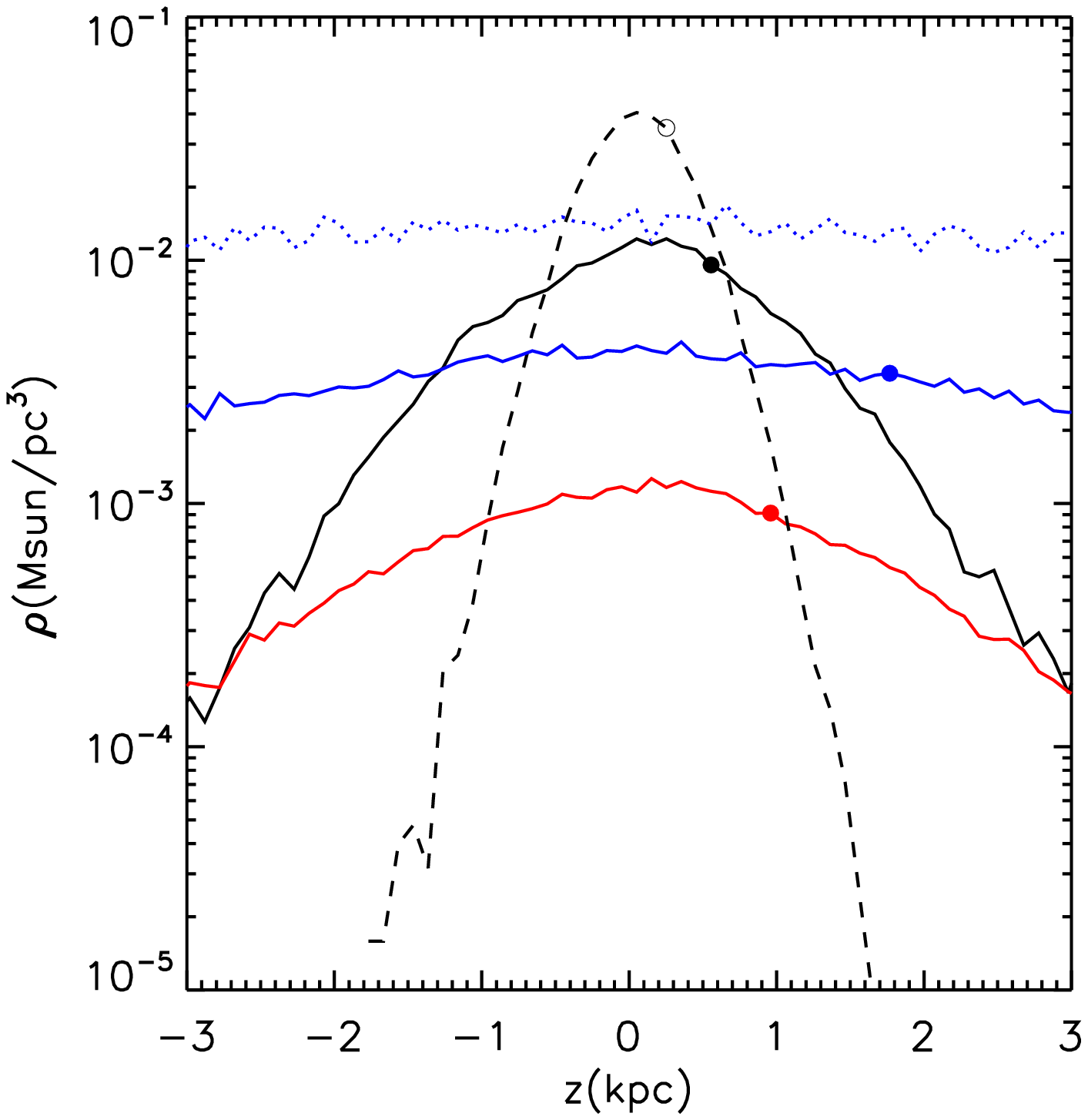}\hspace{\myfspace}
\includegraphics[width=0.22\textwidth]{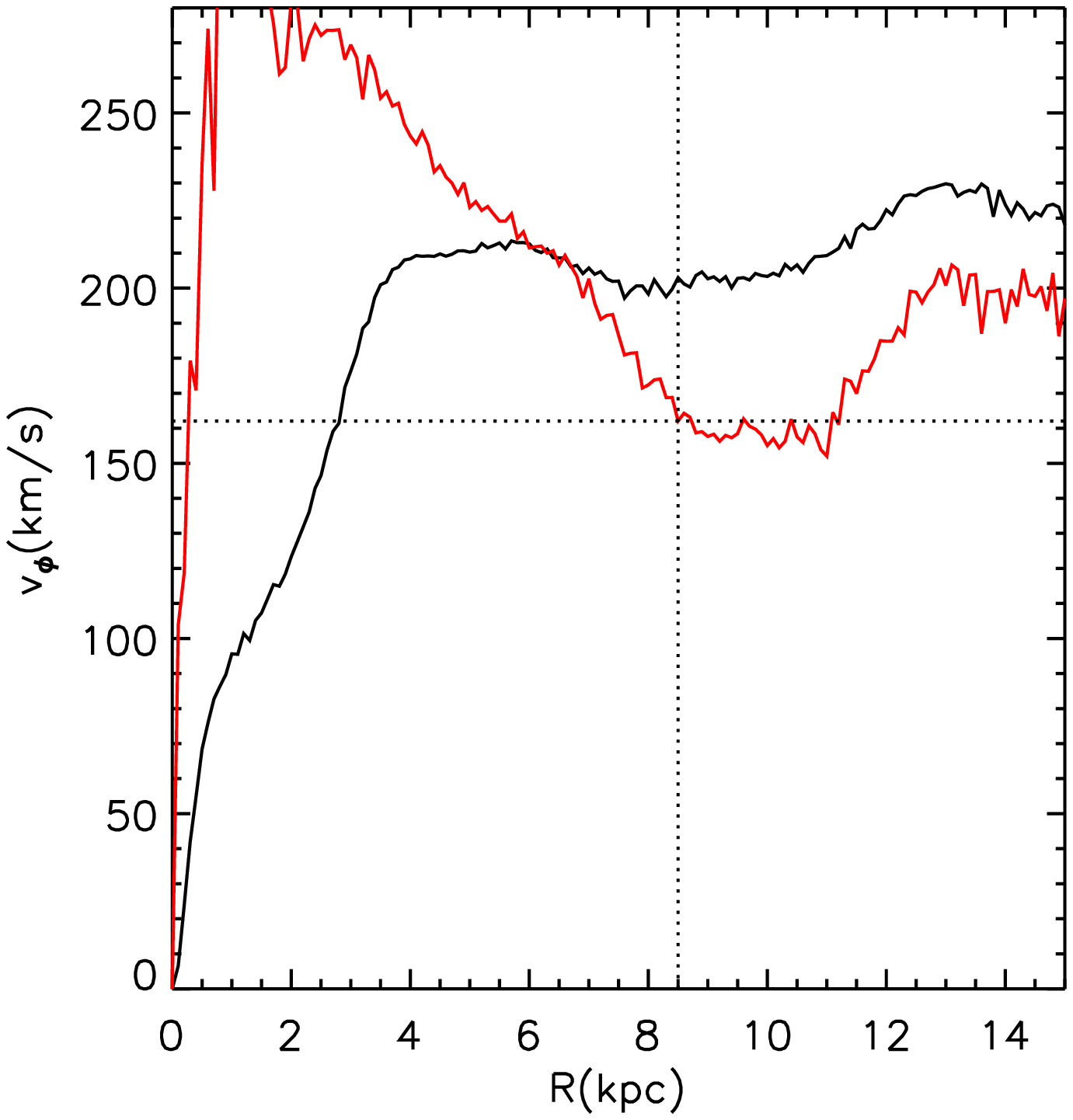}\hspace{\myfspace}
\includegraphics[width=0.22\textwidth]{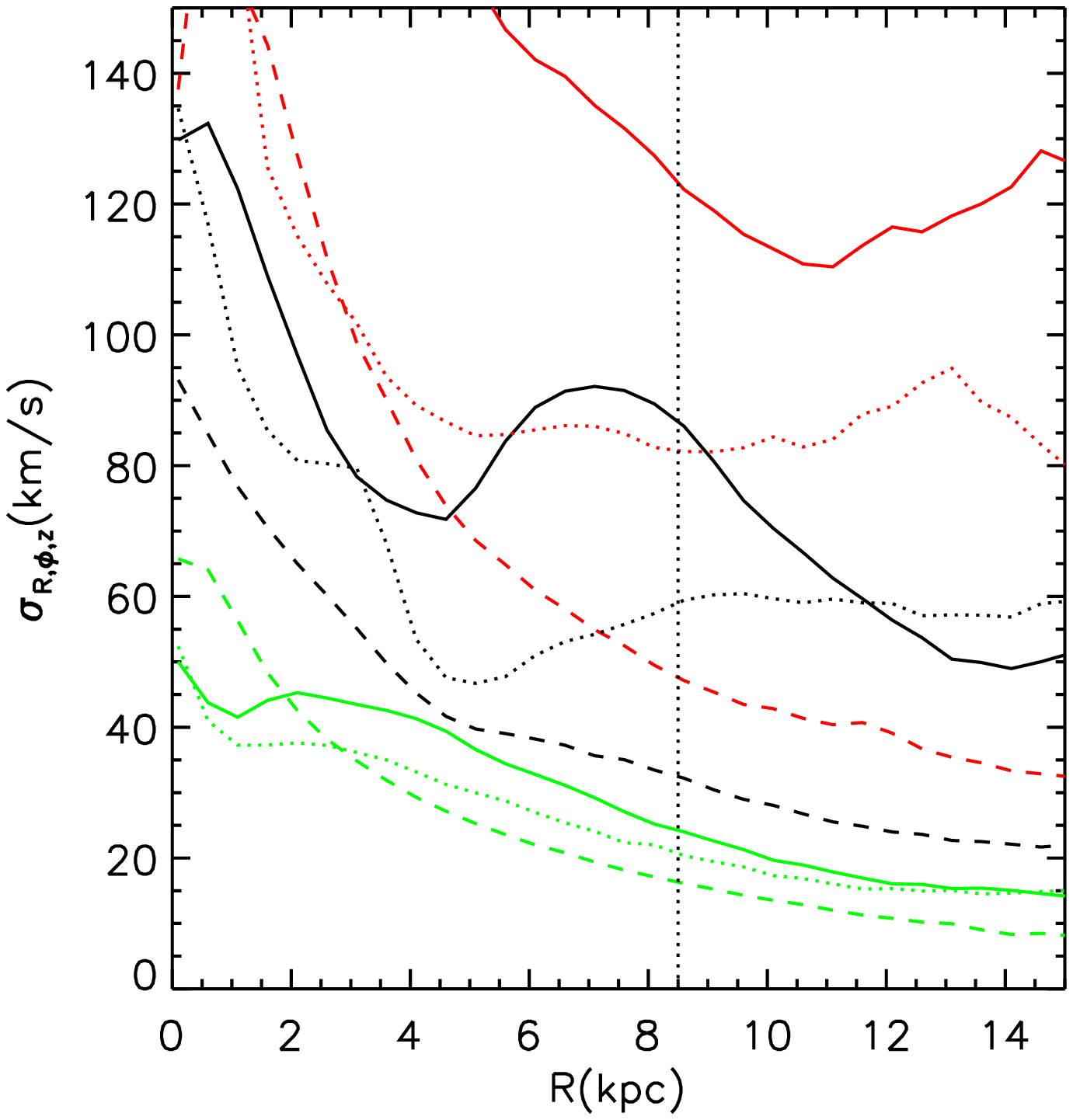}\hspace{\myfspace}\\

\includegraphics[width=0.22\textwidth]{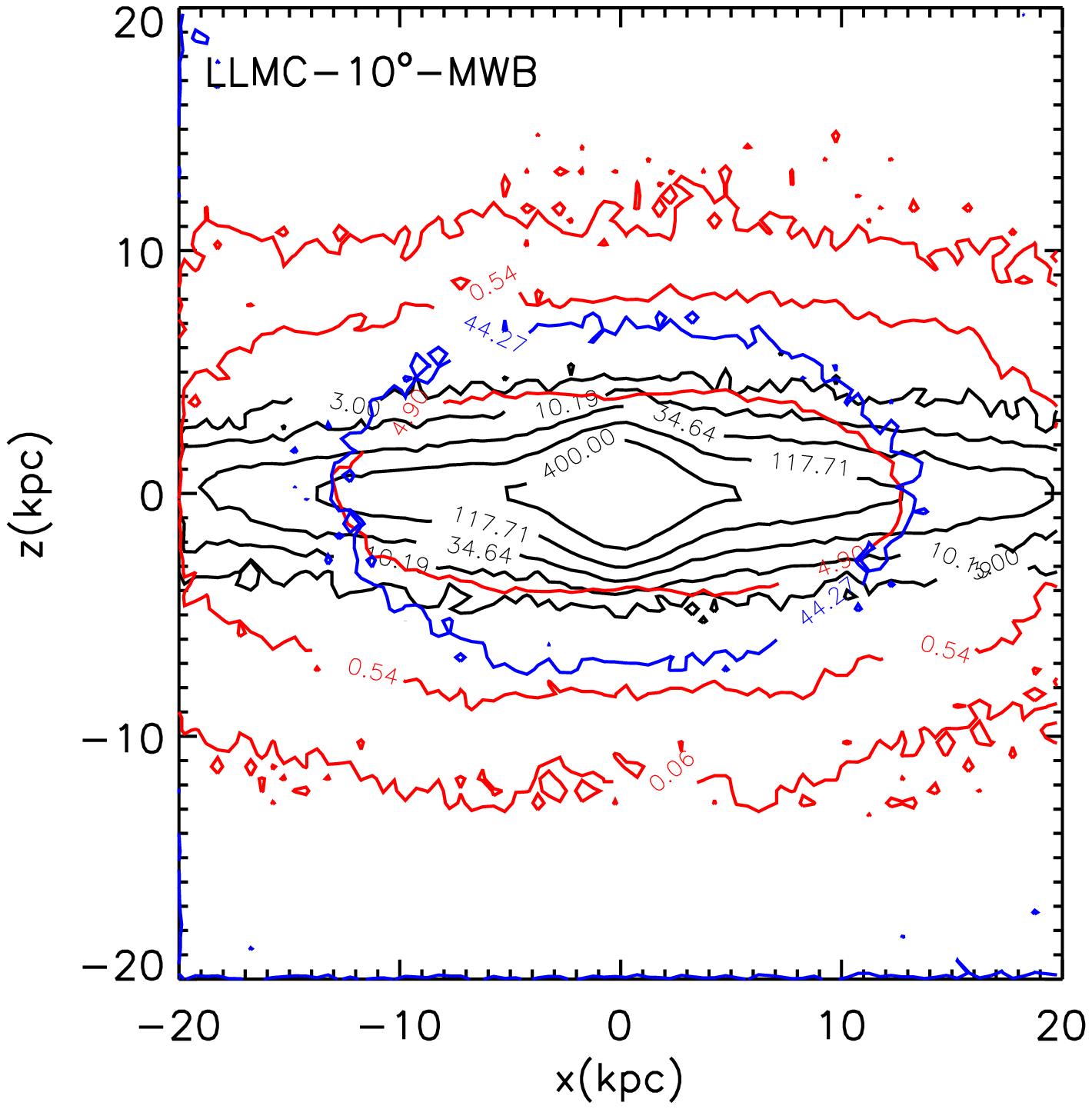}\hspace{\myfspace}
\includegraphics[width=0.22\textwidth]{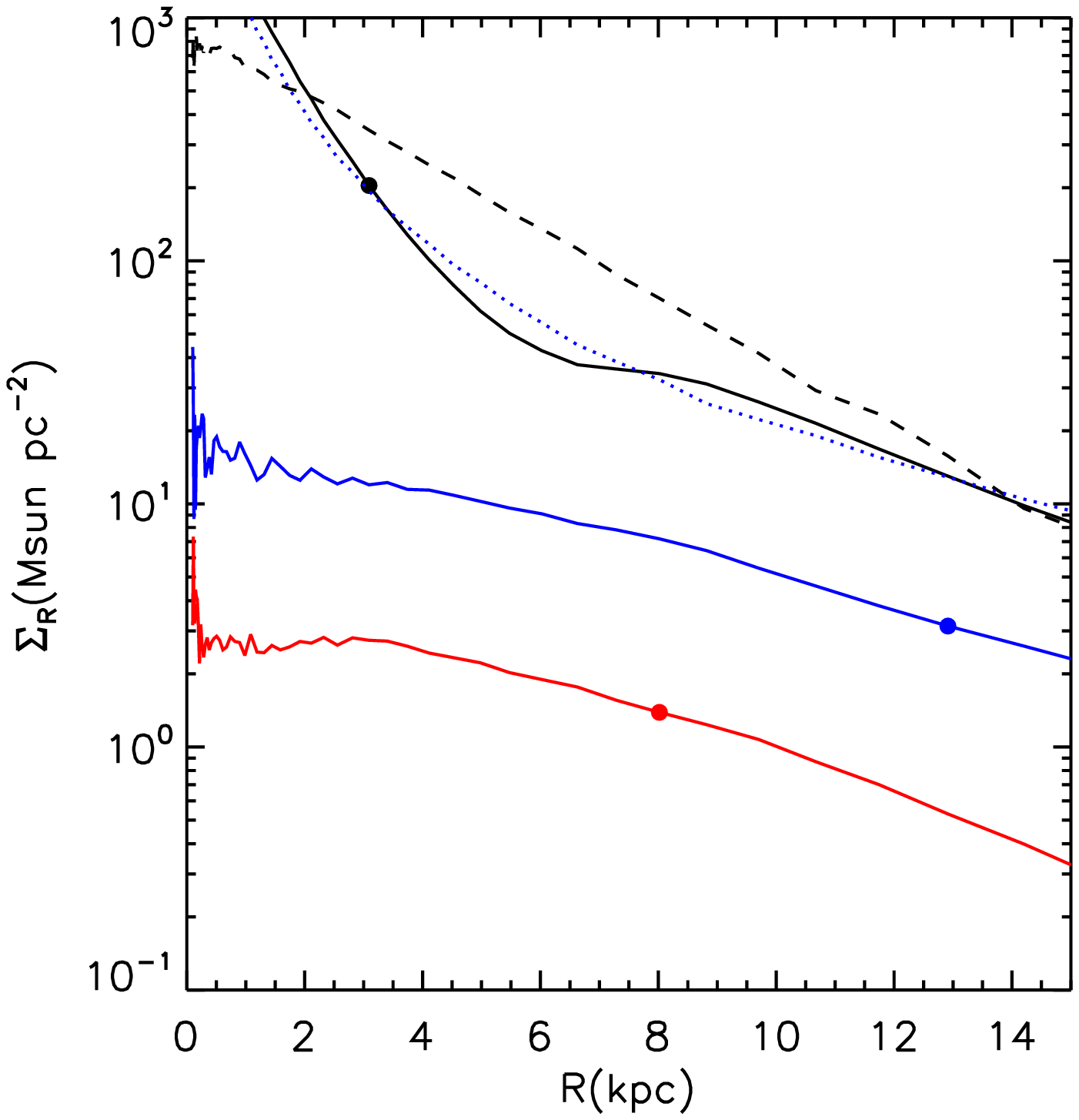}\hspace{\myfspace}
\includegraphics[width=0.22\textwidth]{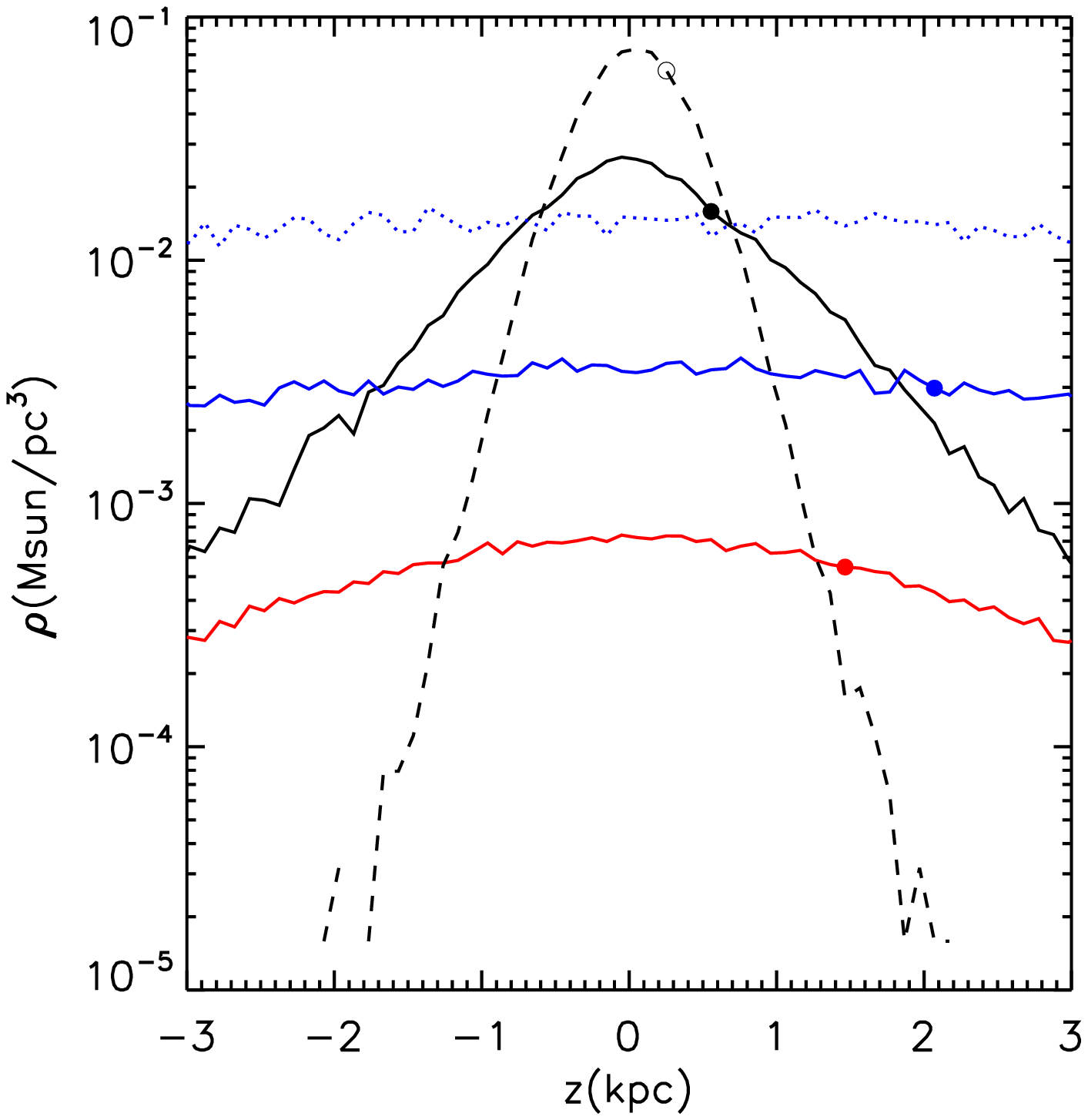}\hspace{\myfspace}
\includegraphics[width=0.22\textwidth]{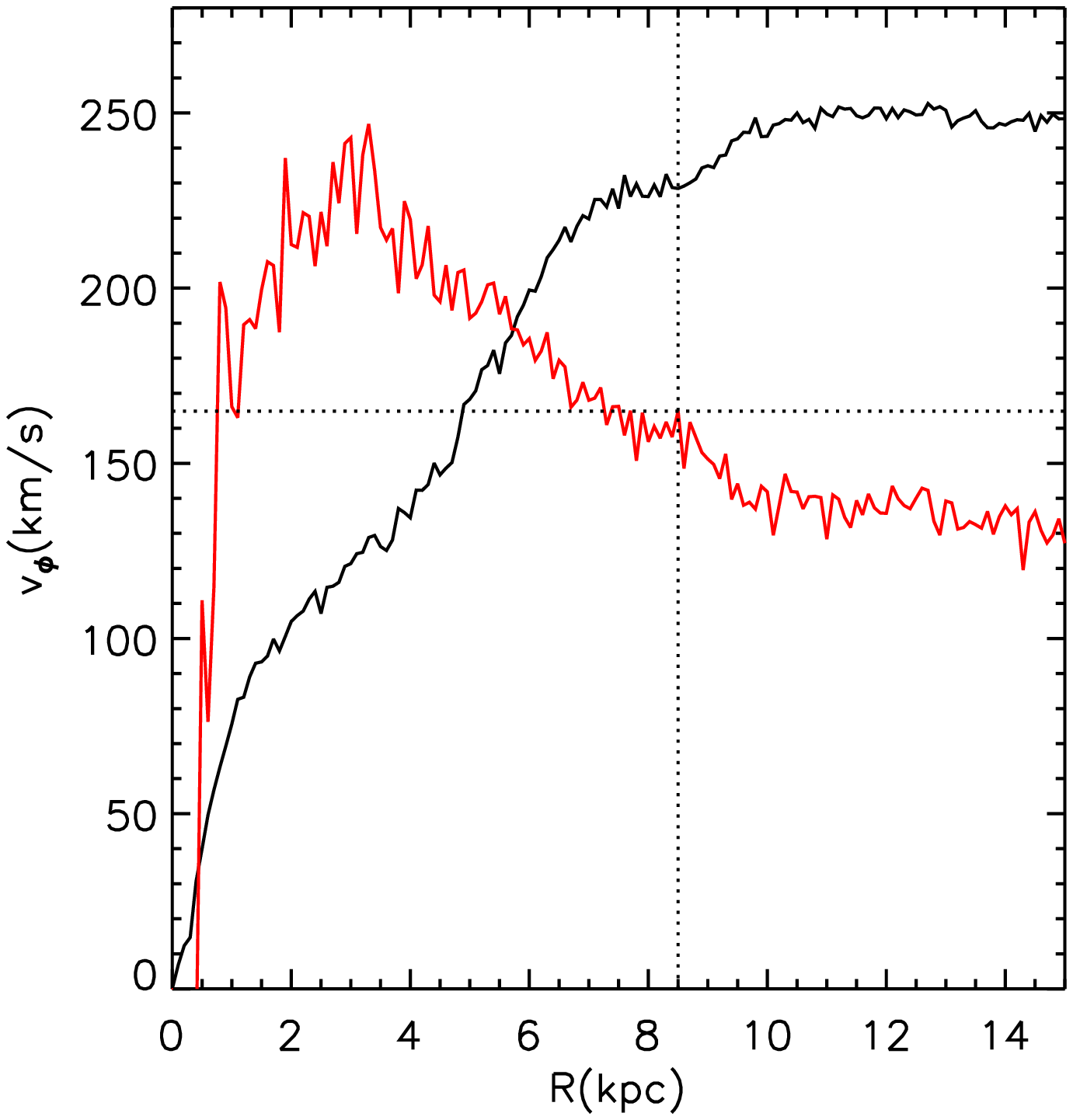}\hspace{\myfspace}
\includegraphics[width=0.22\textwidth]{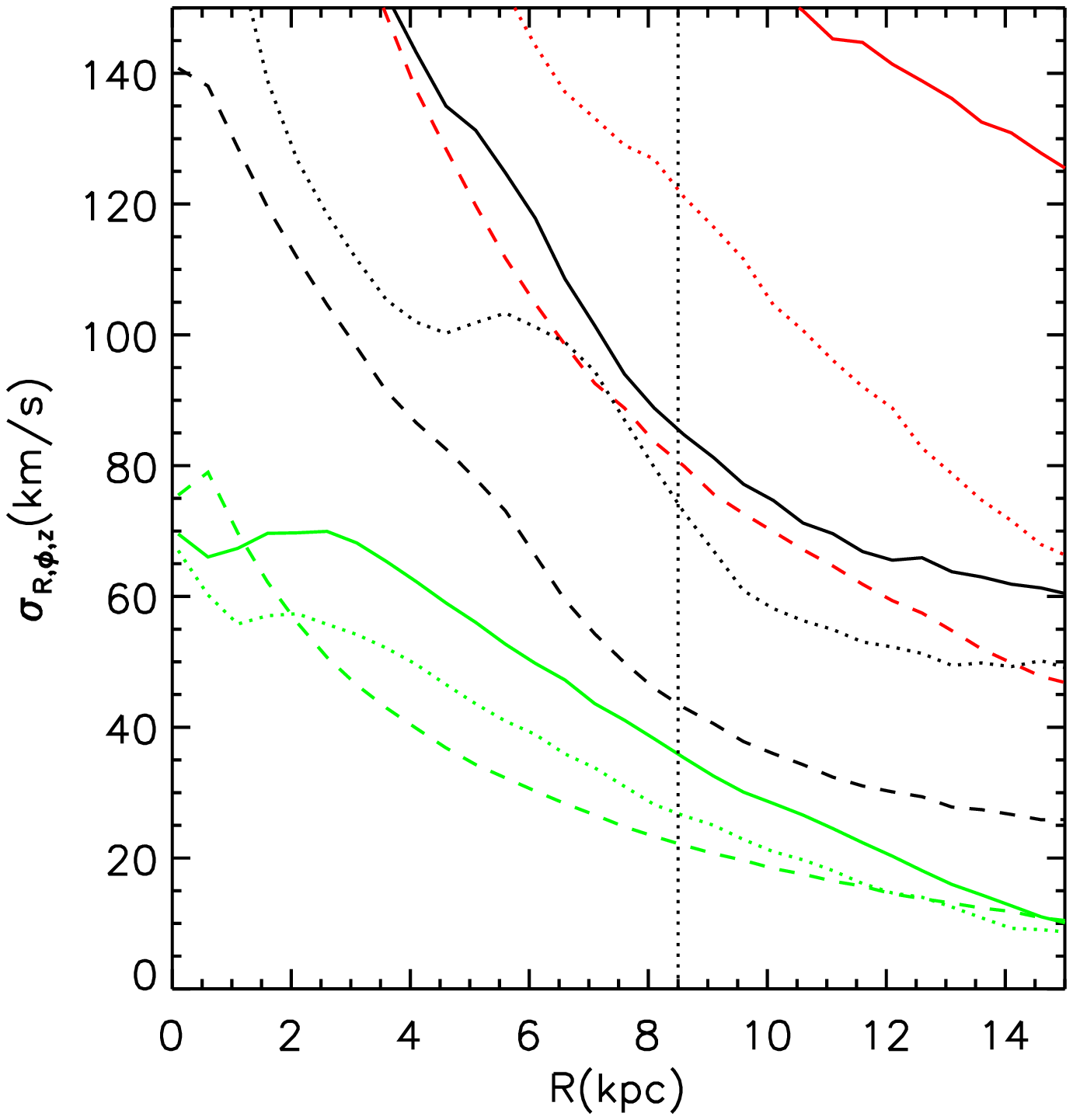}\hspace{\myfspace}\\

\caption{The effect of changing the satellite and host galaxy properties: simulations: \Fornax, \LLMC\ and \LLMCMWB. The lines and panels are as in Figure \ref{fig:impact1}. The low mass satellite, \Fornax, does not perturb the Milky Way disc at all, nor does it form anything which may be reasonably described as a thick disc. The high mass satellite, \LLMC\ produces results similar to \LMCten, but does significant damage to the Milky Way disc which now has density comparable to the dark halo at the solar neighbourhood. The more massive Milky Way disc run, \LLMCMWB, produces results similar to \LLMC, but the increased disc mass causes a prominent bar to form, while the satellite disrupts more rapidly leading to a thick disc of larger scale height and lower density interior to $|z|<1.1$\,kpc.}
\label{fig:satchange1}
\end{center}
\end{figure*}

Figure \ref{fig:satchange1} investigates the effect of varying the satellite and host galaxy properties. We consider a low mass satellite (\Fornax); a high mass satellite (\LLMC); and a more massive Milky Way disc (\LLMCMWB). 

The Fornax merger (\Fornax) does not produce a structure resembling a thick disc. It disrupts too far out and is destroyed completely by $R \sim 5$\,kpc. Its rotation velocity leads rather than lags the thin disc (because of the large hole interior to 5\,kpc); its dispersions are too high; and the accreted stellar density is two orders of magnitude lower than \LMCten. Summing over many such mergers could alleviate some of these discrepancies by filling in the hole with accreted material. However, nearly half of such mergers should be retrograde, while from \S\ref{sec:cosmology}, we expect $\sim 15$ such mergers within 20$^\mathrm{o}$ of the disc plane. The integrated light and dark matter from all such mergers will still be an order of magnitude lower than that contributed by \LMCten. We conclude that only the most massive mergers are of relevance for contributing accreted stars to the thick disc. 

While not forming thick discs, small-satellite encounters do give rise to interesting fine-structure, in this case in the form of a ring\footnote{Rings can also be produced by low density satellites on more circular orbits, though such orbits seem unlikely in our current cosmology (see \S \ref{sec:cosmology} and Figure \ref{fig:cosmoresults}). High density satellites do not produce rings because they fall in via dynamical friction faster than they disrupt.}. Since the satellite disruption radius will depend more on the satellite mass than on its orbit, ring features could originate from a super-position of satellites. Such super-positions will broaden the ring and make it appear disc-like at large radii. Indeed, this could provide an explanation for the extended disc seen in Andromeda (\bcite{2005ApJ...634..287I} and \S \ref{sec:introduction})\footnote{Alternative explanations in the literature are to have one large satellite merge in the disc-plane \citep{2006ApJ...650L..33P}; or to have a prograde fly-by from a massive satellite on an eccentric orbit \citep{2007arXiv0708.1949K,2008arXiv0802.0872Y}.}. From \S \ref{sec:cosmology}, we expect $\sim 15$ such low inclination, low mass, mergers to occur in the Milky Way. Some of these will have occurred before the disc of the Milky Way was fully formed and may be essentially undetecable today. However, one such structure does appear to have been observed (\bcite{2007MNRAS.376..939C}). There may be more to be found. 

\LLMC\ investigates the effect of increasing the satellite mass. Many of the properties of the accreted material are reminiscent of our reference run -- \LMCten: it lags the rotation of the thin disc by $\sim 40$\,km/s, is hotter with $\sigma_z \sim 80$\,km/s at the solar neighbourhood and is of larger scale height ($z_{1/2} = 0.9$\,kpc) and longer scale length ($R_{1/2} = 8$\,kpc) than the thin disc. Such mergers -- $\vmaxd = 80$\,km/s -- are common; we expect on average one such merger to occur within 20$^\mathrm{o}$ of the disc plane (assuming isotropy). Furthermore, what is particularly interesting about this simulation is the heating of the thin disc. The final thin disc scale height and kinematics for this run -- unlike all of the previous runs -- now matches the thick disc of the Milky Way with $z_{1/2} = 0.6$\,kpc and $\sigma_z = 40$\,kpc at the solar position. We will discuss this in more detail in \S\ref{sec:milkyway} and \S\ref{sec:thickdisc}.

Finally, \LLMCMWB\ investigates the effect of increasing the mass of the Milky Way disc. With the more massive Milky Way disc, the satellite now disrupts further out than in \LLMC, leading to a thick disc of larger scale height and lower density interior to $|z|<1.1$\,kpc. The more massive disc goes bar unstable without any encounter, forming a prominent bar that appears bulge-like when viewed from the side in projection (a detailed account of this type of instability is given in \bcite{2006ApJ...645..209D}).  

\subsubsection{The Milky Way thin disc: a bar, flare, warp and heated thick disc}\label{sec:milkyway}

\begin{figure}
\begin{center}
\includegraphics[width=0.49\textwidth]{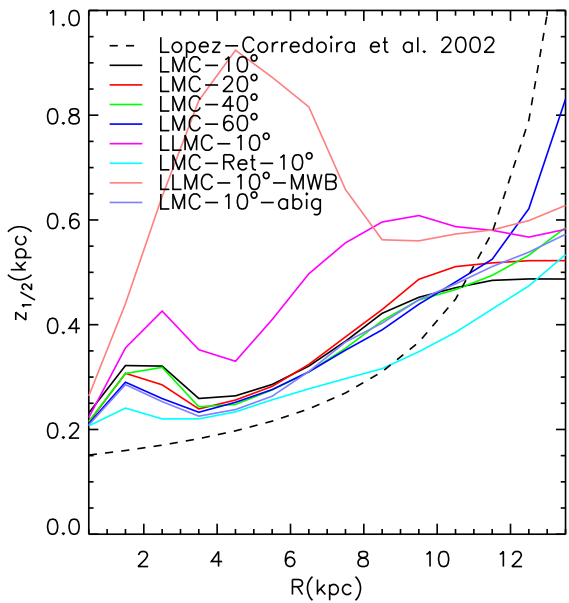}
\caption{The half mass scale height of the Milky Way disc $z_{1/2}$ as a function of $R$. The different coloured lines show selected simulations; the dashed line shows a fit to data from \citet{2002A&A...394..883L} for the Milky Way flare. The first peak in the simulated distributions is due to the bar, the secondary rise is due to the flared disc.}
\label{fig:flare}
\end{center}
\end{figure}

\begin{figure}
\begin{center}
\includegraphics[width=0.49\textwidth]{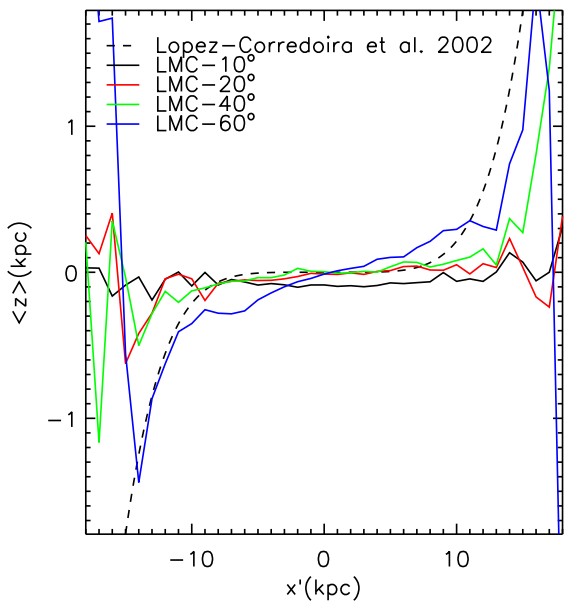}
\caption{The mean height of the disc along $x'$ -- the direction of maximal height variation. We average over $|y'|<0.5$\,kpc and all $z$. The different coloured lines show selected simulations; the dashed line shows a fit to data from \citet{2002A&A...394..883L} for the Milky Way warp.}
\label{fig:warp}
\end{center}
\end{figure}

Figures \ref{fig:flare}, \ref{fig:warp} and the left-most panels of Figures \ref{fig:impact1}, \ref{fig:orbchange1} and \ref{fig:satchange1} show the effect of the mergers on the underlying Milky Way disc. These results should be treated with some caution since we do not know what the Milky Way disc looked like at $z=1$ and we have assumed near-present day properties for our initial conditions. Nonetheless, the mergers produce a wealth of interesting substructures in the disc that match observations of our own Galaxy, and are long-lived (recall that these simulations have been evolved for $\sim 5$\,Gyrs or longer). 

Figure \ref{fig:flare} shows the half mass scale height of the Milky Way disc $z_{1/2}$ as a function of $R$. The different coloured lines show selected simulations; the dashed line shows a fit to data from \citet{2002A&A...394..883L} for the Milky Way flare: 

\begin{equation}
z_{1/2}(R) = 3.6\times 10^{-2}R_\odot\exp\left(\frac{R-R_\odot}{12 - 0.6R(\mathrm{kpc})\mathrm{kpc}}\right) 
\end{equation}
with $R_\odot = 7.9$\,kpc. 

The first peak in the distribution is due to the bar, the secondary rise is due to the flared disc. Increasing the impact angle makes very little difference to the size of the bar in scale length or height and in all cases we obtain $z_{1/2}\sim0.3$\,kpc; $R\sim 2$\,kpc. (Changing the satellite orbit also makes almost no discernible difference and we omit these results for clarity.) Increasing the satellite mass increases the bar size by only a modest amount ($R\sim2.5$\,kpc). This is interesting because the Milky Way bar appears to be longer than this. \citet{2007AJ....133..154L} have recently confirmed the existence of the Milky Way bar using 2MASS star counts. They find dimensions of $3.9 \times 1.2 \times 0.2$\,kpc. Similarly, \citet{2002MNRAS.330..591B} have recently found a bar length of 3.5\,kpc. The only bar we form that is this long is in \LLMCMWB\ ($R\sim4$\,kpc). This bar forms as a result of secular evolution in the disc and is not driven by the merger. It is also significantly fatter ($z_{1/2} = 0.9$\,kpc) than the Milky Way bar, though many of these stars would likely be classified as bulge, rather than bar stars\footnote{The half mass scale height for the Milky Way bulge quoted in Table \ref{tab:milkyway} is 0.25\,kpc which seems much smaller than the 0.9\,kpc for the secular bugle formed in \LLMCMWB. However, the difference is not so acute. The data for the Milky Way bulge probe only the central $\sim 1$\,kpc, where our secular bulge has a comparable $z_{1/2}\sim0.3$\,kpc.}. A final interesting point is that the retrograde orbit \LMCRet\ hardly excites a bar at all. Such results have been seen in the literature dating back to \citet{1941ApJ....94..385H}, who presented the first study of prograde versus retrograde interactions. \citet{1972ApJ...178..623T} attribute the phenomenon to increased resonance in the prograde case, but one can also think of it in terms of increased tidal forces for the prograde interaction (see e.g. \bcite{2006MNRAS.366..429R}).  

The flare and warp are much stronger indicators of merger activity than the bar. From Figure \ref{fig:flare}, we can see that the strength of the flare grows with satellite impact angle and is not affected by any other parameter -- particularly at large radii ($R>10$\,kpc). \LMCsix\ provides the best match to the observed Milky Way flare. This is encouraging since such mergers should be twice as likely as the low inclination mergers that produce the accreted thick disc.  

In Figure \ref{fig:warp} we show the mean height of the disc along $x'$ -- the direction of maximal height variation; the dashed line shows a fit to data from \citet{2002A&A...394..883L} for the Milky Way warp: 
\begin{equation}
<z> = 1.2\times10^{-6}R(\mathrm{kpc})^{5.25}\sin(\phi+5^\mathrm{o})\mathrm{kpc}
\end{equation}
We average over $|y'|<0.5$\,kpc and all $z$. As with the flare, only \LMCsix\ provides a good match to the observed Milky Way warp. 

\citet{2006astro.ph..9554L} have recently suggested that the Milky Way warp is a result of resonant driving by the LMC, but it is intriguing that high impact mergers can produce similar effects, while simultaneously providing a good match to the Milky Way flare. A natural candidate for producing the warp is the Sagittarius dwarf which recently fell in on a polar orbit (\bcite{2003ApJ...583L..79B}; \bcite{2005nfcd.conf..207B}). However, since such merger produced warps are long-lived, earlier mergers could have been more important. 


The longevity of the warp is perhaps surprising. \citet{1995MNRAS.275..897N} calculated that dynamical friction ought to rapidly damp such warps away (see also \bcite{1998MNRAS.297.1237B}). However, \citet{2006MNRAS.370....2S} show that this does not occur in practice because the inner halo moves with the disc. They found, like us, long lived warps that survive over several Gyrs (see also \bcite{1999ApJ...513L.107D} and \bcite{2007iuse.book...67B} for a review). 

Finally, from Figure \ref{fig:flare} notice that the two largest mergers (\LLMC and \LLMCMWB; $\vmaxd \sim 80$\,km/s) heat the thin disc enough that it resembles the thick disc of the Milky Way ($z_{1/2}(R_\odot) \sim 0.6$\,kpc). None of the lower mass satellites produce enough vertical heating. We discuss the contribution of these heated thin disc stars to the thick disc in \S\ref{sec:thickdisc}. 

Our results for the response of the Milky Way disc agree very well with other recent studies in the literature. \citet{2008arXiv0802.3997D} and \citet{2006ApJ...653.1180G} merge 100 satellites with a simulated M31 galaxy. They find, like us, a warp, bar and flare that are excited by LMC-mass interactions. \citet{2007arXiv0708.1949K} also find a warp, bar and flare form naturally as a result of disc-satellite interactions in a $\Lambda$CDM cosmology. Both of these studies are complementary to ours. \citet{2008arXiv0802.3997D} and \citet{2006ApJ...653.1180G} consider the effect of many mergers that occur simultaneously, \citet{2007arXiv0708.1949K} consider the cumulative effect of many interactions that each pass the disc only once, and we follow single interactions over many orbits until they merge. An important difference between \citet{2008arXiv0802.3997D} and \citet{2006ApJ...653.1180G} and our study is that they merge the {\it present surviving} satellite distribution with M31 (this is also the case for most previous studies, e.g. \bcite{2001ApJ...563L...1F}). \citet{2007arXiv0708.1949K}, like us, use the distribution of accreted satellites that is systematically more massive and more destructive. As a result, these earlier studies underestimate the vertical heating of the disc. A second important difference is that \citet{2008arXiv0802.3997D} and \citet{2006ApJ...653.1180G} do not account for the larger group environment around the subhalos. They find that dynamical friction is largely unimportant, as in our \LMCabig\ simulation. Once the larger group environment is taken into account, dynamical friction does play an important role and allows massive subhalos to successfully merge with the disc at high redshift. Finally our results for retrograde mergers agree well with a study by \citet{1999MNRAS.304..254V}. They found, like us, that retrograde interactions give very little disc heating. This demonstrates that the heating is a result of resonant driving from the satellite that is strongly suppressed in retrograde interactions. 

\subsubsection{Accreted versus heated thick discs}\label{sec:thickdisc}

Combining the results from the previous sections, we find that an accreted thick disc of stars forms for impact angles $\simlt 20^\mathrm{o}$. For an impact angle of $10^\mathrm{o} <\theta <20^\mathrm{o}$, an eccentricity of $0.36 < e < 0.8$, and a satellite mass of $70-90$\,km/s, the accreted material has almost all of the observational properties of the Galactic thick disc at the solar neighbourhood ($8 < R < 9$\,kpc). For our reference simulation, \LMCten, we find a thick disc that lags the thin disc rotation by $\sim 20$km/s, is hotter with $\sigma_z \sim 40$\,km/s, compared with $\sigma_z \sim 20$\,km/s for the thin disc, and has a larger scale height ($z_{1/2} = 0.7$\,kpc) and longer scale length ($R_{1/2} = 7.3$\,kpc; $R_0 = 4.3$\,kpc) than the thin disc. Increasing $\theta$ increases $\sigma_z$ and the rotation lag; reducing $e$ reduces $\sigma_R$ and the rotation lag; and increasing the satellite mass increases the velocity dispersion of the thick disc in all directions. 

However, as in previous studies (e.g. \bcite{1996ApJ...460..121W}), our accreted thick discs are significantly less massive than that of the Milky Way. In \LMCten, we find $\rhothick/\rhothin = 0.4\%$ (assuming the {\it observed value} of $\rhothin = 0.09$\,M$_\odot$\,pc$^{-3}$), which is too low by a factor $\sim 30$ (see Table \ref{tab:milkyway}). Our other runs produce similar numbers: \LLMC\ gives the most massive thick disc, with $\rhothick/\rhothin = 1.3\%$, but even this is too small by a factor $\sim 10$. In fact, these normalisations are a much closer match to the local $\rhohalo/\rhothin = 0.5$\% \citep{2008ApJ...673..864J}. 

\begin{figure}
\begin{center}
\includegraphics[width=0.49\textwidth]{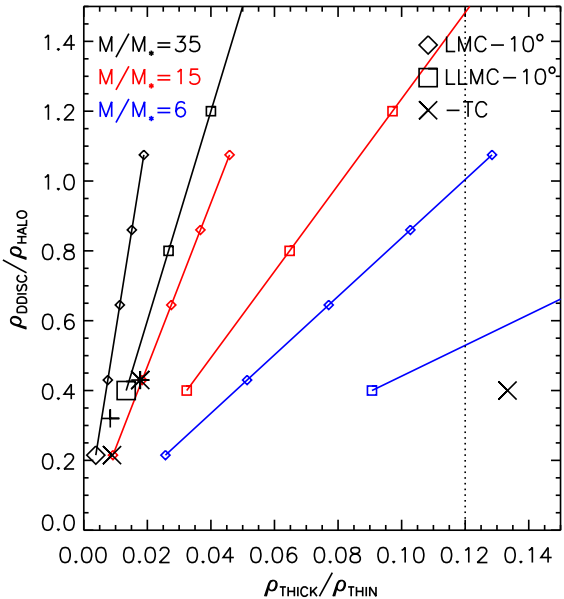}
\caption{Current constraints on accreted stellar and dark discs at the solar neighbourhood. The plot shows the dark disc to dark halo ratio $\rhodd/\rhoh$ as a function of the thick to thin disc density ratio $\rhothick/\rhothin$; we assume for this plot $\rhoh = 0.01$\,M$_\odot$\,pc$^{-3}$ and $\rhothin = 0.09$\,M$_\odot$\,pc$^{-3}$. Results are extrapolated from two simulations: \LMCten\ and \LLMC, marked by the diamond and square, respectively. The star and plus show two additional simulations: \LLMCMWB\ and \LLMCBMWB, respectively. The different colours extrapolate to lower mass to stellar mass ratio, $M/M_*$. The solid lines extrapolate to multiple mergers of identical satellites. The X's show the effect of including heated thin disc stars in the thick disc for \LMCten\ (60\%) and \LLMC\ (90\%; see text for details). The vertical dotted black line marks the observed $\rhothick/\rhothin$ at the solar neighbourhood.}
\label{fig:siglimits}
\end{center}
\end{figure}

The thick disc density normalisation $\rhothick/\rhothin$ can be increased by: (i) increasing the satellite stellar mass; (ii) integrating over many near disc plane mergers; and (iii) including heated thin disc stars in the thick disc. A possible fourth mechanism for driving up the mass at the solar position would be to put the stars and dark matter in a ring, rather than a disc. In practice this does not work because dynamical friction drives the satellite inwards as fast as it loses mass and the resulting density at the solar neighbourhood remains low. (We tested this explicitly using test simulations where the satellite was placed on a circular orbit; these are omitted for brevity.)

\begin{figure}
\begin{center}
\includegraphics[height=0.49\textwidth]{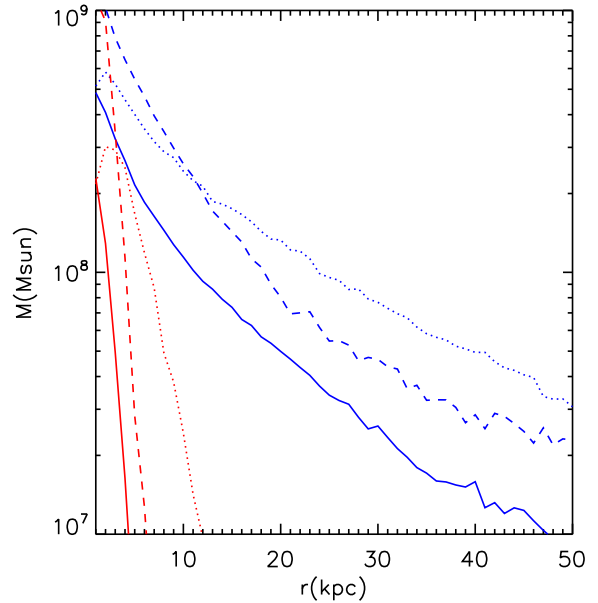}
\caption{The mass in stars (red) and dark matter (blue) that contribute to the thick discs interior to $|z|<1.1$\,kpc, as a function of their initial radius within the satellite (in 1\,kpc bins) for \LMCten\ (solid lines); \LLMC\ (dotted lines); and \LLMCBMWB\ (dashed lines).}
\label{fig:georgeplot}
\end{center}
\end{figure}

In Figure \ref{fig:siglimits}, we summarise each of the effects (i) to (iii), above by extrapolating the results for \LMCten\ and \LLMC. Our extrapolations were performed in the following way: 
\\\\
{\it (i) Increasing the satellite stellar mass}\\
\noindent
We increase the satellite stellar mass while holding the dark matter mass fixed. This is shown by the black red and blue lines in Figure \ref{fig:siglimits} which are for $M/M_* = 35,15,6$, respectively. 

To test the extrapolation, we ran an additional simulation \LLMCBMWB\ (shown by the star in Figure \ref{fig:siglimits}). This was identical to \LLMCMWB\ (shown by the plus) except that the satellite had twice the stellar mass. The mass in accreted stars doubled, as expected. However, the mass in the dark disc also increased. The reason for this is given in Figure \ref{fig:georgeplot}. This shows the mass in stars (red) and dark matter (blue) that contribute to the thick discs interior to $|z|<1.1$\,kpc, as a function of their initial radius within the satellite for \LMCten\ (solid lines); \LLMC\ (dotted lines); and \LLMCBMWB\ (dashed lines). Most of the mass that contributes originates from the central 10\,kpc in the satellite. This is the majority of the stars, but only a fraction of the dark matter. For \LLMCBMWB, the increased disc mass caused the central satellite dark matter to adiabatically contract, causing more dark matter to end up in the thick disc too. This test demonstrates that it is the satellite mass interior to $\sim 10$\,kpc -- both stars and dark matter -- that contributes to the thick disc. It also explains why the mass to light ratio of our accreted thick discs is lower than that of the satellites from which they form.\\\\
{\it (ii) Multiple mergers of like satellites}\\
\noindent
The solid lines in Figure \ref{fig:siglimits} show the effect of multiple mergers of like satellites. We assume that the accreted material simply adds to that already present. From Figure \ref{fig:cosmoresults}, left panel, we can see that cosmic variance gives an upper bound of $\sim 10$ mergers above LMC mass. However, for multiple mergers to contribute to the thick disc, they must all be low inclination and prograde. Such a scenario seems unlikely unless accretion occurs along filaments and the disc is favourably aligned.  
\\\\
{\it (iii) The effect of heated thin disc stars}\\
\noindent
For our most massive mergers -- \LLMC\ and \LLMCMWB\ -- the originally thin disc is heated sufficiently that it resembles the Milky Way thick disc at the solar neighbourhood: $\sigma_R \sim 80$\,km/s; $\sigma_\phi \sim 60$\,km/s; $\sigma_z \sim 40$\,km/s; rotation lag $\sim 20$\,km/s; and scale height $z_{1/2}\sim 0.6$\,kpc. 

Here we consider the effect of adding these heated thin disc stars to the thick disc, both for the most massive mergers, where almost all of the stars can contribute, and for the less massive merger \LMCten. We populate the thick disc with thin disc stars at the solar neighbourhood by comparing thick and thin disc distribution functions for each star: 

\begin{equation}
f = A e^{-\frac{v_r^2}{2\sigma_r^2}}e^{-\frac{(v_\phi-<v_\phi>)^2}{2\sigma_\phi^2}}e^{-\frac{v_z^2}{2\sigma_z^2}} e^{-\frac{z^2}{z_e^2}}
\end{equation}
with $\sigma_r = 85$\,km/s, $\sigma_\phi = 60$\,km/s, $\sigma_z = 40$\,km/s, $<v_\phi> = 183$\,km/s, and $z_e = 1.48$\,kpc for the thick disc; and $\sigma_r = 60$\,km/s, $\sigma_\phi = 40$\,km/s, $\sigma_z = 19$\,km/s, $<v_\phi> = 200$\,km/s, and $z_e = 0.85$\,kpc for the thin disc. These choices are motivated by our results in \S\ref{sec:impact}, but our results are not sensitive to these parameters. We assign thin disc stars to the thick disc if $f_\mathrm{THICK}/f_\mathrm{THIN} > 1$ for that star. We then vary the fraction of thin disc contaminants by varying the relative normalisation of each distribution, $A$. 

For our reference simulation, \LMCten, we find that we can populate up to 60\% of the thick disc with hot thin disc stars without significantly reducing the thick disc scale height. However, greater contamination than this leads to a thick disc that is too cold vertically. For \LLMC, the more massive satellite heats the thin disc enough that it resembles a thick disc. A decomposition is still required, however, since the heated thin disc at the end of the simulation does not show enough rotation lag (see Figure \ref{fig:satchange1}(d)). We find that for this simulation 90\% contamination gives a good qualitative match to the Milky Way thick disc and recovers the full observed thick disc mass. We show the effect of 60\% thin disc contamination for \LMCten\ and 90\% for \LLMC\ by the X's in Figure \ref{fig:siglimits}.

From Figure \ref{fig:siglimits}, we can see that while multiple mergers and lower $M/M_*$ help to increase the thick disc mass, heated thin disc stars due to a very massive merger ($\vmaxd \sim80$\,km/s) seem essential for reaching the high thick disc mass observed in the Milky Way. Accreted stars likely contribute $10-50$\% of the observed thick disc. This is in reasonable agreement with the recent cosmological simulation of \citet{2003ApJ...597...21A}, where they find a thick disc that is equally composed of heated thin disc stars and accreted material. \citet{Villalobos:2008rw}, also find that accreted stars can make up only a small fraction of the thick disc in the disc plane. 

The above seems at odds with the recent extra-galactic observations of FGC1415 \citep{2005ApJ...624..701Y}, which shows a massive counter-rotating thick disc. A possible solution is that FGC1415 had a rather more massive (and more rare) retrograde merger than those studied here.

In principle, heated and accreted stars should be separable by their chemistry. The high $\rhothick/\rhothin$ obtained at the solar neighbourhood comes from a fit to star counts for stars that are well separated in colour, but for which there is currently no spectroscopic information \citep{2008ApJ...673..864J}. But this work will be difficult. State-of-the-art spectroscopic studies have too few stars ($\sim 200$) to obtain an accurate density distribution \citep{2007ApJ...663L..13B}. 

As a final note for this section, recall that the Milky Way thick disc has stellar mass $\sim10^{10}$\,M$_\odot$, while the stellar halo is $\sim10^9$\,M$_\odot$ (see Table \ref{tab:milkyway}). If we put too many accreted stars into the thick disc then we have a problem: the stellar halo will become too massive because there will be two high inclination mergers for every merger near the disc plane. This suggests that either the Milky Way had its most massive merger near the disc plane, or -- which is more likely -- a significant fraction of the Milky Way thick disc comprises heated thin disc stars (or the thick disc formed through some other mechanism).

\subsubsection{A dark matter thick disc}\label{sec:darkdisc}

A key new idea presented in this work is that near-disc accretion -- that must occur in a $\Lambda$CDM cosmology -- leads to the formation of a dark matter disc, as well as an accreted stellar thick disc. We can see this in the left three panels of Figure \ref{fig:impact1} (blue lines). For impact angles $< 20^\mathrm{o}$, a dark disc forms. For \LMCten\ it has scale length and height given by:  $R_{1/2} = 11.7$\,kpc, $z_{1/2} = 1.5$\,kpc, while its density at the solar neighbourhood is $\rhodd(R_\odot) = 0.22\rhoh$ (assuming $\rhoh = 0.01$\,M$_\odot$\,pc$^{-3}$). For \LLMC\ it has $z_{1/2} = 1.7$\,kpc and $\rhodd(R_\odot) = 0.42\rhoh$. At these low densities, the dark disc is not dynamically interesting in the disc plane. Integrating over the total expected mergers brings the density up, but it is hard for it to exceed the density of the dark halo at the solar neighbourhood without having an unrealistically high number of mergers (see \S \ref{sec:cosmology}). The extrapolated constraints for \LMCten\ and \LLMC\ are shown in Figure \ref{fig:siglimits}. 

We can obtain an upper bound on $\rhodd$ for $|z|<1.1$\,kpc from the kinematics of stars at the solar neighbourhood. \citet{2000MNRAS.313..209H} find a local dynamical mass density of $0.102\pm 0.01$M$_\odot$\,pc$^{-3}$ with $0.095$\,M$_\odot$\,pc$^{-3}$ in visible disc matter. This leaves room for a maximal dark density of $0.008$\,M$_\odot$\,pc$^{-3}$ which seems consistent with a dark halo and no dynamically significant dark disc. However, \citet{1989ApJ...344..217S} have argued that model degeneracies in this analysis could lead to systematic errors as large as 30\%. If we take this more generous limit, we can have a maximal density of $0.038$\,M$_\odot$\,pc$^{-3}$, which leaves room for a dynamically significant dark disc up to $\rhodd(R_\odot) = 3\rhoh$. These limits will improve dramatically with the improved data from large kinematic surveys like RAVE and GAIA. 

Interestingly, \citet{2003ApJ...588..805K} and \citet{2007A&A...469..511K} have recently argued for a dark disc in the Milky Way to explain the observed flaring in the HI gas. The dark disc they require has a scale height and length of $z_{1/2} = 2.8$\,kpc, $R_{1/2} = 12.6$\,kpc and total mass $\sim 2\times 10^{11}$\,M$_\odot$. For \LLMC, our dark disc has total mass $8.6\times 10^{9}$ for $|z|<1.1$\,kpc and $3\times 10^{10}$\,M$_\odot$ for $|z|<5$\,kpc; theirs has $4-6$ times more mass than this within the same $z$-limits. While the half mass scale length and height of our dark disc match well those of \citet{2007A&A...469..511K}, our disc appears significantly less massive.

\begin{figure*}
\begin{center}
\includegraphics[height=0.32\textwidth]{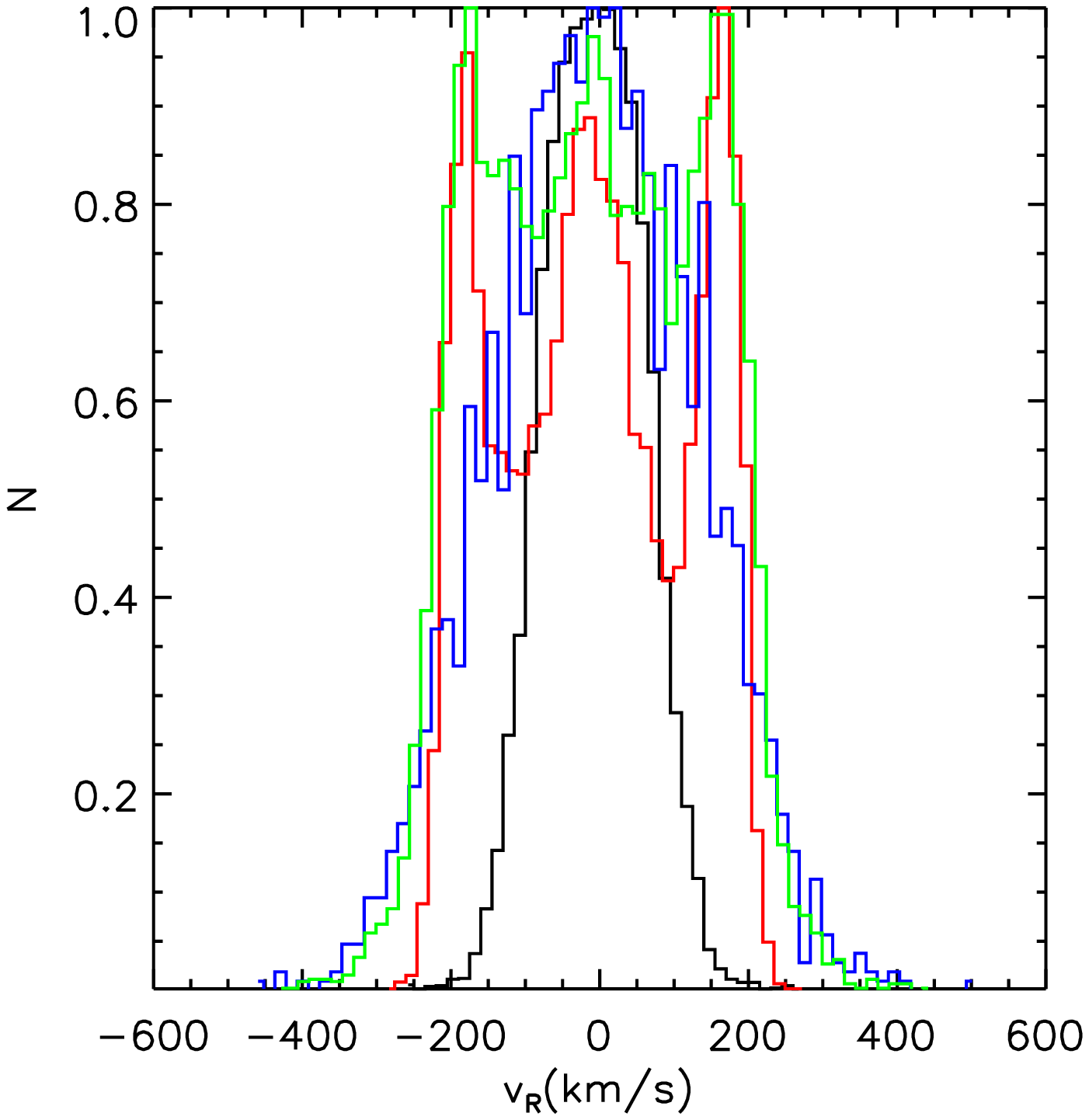}
\includegraphics[height=0.32\textwidth]{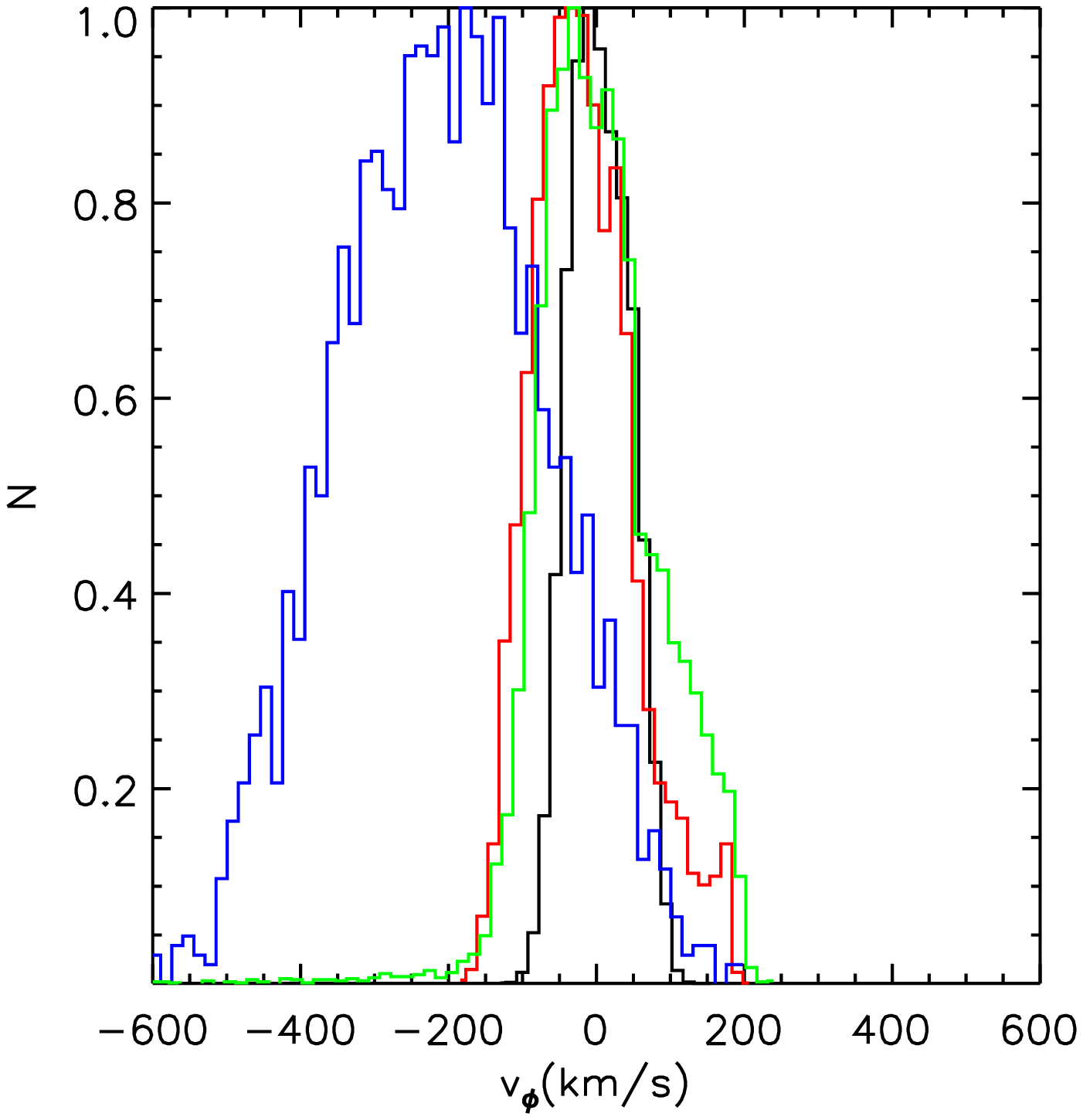}
\includegraphics[height=0.32\textwidth]{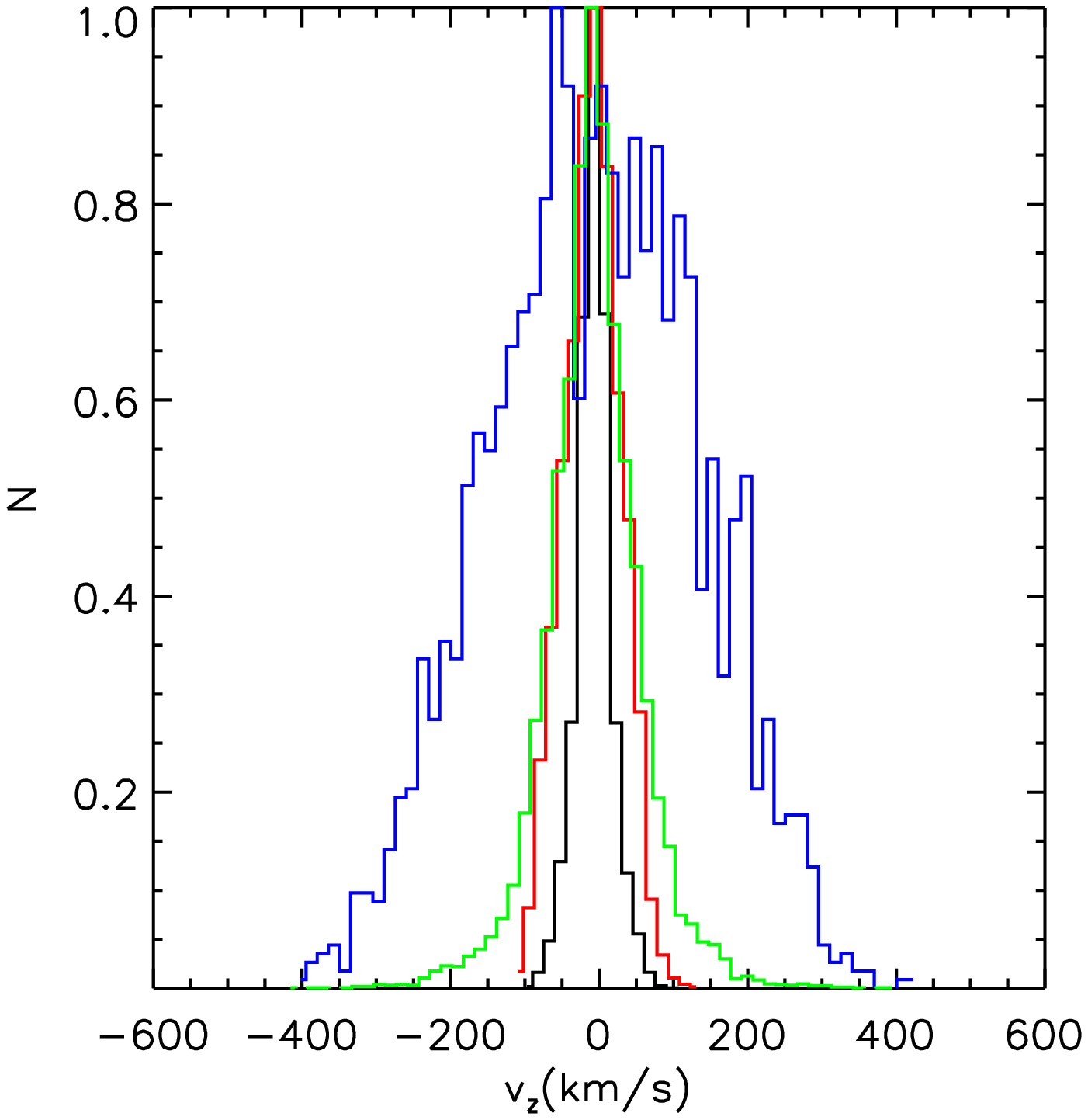}
\caption{Distribution functions for the thin disc (black), thick disc (red), dark disc (green) and dark halo (blue) for \LMCten at the solar neighbourhood ($8 <R <9$\,kpc; $|z|<0.35$\,kpc); we plot results from the {\it Earth frame} so that the dark halo appears to rotate, while the discs are stationary. The distributions are normalised to peak at 1 and do not represent the mass in each component. Notice that the dark disc has very similar kinematics to the stellar thick disc, except in that it has slightly larger $\sigma_z$. Similar results were found for our other runs. The radial velocity distribution is triple-peaked because the stars and dark matter are not fully phase mixed yet, even after 6.5\,Gyrs.}
\label{fig:distributions}
\end{center}
\end{figure*}

Even if dynamically uninteresting, the dark disc has important implications for the direct detection of dark matter because it rotates much more slowly with respect to the Earth than the Milky Way halo. The solar neighbourhood distribution functions for the thin disc, thick disc, dark disc and dark halo in \LMCten\ are shown in Figure \ref{fig:distributions}; we plot results from the {\it Earth frame} so that the dark halo appears to rotate, while the discs are stationary. Notice that the dark disc has very similar kinematics to the stellar thick disc. If the thick disc of the Milky Way is mostly accreted rather than heated stars, then this is useful. It means that we can obtain a good measure of the kinematics of the dark disc from observations of the stellar thick disc at the solar neighbourhood. 

The effect of the dark disc on the dark matter detection rate, the annual modulation signal, and on the dark matter capture rate by the Sun and Earth will be quantified in detail in forthcoming papers (\bcite{2008arXiv0804.2896B}; Bruch, Read, Baudis and Lake in prep. 2008).

\section{Conclusions}\label{sec:conclusions}

We have presented a detailed study of the effect of the Milky Way stellar disc on merging satellites in a $\Lambda$CDM cosmology. We have used both cosmological dark matter simulations of structure formation to assess the likelihood of near-disc plane accretion events, and collisionless simulations of satellite mergers to quantify the final state of the accreted material, and the effect on the Milky Way stellar disc. Our main findings are as follows: 

\begin{enumerate}

\item On average, a Milky Way-sized galaxy has 3 subhalos with peak circular speed $\vmaxd>80$\,km/s; 7 with $\vmaxd>60$\,km/s; and 15 with $\vmaxd>40$\,km/s merge at redshift $z \simgt 1$. Assuming isotropy, a third of these merge at an impact angle $\theta <20^\mathrm{o}$ and are dragged into the disc plane by dynamical friction. 

\item For an impact angle of $10^\mathrm{o}<\theta<20^\mathrm{o}$, an eccentricity of $0.36 < e < 0.8$, and $\vmaxd = 60-80$\,km/s, the accreted material has almost all of the observational properties of the Galactic thick disc at the solar neighbourhood ($8 < R < 9$\,kpc), but the resultant accreted thick disc is under-massive by a factor $\sim 2-10$. 

\item The most massive mergers at high redshift -- $\vmaxd \simgt 80$\,km/s -- heat the thin disc enough to produce a heated thick disc; none of the lower mass mergers provide enough heating. The heated thick disc is essential for obtaining a thick disc as massive as that observed in the Milky Way and likely contributes $\sim 50-90$\% of the thick disc stars at the solar neighbourhood. 

\item A key new point we make is that low inclination mergers -- that must occur in a $\Lambda$CDM cosmology -- also give rise to a thick disc of {\it dark matter}. The dark disc is of longer scale length and height than the accreted stellar thick disc ($R_{1/2} \sim 12$\,kpc, $z_{1/2} \sim 1.5$\,kpc), and provides $\sim 0.25-1$ times the density of the dark halo within $|z| < 1.1$\,kpc at the solar neighbourhood. The precise number depends on the satellite properties and the number of mergers. At all but the highest of these densities, the dark disc is not likely to be dynamically interesting. However, the dark disc does have important implications for the direct detection of dark matter because of its low velocity with respect to the Earth. A full quantitative analysis is presented in a companion paper \citep{2008arXiv0804.2896B}. 

\item Higher inclination encounters $\theta > 20^\mathrm{o}$ are twice as likely as low inclination ones. These lead to structures that are hotter ($\sigma_z > 100$\,km/s), rotate more slowly ($v_c < 100$\,km/s), extend $10-20$\,kpc above the disc, and closely resemble the inner/outer stellar halos recently discovered by \citet{2007arXiv0706.3005C}. The integrated light from such encounters is of order the total mass in the Milky Way stellar halo.

\item We quantify the effect these mergers have on the Milky Way disc. All encounters excite a bar, flare and warp in the disc that are long-lived. The bar is a poor indicator of merger activity because a bar formed by secular evolution (due to a more massive stellar disc) is longer and provides a better fit to the Milky Way bar than those induced by mergers. By contrast, the flare and warp are strong indicators of merger activity. Their strength grows with satellite impact angle and is not strongly affected by any other parameter -- particularly at large radii ($R>10$\,kpc). Of the simulations that we consider, a $\vmaxd = 60$\,km/s merger at $\theta = 60^\mathrm{o}$ to the Milky Way disc plane provides the best match to the observed Milky Way warp and flare.

\item Finally, we show that even $\vmaxd = 70$\,km/s galaxies have dynamical friction times longer than the age of the Universe once mass loss due to tides is taken into account. In practice, in our cosmological simulations, they merge much faster than this because they `ride in' inside loosely bound groups that are ten times more massive. This should be taken into account when calculating the orbits of Local Group satellite galaxies such as the Large and Small Magellanic Clouds, and the Sagittarius dwarf galaxy.

\end{enumerate}

\section{Acknowledgements}
We would like to thank Joachim Stadel and Jonathan Coles for very useful discussions. Thanks go to Doug Potter for making the zBox2 -- the computer on which almost all the simulations presented here were run -- fly. One simulation used the Arctic Region Supercomputer Center, for which we are grateful. We would like to thank Juerg Diemand for making his simulation output available to us. Finally, we would like to thank Chris Brook and Juerg Diemand for useful comments. 

\bibliographystyle{mn2e}
\bibliography{/Users/justinread/Documents/LaTeX/BibTeX/refs}
 
\end{document}